\documentclass{article}
\usepackage{blindtext}
\usepackage[a4paper,
            bindingoffset=0.2in,
            left=1in,
            right=1in,
            top=1in,
            bottom=1in,
            footskip=.25in]{geometry}

\usepackage[T2A]{fontenc}
\usepackage{graphicx}
\usepackage{makecell} 
\usepackage{pdfpages}
\usepackage{subcaption}
\usepackage{dcolumn}
\usepackage{bm}
\usepackage{amsmath} 
\usepackage{verbatim}
\usepackage{multirow}
\usepackage{tabularray}
\usepackage{threeparttable}
\usepackage{tablefootnote}
\usepackage{tcolorbox}
\usepackage{amsmath}
\usepackage{subfiles} 

\usepackage{float}
\usepackage[noend,boxed,linesnumbered,algonl]{algorithm2e}

\SetKwProg{Fn}{Function}{}{end}
\usepackage{algpseudocode}
\title{algorithm}

\usepackage{setspace}
\usepackage{soul}

\usepackage{pdfpages}  
\usepackage{pgffor}
\makeatletter
\AtBeginDocument{\let\LS@rot\@undefined}
\makeatother

\usepackage{amsmath}

\usepackage[sorting=none, style=numeric-comp]{biblatex}
\usepackage{authblk}
\title{PINNs-MPF: A Physics-Informed Neural Network Framework for Multi-Phase-Field Simulation of Interface Dynamics} 
\author[*]{Seifallah Elfetni} \author[**]{Reza Darvishi Kamachali} 
\affil[ ]{Federal Institute for Materials Research and Testing (BAM), Unter den Eichen 87, 12205 Berlin, Germany} 
\affil[*]{seifallah.elfetni@bam.de; $^{**}$reza.kamachali@bam.de}

\date{}
\begin{document}

\maketitle
\begin{abstract}
We present an application of Physics-Informed Neural Networks (PINNs) to handle MultiPhase-Field (MPF) simulations of microstructure evolution. To handle such problems, it has been showcased that a combination of optimization techniques extended and adapted from the PINNs literature, as well as the introduction of specific techniques inspired by the MPF Method background, is required.
The numerical resolution is realized through a multi-variable time-series problem by using fully discrete resolution. 
Within each interval, space, time, and phases/grains are treated separately, constituting discrete subdomains. 
An extended multi-networking concept is implemented to subdivide the simulation domain into multiple batches, with each batch associated with an independent Neural Network (NN) trained to predict the solution. 
To ensure efficient interaction across different phases/grains and in the spatio-temporal-phasic subdomain, a Master NN handles efficient interaction among the multiple networks, as well as the transfer of learning in different directions. 
A set of systematic simulations with increasing complexity was performed, that benchmarks various critical aspects of MPF simulations, including different geometries, types of interface dynamics and the evolution of an interfacial triple junction.
A comprehensive approach is adopted to specifically focus the attention on the interfacial regions through an automatic and dynamic meshing process (introduced as an adaptive Mesh-free optimization), significantly simplifying the tuning of hyper-parameters and serving as a fundamental key for addressing MPF problems using Machine Learning. The pyramidal training approach is proposed to the PINN community as a dual-impact method: it facilitates the initialization of training when dealing with multiple networks, and it is proposed as a method to unify the solution through an extended transfer of learning.
The proposed PINNs-MPF framework successfully reproduces benchmark tests with high fidelity and Mean Squared Error (MSE) loss values ranging from 10$^{-4}$ to 10$^{-6}$ compared to ground truth solutions. \\

\textit{\textbf{Keywords}}: PINNs, Phase-field method, Parallel training, Machine learning, Neural networks, Adaptive mesh refinement

\end{abstract}

\tableofcontents

\clearpage

\section{Introduction}
\par Today's microstructure modeling has grown into an extensive branch of materials research, on the one hand, revealing the multi-physics of the concurrently evolving microstructural constituents across scales and, on the other hand, bridging these with the process, property and performance of materials.
Despite remarkable advances over the last two decades, the rapidly growing complexities of materials' chemistry and processing are constantly surpassing our actual microstructure modeling capacity, thus, decelerating materials innovation. 
A potential solution for this growing challenge is to utilize the recent developments in machine learning (ML) and deep learning (DL) for microstructure modeling \cite{choudharyrecent2022}.
The recently introduced Physics-Informed Neural Networks (PINNs) are showing promising capabilities in addressing physical phenomena \cite{cuomoscientific2022}. 
By enabling non-linear approximations of solutions for Partial Differential Equations (PDEs), the PINNs counterpart of the conventional approaches, such as phase-field, fluid dynamics and finite-element methods, would allow for a mesh-free modeling framework, capable of transfer of the learning and without the necessity of local approximations or simplifications in the underlying physics \cite{cuomoscientific2022,HAGHIGHAT2021113741,Ramuhalli1528518}.
These promote new perspectives in microstructure modeling, motivating us to pursue the current line of studies.

\par First applications of PINNs to solve 1D PDEs (e.g., the heat equation, wave equation and Burgers' equation) were quite successful, demonstrating its simplicity and efficiency \cite{raissiphysicsinformed2019,YANG2019136,ZHANG2019108850}. Since then, deeper applications have been increasingly targeted in various areas including heat transfer \cite{Sharmaheattransfer2023,ZOBEIRY2021104232}, solid mechanics \cite{HAGHIGHAT2021113741,DIAO2023116120,HENKES2022114790,GUO2023113334,bai2023introduction,S0219876223500135,GOSWAMI2020102447,khorrami2023artificial}, electromagnetic \cite{HAGHIGHAT2021113741,Ramuhalli1528518,CAO2023123622}, turbulence modeling \cite{Shirui2020,HANRAHAN2023109232}, electro-chemistry \cite{CHEN2022116918,Hofmann2023} and Energy \cite{ HAN20233450,PRIYADARSHI2024110231}.

However, as the literature on resolving PDEs using PINNs has grown, several challenges have been revealed, highlighting the necessity for optimization techniques tailored to specific physical contexts. Jagtap et al. \cite{Jagtap2020} proposed Extended Physics-Informed Neural Networks (XPINNs) to address moving boundaries. This method subdivides the domain into different mini-batches to locally compute the solution using distinct neural networks. XPINNs were tested on the one-dimensional viscous Burgers equation, employing various optimization techniques such as subdomain decomposition (in time and space), adaptive activation functions, independent hyper-parameter adjustment for each network, and a specific data processing method to characterize the average behavior of predictions along the subdomain interfaces.
Kharazmi et al. \cite{KHARAZMI2021113547} introduced hp-Variational Physics-Informed Neural Networks (hp-VPINNs), utilizing domain decomposition and h-refinement with dynamic mesh resolution adjustment to target the advection–diffusion equation (1D) and Poisson equation (1D-2D). 
Meng et al. \cite{MENG2020113250} propose a Parareal PINNs (PPINN) framework for solving time-dependent PDEs using domain decomposition. The examples include ordinary differential equations (ODEs) and a 2D nonlinear diffusion-reaction equation. They employ various optimization techniques like alternating between Adam and L-BFGS optimizers, Quasi-Monte Carlo sampling for stochastic ODEs, and principally a parallel-in-time training approach with a serial update at the subdomain interfaces using the fine PINN solutions as a parallel prediction-correction process.
Penwarden et al. \cite{PENWARDEN2023112464} further enhanced these techniques by introducing stacked-decomposition inspired by causality literature (time-causality enforcement methods  \cite{krishnapriyan2021characterizing,BIHLO2022111024}), window-sweeping, alternance of optimizers and transfer of learning  methods proposed to overcome training challenges, respect causality, and improve scalability by limiting computation per optimization iteration. The authors consider the 1D convection equation, 1D Allen-Cahn equation and 1D Korteweg–de Vries. While the reported methods have demonstrated promising results, further testing and extension are necessary to address more complex scenarios, such as two-dimensional configurations involving systems of equations, Multi-Phase Field problems and three-dimensional domains. These scenarios introduce additional challenges that require innovative approaches and optimization techniques.
\par 
A few applications of PINNs to handle diffuse interfaces were recently presented, especially to study evolution of multiple interacting phases or components \cite{ROJAS2023100450, ZHANG202364,Haghighat2022,ZHANG2023111919,AMINI2023112323,wight2020solving}. 
Zheng et al. \cite{Zheng2022} have developed a physics-constrained neural network (PCNN) to predict the sequential motions of flow simulations. Due to the challenging nature of multi-phase problems, the applied approach was based on the satisfaction of the mass conversation constraint when encoding the observation of the recurrent NN. A conservative boundedness mapping algorithm (MCBOM) was then called to correct the phase prediction, reported to be efficient for handling sequential notions in some testing benchmarks in 2D, e.g., the evolution of three phases in the shear layer and dam break problems.
Haghighat et al. \cite{Haghighat2022} proposed a dimensionless form of the PDEs combined with a sequential training strategy based on stress-split algorithms and multi-networking, applied for solving several problems in poroelasticity. It was found that both the size and structure of the NN, along with the hyperparameters of the optimizer, such as the learning rate, can significantly influence the quality of PINN solutions.
Amini et al. \cite{AMINI2023112323} introduced inverse modeling of nonisothermal multiphase poromechanics using PINNs, successfully applying the method to both single- and multi-phase flow in porous media.

\par During these remarkable advances, several challenges were also spotted in the field that remain to be addressed. A primary concern with PINNs involves the gradient pathology in the total loss function, due to larger gradients in higher-derivative terms, affecting the accuracy of the predictions \cite{Haghighat2022}.
Furthermore, ensuring systematic convergence of training becomes difficult, especially with the high computational cost of multi-networking. 
The use of a multi-networking approach can generate differing architectures or training data among separate NNs, which leads to imperfect alignment and thus discontinuities or artifacts along boundaries. 
Adjusting the size of training sets, including initial conditions, collocation points, and boundary conditions, was shown to help resolve this problem, but at the same time, this adds to the complexity and computational costs \cite{cuomoscientific2022,ZHU201956}. 
The assembly and transfer of collective learning from multiple networks (transferring the learning) is another major challenge in PINNs. 
Here tuning hyper-parameters, even with a single NN, remains challenging: Adaptive weighting, while introduced to address feature dependencies, reflects non-systematic learning of solution patterns \cite{CHEN2022116918}. 
In addition to these, the profound sensitivity of the solution to the hyper-parameters, encompassing variables such as the number of neurons, network layers, learning rate, and the composition of training datasets, emerges as a major issue that requires more attention when dealing with PINNs \cite{CHEN2022116918,ARZANI2023111768}. 
Regarding Multi-phase Field (MPF) equations, they involve a coupled system of nonlinear PDEs with multiple phases and several parameters. The presence of higher-order derivative terms and a non-convex potential term further contributes to additional non-linearities and the emergence of several local minima. Additionally, the generated solutions often require correction algorithms to satisfy conservation laws, rendering classical PINNs unable to make accurate predictions. Therefore, incorporating conservative correction algorithms and adequate optimization methods becomes necessary \cite{HUANG2021103727}. 
\par Multi-phase-field methods (MPFM) for microstructure modeling have proven to be powerful tools for capturing intricate multi-physics phenomena in multi-phase, multi-component materials \cite{chen2002phase,steinbach2009phase,steinbachphasefield2009}. Being based on a generalizable free energy functional, a key feature of the MPF framework is its capacity to be flexibly expanded to different degrees of complexity which can be specific to the physics of the problem at hand. A representative demonstration of this flexibility has been explored in dealing with polycrystalline microstructures: Starting from curvature-driven interface kinetics, the MPF framework has gradually expanded to cover the chemical \cite{GROSE2022111570}, mechanical \cite{SINGH2021107348} and even the physics of magnetic and electric contributions \cite{Huo2023}. These led to the accurate investigation of complex phenomena such as precipitation, recrystallization, and phase transformation as well as various chemo-mechanically coupled phenomena \cite{darvishikamachali2013grainPhD,kamachali2015texture,schwarze2018computationally}.
Despite successful implementations and applications \cite{kamachali2018numerical,tegeler2017parallel}, an intrinsic challenge of the MPF approach is that with any further advancement and raising scientific questions, immediate needs for extensive programming and optimization efforts are required, which often spread to deep levels of redefining key parameters and functions as well as memory- and storage-structures in the code. These originate from the fact that any update to the model necessitates changes to the underlying free energy functional and corresponding equations of motion. As a result, it becomes difficult to keep the software updates up with the pace of emerging topics in materials research, for instance, the unanswered questions regarding non-equilibrium microstructure evolution in additively manufactured materials \cite{WANG2020101538,PARSAZADEH2023101102} or the complex diffusion-interface couplings in novel high-entropy alloys \cite{QIAO2021160295,Ziyuan2022HEA}. 

\par This work presents a comprehensive framework, termed PINNs-MPF, specifically designed to tackle the challenges of solving MPF equations using PINNs. By identifying the limitations of existing PINN methods when applied to MPF equations and recognizing the additional complexities introduced by these non-linear, convex, and constrained systems, we propose a unified approach that combines and extends relevant PINN techniques with MPF method strategies.
This strategy is inspired by the core development of the OpenPhase software for MPF simulations, which has demonstrated success in studying diffuse interface problems, but aiming to overcome reported limitations, by enabling a learnable solution while efficiently handling non-linearities, to integrate more physics.
\cite{darvishikamachali2013grainPhD}.
\newline
This framework introduces key novelties through its optimization strategies: (i) A extended domain decomposition approach is implemented that uses a central coordinating network (referred to as the "Master") to manage tasks and ensure continuity among subnetworks (referred to as the "workers") during full parallel training in space, time and phases. (ii) An extreme mesh refinement approach, with a distinct focus on the phase-field interfacial zones, is applied. (iii) We simplify and reduce the hyper-parameters as elaborated in the following. (iv) A synchronization step to ensure efficient interactions between the evolving phase-fields is applied. Finally, (v) a pyramidal training approach is applied as a robust alternative to adaptive weighting and an efficient solution for transferring learning of multiple-to-multiple and single-to-multiple networks. 
Inspired by the MPF benchmark philosophy \cite{darvishikamachali2013grainPhD}, we test our PINNs across four core applications that built the MPF concept with a progressive complexity: First, we model the motion of a diffuse interface (propagation of a stable phase-field profile), second and third, we study the curvature-driven shrinkage of a grain in the presence and absence of a driving force, and fourth, we investigate the evolution of a triple junction that demonstrates the core principles of a multi-phase evolution. The results are compared against the outputs from OpenPhase and discussed in the context of our hyper-parameters and concepts implemented within our PINNs framework. 
\section{The PINNs-MPF Framework}
\par The MPF temporal evolution is governed by the set of PDEs as follows:
\begin{equation}
\label{eq:phiPunktMP}
    \dot\phi_\alpha = \frac{\mu \sigma}{N} \sum_{\beta=1}^N \left( \nabla^2 \phi_\alpha - \nabla^2 \phi_\beta + \frac{\pi^2}{2 \eta^2} (\phi_\alpha - \phi_\beta) \right)  + 
    \frac{\mu}{N} \sum_{\beta=1}^N \Delta G \left(\phi_\alpha, \phi_\beta \right),
\end{equation}
with phase-fields $\phi_\alpha \in [0, 1]$, $\mu$ the interface mobility, $\sigma$ the interface energy and $\eta$ the interface width. 
The pairwise term $\Delta G$ counts for any additional driving force (e.g. chemical, elastic, etc.) influencing the interface dynamics.
At every point in space and time, the summation of phase-fields/order parameter reads:
\begin{equation}
    \sum_{\alpha=1}^N \phi_\alpha = 1,
    \label{eq:MPF_sum}
\end{equation}
with $N$ the number of phase-fields present. 
Further details of the MPF model are given in the Methods section \ref{method_sec}.
The PINNs-MPF framework is targeted to resolve Eqs.~(\ref{eq:phiPunktMP}) and (\ref{eq:MPF_sum}) in time and space and for various initializations.
In doing so, PINNs aim to minimize a loss function \(\mathcal{L}(\theta)\) with respect to the hyper-parameters \(\theta\).
With the MPF model in hand, this is composed of several terms:
\begin{equation}
\label{theta_theori}
\theta^* = \arg\min_{\theta} \left[\mathcal{L}(\theta) \right] = \arg\min_{\theta} \left[ \lambda_{\mathrm{pde}} \mathcal{L}_{\mathrm{PDE}}(\theta) + \lambda_{\mathrm{bc}} \mathcal{L}_{\mathrm{BC}}(\theta) + \lambda_{\mathrm{ic}} \mathcal{L}_{\mathrm{IC}}(\theta) + \lambda_{\sum \phi} \mathcal{L}_{\sum \phi}(\theta)\right]
\end{equation}
where $\mathcal{L}_{\mathcal{PDE}}$ is the loss associated with enforcing the governing physics through PDEs: 
\begin{equation}
\label{eq:theta_PDE}
\mathcal{L}_{\mathrm{PDE}}(\theta) = \frac{1}{N_r} \sum_{i=1}^{N_r} \left[\boldsymbol{r}\left(\boldsymbol{x}_r^i, t_r^i, \theta\right)\right]^2
\end{equation}
in which, \textit{r} represents the residual of the PDEs or the physical loss (Eq. \ref{eq:phiPunktMP}) and ${N_r}$ the number of collocation points. 
$\mathcal{L}_{\mathcal{BC}}$ is the loss associated with enforcing boundary conditions (BC) on the simulation domain:
\begin{equation}
\label{theta_BC}
\mathcal{L}_{\mathrm{BC}}(\theta) = \frac{1}{N_{\mathrm{BC}}} \sum_{i=1}^{N_{\mathrm{BC}}} \left[\boldsymbol{u}\left(\boldsymbol{x}_{\mathrm{BC}}^i, t_{\mathrm{BC}}^i, \theta\right) - g_{\mathrm{BC}}^i\right]^2 
\end{equation}
with $\boldsymbol{u}$ representing the approximated solution at boundaries $\boldsymbol{x}_{\mathcal{B}}^i$ and time $t_{\mathcal{B}}^i$ using the neural network with parameters $\theta$ and, $g_{\mathcal{B}}^i$ is the prescribed or observed value at the corresponding boundary and time.
$\mathcal{L}_{\text{IC}}$ is the loss associated with enforcing initial conditions (IC), i.e., the spatial configuration of the phase-fields:
\begin{equation}
\label{theta_IC}
\mathcal{L}_{\mathrm{IC}}(\theta) = \frac{1}{N_{\mathrm{IC}}} \sum_{i=1}^{N_{\mathrm{IC}}} \left[\boldsymbol{u}\left(\boldsymbol{x}_{\mathrm{IC}}^i, 0, \theta\right) - h_{\mathrm{IC}}^i\right]^2 
\end{equation}
that compares the approximated solution at initial state $\boldsymbol{x}_{\text{IC}}^i$ and time $t=0$ against the prescribed or observed value at the corresponding (initial) state $h_{\text{IC}}^i$.
Finally, $\mathcal{L}_{\sum \phi}$ is the loss associated with the constraint Eq.~(\ref{eq:MPF_sum}) of the phase summation:
\begin{equation}
\label{eq:theta_sum_phi}
\mathcal{L}_{\sum \phi}(\theta) = \left| \sum_{\alpha=1}^{N} \phi_\alpha \left(\boldsymbol{x}, t, \theta\right) - 1 \right|^2
\end{equation}
where $N$ is the total number of phases in the given point. 
Unlike the other losses which are computed over a subpart of the domain, the loss over the sum constraint is computed for the entire simulation domain.

\par The coefficients $\lambda_{PDE}$, $\lambda_{BC}$, $\lambda_{IC}$ and $\lambda_{\sum \phi}$ in Eq.~(\ref{theta_theori}) are weight factors associated with different loss terms, to balance the importance of each term.
The objective of Eq.~(\ref{theta_theori}) is to find the optimal parameters $\theta^{*}$ that minimize the whole loss function, resulting in \textit{learned} neural networks that are then capable of approximating the MPF solutions of a problem with the given initialization and boundary conditions.
We note that since the MPF method is based on a free energy functional, reinforcing terms to account for the energy dissipation \cite{SECCI2024105494} could be directly added to the loss function, further elaborated in the Discussion (section \ref{discussion_part}).

\par
\begin{figure*}
    \centering
    \begin{subfigure}{\textwidth}
        \includegraphics[width=\linewidth]{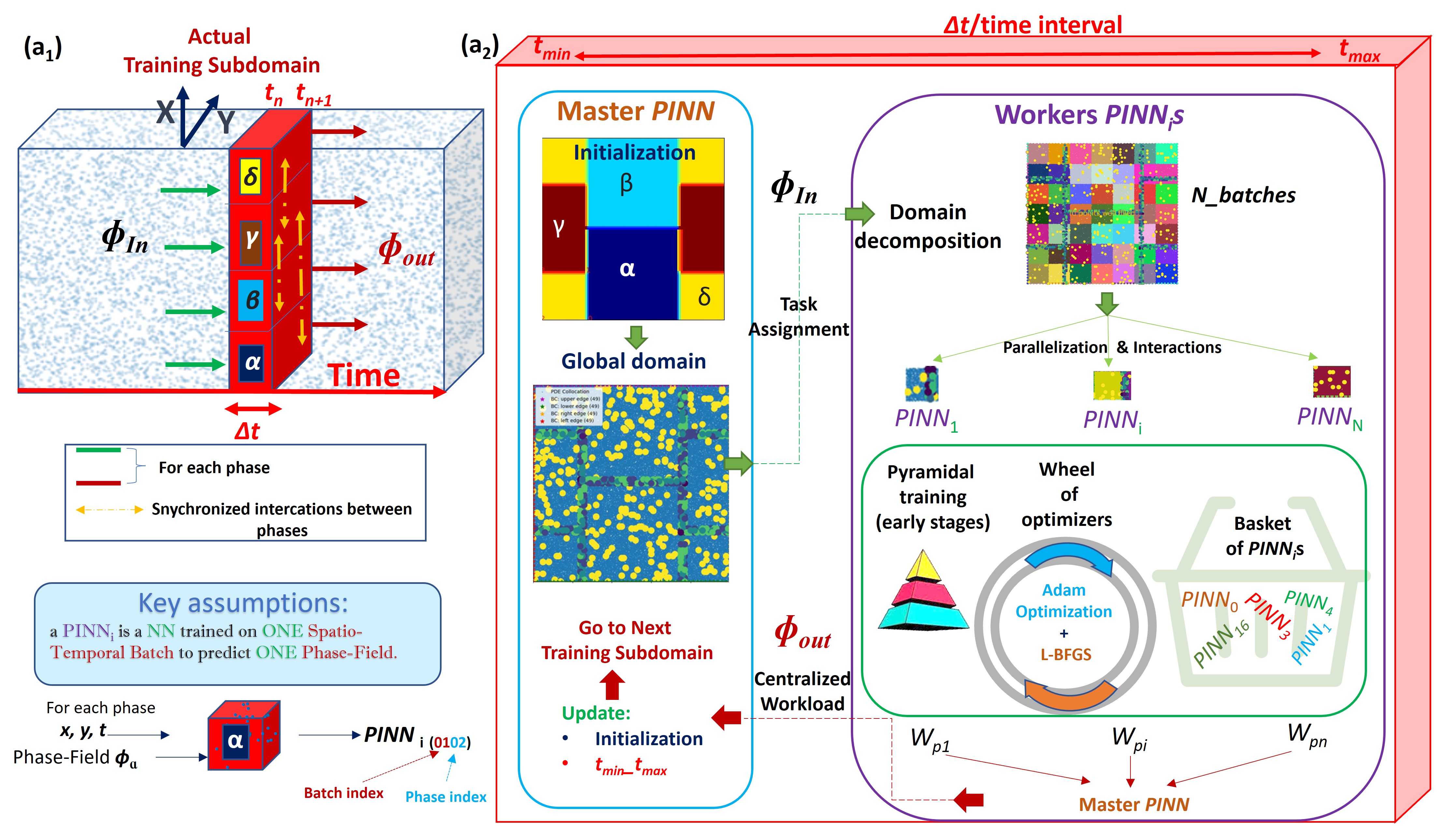}
        \caption{Global architecture of the PINNs-MPF framework. A general description of the discrete training subdomain is provided in (a$_1$), while a detailed illustration of the training process within each time interval is shown in (a$_2$).}
    \end{subfigure}
    
    \begin{subfigure}{\textwidth}
        \includegraphics[width=\linewidth]{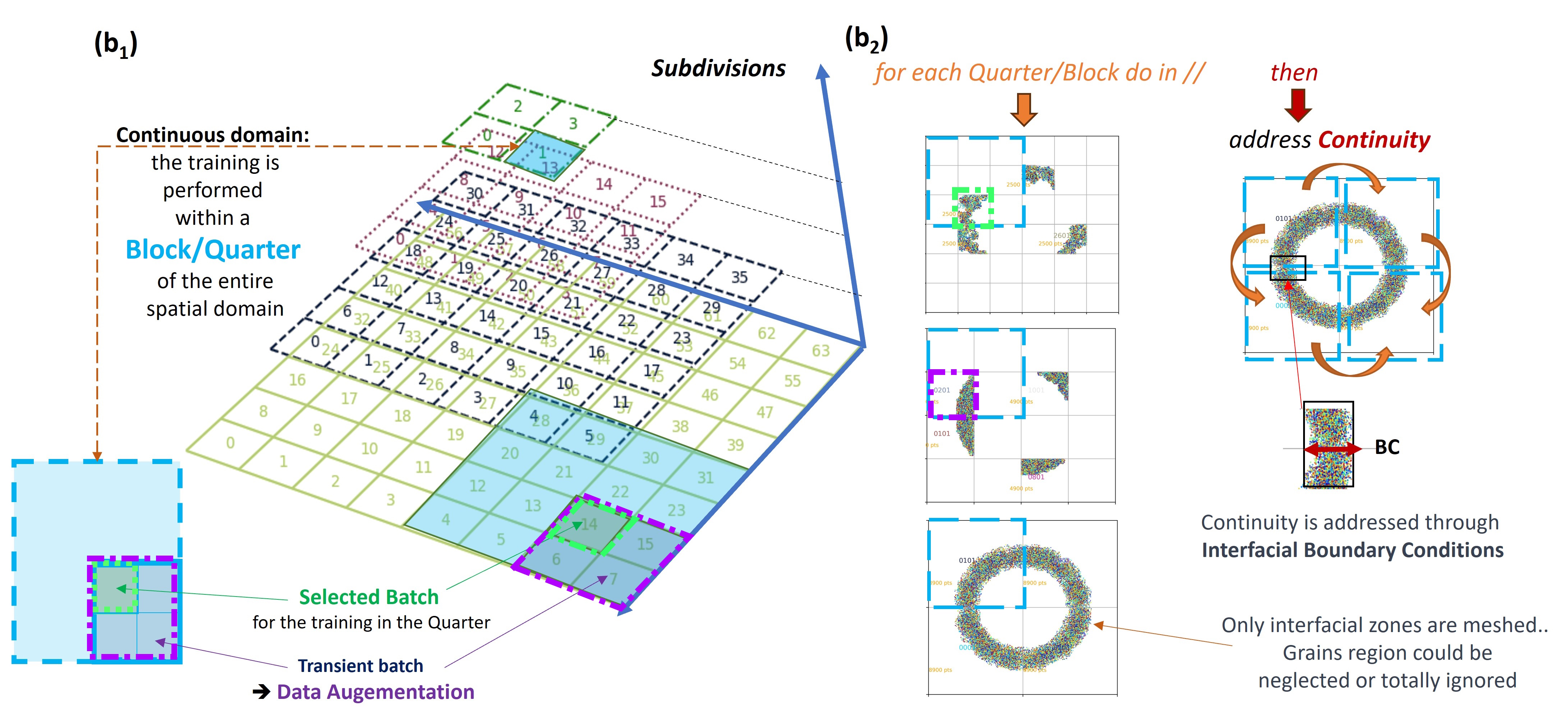}
        \caption{Pyramidal training and continuity across domains: A comprehensive description of the pyramidal transfer of learning from one subdivision level to another is depicted in (b$_1$), while an illustrative example of this hierarchical approach, followed by addressing continuity between domains, is provided in (b$_2$)}
    \end{subfigure}
    
     \caption{The architecture of PINNs-MPF framework. Panel (b) adds some details to panel (a). The pyramidal training enables continuity and multiple-to-multiple transfer of learning. The continuity of the prediction is ensured by the propagation of boundary predictions. The set of techniques employed in this model is accessible within the code. The user has the flexibility to reserve the entire or a specific subset of the options, based on the simulation requirements and complexities.}
    \label{fig:flowchart}
\end{figure*}

 The architecture of the current PINNs-MPF framework is modular and composed of various implementations. To better illustrate the PINNs-MPF framework, Figure \ref{fig:flowchart} presents a flowchart of the framework. 

Panel (a) in Figure \ref{fig:flowchart} presents the global architecture:
Similar to the conventional MPF method, full discrete training in space and time is considered.
Our experiments demonstrate that this approach not only eases the comparison to our reference MPF solutions but also allows for maintaining a constant demand for resources throughout the entire simulation, in contrast to continuous training, which demands growing consumption \cite{cuomoscientific2022}. 
The general control of the training is handled through a master network, referred to as the \textit{MASTER-PINN}, while the workload is shared among multiple parallel NNs (subdomains), referred to as WORKER PINNs (c.f. paragraph \ref{Multi_Networking}).
The subdividing procedure is automated.
Within each subdomain, the training methodology is carried out independently, employing all distinctive techniques as introduced in the following. 
This includes `pyramidal training', a structured approach designed for handling the transfer of learning along the spatial domain (c.f. paragraph~\ref{pyramidal_training}). The term `basket of PINNs' (c.f. paragraph~\ref{wheel_optimizers})
 refers to an ensemble of networks dynamically selected for optimization to enhance efficiency. 
Indeed, Without an efficient strategy, the resource demand can grow exponentially, leading to impractical training times and excessive use of computational resources. 
Additionally, the concept of the 'wheel of optimizers' (c.f. paragraph \ref{wheel_optimizers}) is introduced, denoting the interaction between the optimizers.

The training is in two directions:
First in space, where the domain is decomposed into regular boxes (denoted as N$_{batches}$).
The degree of this decomposition depends on the complexity of the problem. 
The second direction of training concerns the phase-fields.
For each of these spatial-temporal blocks, referred to as a batch, a single WORKER PINN, referred to as \textit{PINN}$_i$, is trained to locally predict the solution. 
Hereafter, the term "blocks" is employed interchangeably with "quarters," as the spatial domain is consistently divided into four blocks.
Each \textit{PINN}$_i$ handles a given phase in a given batch and, depending on the domain decomposition, interacts along boundaries for spatial continuity. 

\par At the beginning of each training (time) step, the batch structure is separately built for each phase-field $\alpha$, i.e., N$_{batches}$ PINNs are created and inherit the same architecture as the \textit{MASTER-PINN}. 
During the training, several optimization techniques could be applied (based on the initial user selection), including pyramidal training, the wheel of optimizers, the basket of \textit{PINN}$_i$s, and the concept of Quarters-based training. 
An upper view of the global algorithm is shown in Algorithm \ref{general_algorithmic}, where $N_{\text{batches min}}$, $scipy_{\text{epoch}}$ denote the minimum number of batches (coarser subdivision in case of pyramidal training) and the periodic epoch for scipy optimization respectively. 
In the following, we describe the details of various implementations of the PINNs-MPF framework and the related algorithm. 
\begin{algorithm*}[htbp]
  \caption{General training scheme for the implemented PINNs-MPF.}
  \label{general_algorithmic}
  \DontPrintSemicolon
    \SetKwInput{Input}{Input1}
    \SetKwInput{InputTwo}{Input2}
    
    \Input{Initialization}
   \InputTwo{USER inputs}
    \BlankLine 
    $Initialize~the~Master~PINN  \gets $~User~inputs \;
    $Preapre~Global~training~data  \gets $~Initialization + USER inputs\;
    \tcp*[l]{}
    \For{epochs in range N$_{epochs}$ (total number of epochs)}
        {
        \textbf{set} $t_{min}$ and $t_{max}$ \;
        \tcp*[l]{$\rightarrow$ training subdomain bounds}  
        \While{$N_{batches} \succeq N_{\text{batches~min}}$ } 
            {   
                \textbf{initialize} the multi-networks/pinns. \;
                \textbf{decompose} the subdomain. \;
                $\rightarrow$ Horizontal subdivision (initialization of \textit{PINN}$_{i}$s  handling first phase). \;
                + Vertical extrusion (initialization  of \textit{PINN}$_{i}$s handling other phases). \;
                 \textbf{affect} IC training data to each \textit{PINN}$_{i}$). \;
                 $\rightarrow$ dynamically generate BC and collocation points for each \textit{PINN}$_{i}$. \;
                 \textbf{divide} the subdomain into different training Blocks/Quarters. \;
                 \textbf{assign} each \textit{PINN}$_i$ to a parent Block. \;
                 \For{each Block} 
                    { 
                    \For{each \textit{PINN}$_i$ in Block } 
                        { \tcp*[l]{Remind: each \textit{PINN}$_{i}$ handles one different phase}
                        \If{first phase}
                            {
                            \textbf{select} a candidate \textit{PINN}$_i$ $\leftarrow$ highest ratio of interfacial points. \;
                            $\rightarrow$ put "selected \textit{PINN}$_i$" in the Basket. \; 
                            \textbf{store} information about "selected \textit{PINN}$_i$" (dictionary).
                            $\rightarrow$ to use if pyramidal training \;
                            }
                        \Else 
                            {
                            \For{each phase in current Block }
                                {
                                    \textbf{get} $\leftarrow$ the NN with the same batch index as the "selected \textit{PINN}$_i$".
                                }
                            }
                        }
                    }
            \tcp*[l]{$\Rightarrow$ Basket of \textit{PINN}$_i$s Ready  \;}     
            \textit{\textbf{Multi-processing}} $\rightarrow$ start Parallel Training     
            $\rightarrow$ parallel optimization using Adam. \;
            \If {epoch \% scipy$_{\text{epoch}}$ }
                {
                \If{pinn.loss $\succeq$ Threshold for pinn in pinns}
                    {
                    \textbf{apply} L-BFGS-B optimization. \;
                    }
                \textbf{resample} collocation points. \;
                }
            \If{\textit{PINN}$_i$.loss $\prec$ Threshold for \textit{PINN}$_i$ in \textit{PINN}$_i$s}
                {
                \If{$N_{batches} \succ N_{\text{batches~min}}$}
                    {
                    reduce $N_{batches}$ $\leftarrow$ Eq. \ref{batches_coarser} \;
                    \tcp*[l]{ $\rightarrow$ Pyramidal training.
                    }
                }
            }     
        }
        \textbf{increase} $t_{min}$ and $t_{max}$ \;
        \tcp*[l]{ $\rightarrow$ go to the next training subdomain} 
        } 
\end{algorithm*}

\subsection{Discrete Resolution in Time}
We consider the MPF equations as solving a multi-variable time-series problem through the NNs. This has also been discussed in previous studies \cite{montes2021accelerating,HU2022115128,oommen2022learning,fetni2023capabilities}.
The computational domain in the upper left corner panel in Figure \ref{fig:flowchart}(a) depicts this idea. 
Here the training process occurs in spatial domains/subdomains at fixed time intervals, analogous to time steps in conventional modeling methods, but with relatively larger intervals to harness the robust linearization capabilities of deep learning (DL). 
In order to proceed to a subsequent time domain, a global loss threshold is predefined to be achieved. 
This discrete resolution ensures a quasi-constant resource consumption, hence preventing any accumulating computational load along the time axis. 
To draw an analogy between discrete resolution in MPF modeling, the notation $\Delta t$ (also denoted ${\Delta} t^{*}$ for dimensionless resolution) is adopted here for both PINNs and MPF resolutions; in the context of PINNs, this notation refers to the length of the time interval. 
In the context of the PINN literature, this approach categorizes our framework within the class of time-marching or temporal PINNs, as discussed in the causality literature \cite{PENWARDEN2023112464,CHEN2024111423,WANG2024116813}. 
\subsection{Extended domain decomposition} \label{Multi_Networking}
This technique extends the domain decomposition approach introduced in previous PINN studies \cite{Jagtap2020,SHUKLA2021110683,MENG2020113250} by adding a third dimension of decomposition and parallelization along the phases direction. This extension aims to handle the increased complexity of the target problem more effectively. A centralized network ensures synchronization among the multi-networks, facilitating a coordinated learning process across the decomposed spatial, temporal and phase domains.
Each phase, managed by a corresponding Neural Network, requires concurrent access to the interaction terms ($I_{\gamma}$) of other phases (Eq. \ref{eq:Inter_term}), which collectively form dual or higher-order junctions. These interaction terms demand a synchronization method (Algorithm \ref{pde_algorithmic} ). Consequently, parallel implementation is crucial for the precise calculation of interaction terms, as per the NN definitions. A serial approach cannot adequately manage these interactions, rendering parallelism indispensable rather than merely an improvement to be quantitatively compared against serial execution or existing methods, if available.
\par One challenging aspect of a PINNs-MPF is the heavy computational costs associated with the size of the simulation domain and the number of phase-fields. Here the concept of multi-networking is essential to distribute this computational load efficiently.
This involves the parallelization of multiple \textit{PINN}$_i$s and their associated subdomains obtained through a structured domain decomposition. 
In this context, a \textit{MASTER-PINN} is designed to distribute tasks, training data, and collate the learned outcomes from the \textit{WORKER-PINN$_i$}s/\textit{PINN}$_i$s. Each \textit{PINN}$_i$ operates independently and concurrently, utilizing a dedicated thread in the computer. It functions indeed as an autonomous entity responsible for processing a specific batch within the spatial domain and managing a given phase-field within the MPF domain.
All \textit{PINN}$_i$s are assumed to have the same architecture (number of layers and neurons per layer, learning rate, etc.).
This facilitates complete parallelization of the training process and ensures spatial continuity throughout the training.

\par As depicted in the flowchart in Figure \ref{fig:flowchart}, the \textit{MASTER-PINN} does not undergo any training but performs several central tasks, including (i) preparing the training data set, (ii) initializing and assigning each WORKER \textit{PINN}$_i$ for training, (iii) initiating and controlling the training process, (iv) facilitating communication among the WORKER PINNs, (v) managing the temporal transition of the training, and (vi) transferring the (ideally) successful learning from the actual training to another simulation.  
From a technical point of view, it is noteworthy that the current multi-networking approach enables the transfer of learning from multiple networks to a single network. This capability is available within the code through the pyramidal training option, wherein the subdivision could be progressively transformed from fine to coarse, ideally converging to a single network (c.f. paragraph \ref{pyramidal_training}). 

\par The communication between \textit{PINN}$_i$s could be classified into two categories: a spatial ‘horizontal’ communication which is required to ensure spatial continuity across the boundary between their subdomains (c.f paragraph  \ref{horizontal_optim}) and a phase-field ‘vertical’ communication that allows the coupled co-evolution of the phase-fields in the context of the MPF model (c.f paragraph \ref{vertical_optim}).
To facilitate the training of the \textit{WORKER}-\textit{PINN}$_i$s and their efficient management by the \textit{MASTER-PINN}, the concept of the trinity batch-\textit{PINN}$_i$-phase is crucial. Each \textit{PINN}$_i$ is assigned a unique identifier by associating it with a specific batch number and phase-field number, as illustrated in Figure \ref{fig:flowchart}a. Consequently, the local neural network incorporates horizontally stacked spatial-temporal coordinates $(x, y, t)$, the output (prediction) of which is a single array representing the fraction of the corresponding phase, ranging from 0 to 1. By simplifying the multi-phase problem into discrete single-phase problems in this manner, the computational approach is streamlined. This necessitates efficient interactions between the phases, which is treated through the ‘vertical’ communication. 
Finally, it is preferred to spatially decompose the simulation into N$_{batches}$ equal batches.

\subsection{Pyramidal Training and Block Training} \label{pyramidal_training} 
\par Merging the hyper-parameters of various \textit{PINN}$_i$s (WORKERS) at once and with equal significance can result in overwriting certain features in the domain. 
For this purpose, we have thought of a gradual pyramidal training and merging. 
The concept of pyramidal training draws inspiration from the data augmentation principle in ML \cite{HUTER2020109488,CARPENTER2002183}. 
It is based on the assumption that adding more data to a pre-trained model serves as a warm starting point, enabling the NN to better capture crucial features before further data processing. 
We propose to implement a pyramidal training paradigm in conjunction with another concept of `training on blocks'. 
These two complementary concepts are showcased in Figure \ref{fig:flowchart}(b) and described as follows: 
Once the spatial domain is subdivided into multiple batches, each one can be further divided into distinct domains, giving a pyramidal structure. 
Each of these finer domains in the pyramidal structure is called a block or a quarter. The procedure of subdivision can continue as depicted in \ref{fig:flowchart}(b).
After the pyramidal subdivision is completed, the training process initiates at the finest subdivision level (i.e., the base of the pyramid). However, instead of training all Neural Networks (NNs) within all batches, within each block/quarter, \textit{PINN}$_i$ with the highest number of interfacial points is selected for training; this selection concerns the first phase and the selected \textit{PINN}$_i$s serve as a reference for other phases. Similarly, for the other phases (in the vertical direction), training is assigned to the \textit{PINN}$_i$ within the same batch as the reference \textit{PINN}$_i$ within each block/quarter. 
This selective handling reduces the computational costs and makes the training focused on hot spots in the simulation domain.
Once target losses are reached for the selected \textit{PINN}$_i$s, the training stops at the finer level and transitions to a coarser level using the same algorithm described above. The training on the coarser level begins with a warm-start approach, where each selected \textit{PINN}$_i$ receives learning from the previously selected \textit{PINN}$_i$ at the finer level, within the same quarter and sharing a spatial intersection area, thereby achieving a data augmentation-like training.
The transition from one level of the pyramid to another is set by a squared relationship to ensure a pyramid-like subdivision:
\begin{equation} \label{batches_coarser}
N_{\text{batches~coarser}} = \left(\sqrt{N_{\text{batches}}}-2\right)^2 
 \end{equation}
where $N_{\text{batches~coarser}}$ is the number of batches in the next upper level until ($N_{batches} = N_{\text{batches min}}$ is reached, that means the continuation of training the \textit{PINN}$_i$s on the whole spatial domain. 
This gradual resolution has two main advantages: On the one hand, it allows a fast identification of the optimal set of hyper-parameters, and therefore key features, without the need for an adaptive weighting strategy. On the other hand, it guarantees the training of the same numbers of NN independently of the number of subdivisions ($N_{batches}$).
For that, it is worth mentioning that such a strategy is highly recommended to be activated for the early stages of the training and then set off to speed up the training.
\subsection{Adaptive mesh-free optimization}  \label{horizontal_optim} 
The Adaptive mesh-free optimization (AMFO) is here introduced as a PINN implementation extending the adaptive mesh refinement (AMR) concept, from the phase-field literature, to dynamically adjust collocation point distributions. AMR variants have indeed proven valuable for phase-field and MPF models, allowing efficient capture of evolving interfaces and steep gradients. By incorporating adaptive collocation point sampling, this approach leverages the mesh-free nature of PINNs while retaining AMR benefits like improved accuracy and robustness \cite{LI20127926,GUPTA2022115347,XU2022108891,FREDDI2023100127,bijaya2023multilevel}. The ease of integrating adaptive sampling with PINNs makes this approach attractive for tackling challenges posed by phase-field models, such as complex geometries and localized high-gradient regions. The proposed technique employs adaptive collocation point sampling to concentrate points around interfacial regions, a denoising loss function to reduce noise in non-interfacial areas, and dynamic resampling to continuously adapt the overall collocation point distribution.
\subsubsection{Adaptive re-meshing}
In the presence of diffuse interfaces, continuity among the phase-fields and across the spatial domains is required. 
Inspired by the storage concept proposed for the OpenPhase \cite{darvishikamachali2013grainPhD}, we adopt a mesh structure that mainly focuses on the interfacial regions, neglecting interior grain zones where phase-fields are equal to 0 or 1.
This strategy not only reduces the computational costs but also reduces the tendency for an error when coping for the continuity across the domains/batches. 
This is further promoted by the high capacity of fitting and extrapolation of ML which we elaborate on later in the Discussion (Section \ref{discussion_part}). 
To this end, some key hyper-parameters are user-predefined as follows. 
First, the minimum and maximum number of IC points per batch N$_{\text{ini min}}$ and N$_{\text{ini max}}$ are defined. 
Once the ratios of 0 and 1 phase-values per batch are given, the number of IC points per batch $N_{\text{ini\ per\ batch}}$ is set, following:
\begin{equation}
N_{\text{ini\ per\ batch}} = \min\left(\text{int}\left(\frac{N_{\text{ini max}} - N_{\text{ini min}}}{1 + e^{-\xi}} + N_{\text{ini min}}\right), N_{\text{ini max}}\right),
 \label{eq:per_int_points}
 \end{equation}
in which $\xi$ donates the percentage of interfacial points per batch. 
The sigmoid-like function in Eq. (\ref{eq:per_int_points}) allows gradual densification of the mesh depending on the importance of the zone, so that the diffuse interfacial area and triple junctions in a multi-phase configuration should be allocated by the higher number of IC points. 
Using Eq. (\ref{eq:per_int_points}), the collocation points are generated dynamically around the IC points (typically within a circle of radius $\eta/2$) through a predefined ratio to get $N_f= C_f *N_{\text{ini}}$, where $N_f$ is the number of collocation points per batch and $C_f$ is a user parameter that could be adjusted depending on the nature of the problem. 
Experimenting with our benchmark studies below, we varied this coefficient between 20 to 40. 
Another potential improvement, to reduce the hyper-parameters in Eq. (\ref{eq:per_int_points}) and to increase the significance to triple junctions, is to define the density of interfacial points dynamically. This would involve computing the percentage of interfacial points per batch $N_{\text{ini\ per\ batch}}$ dynamically, thereby determining the number of interfacial points per batch as follows: 
\begin{equation}
N_{\text{ini\ per\ batch}} = \left[ \frac{|x_{\text{min}} - x_{\text{max}}|\cdot |y_{\text{min}} - y_{\text{max}}|}{\eta^2} \right] \cdot \xi \cdot \chi
\label{eq:per_int_points_chi}
\end{equation}
where \(x_{\text{min}}, x_{\text{max}}, y_{\text{min}}, y_{\text{max}}\) represent the limits of each \textit{PINN}${_i}$. Recall that \( \xi \) corresponds to the associated percentage of interfacial points, emphasizing that \( \eta \) denotes the interfacial width, and these parameters are already available within the code. Additionally, \( \chi \) is an introduced hyper-parameter (user input) representing the local density of the mesh, indicating how much the mesh should be densified based on the number of interfacial points inside. With this improvement, the geometric hyper-parameters are reduced to two variables: the number of subdivisions (N$_{\text{batches}}$) and the local density (\(\chi\)), considering that the $C_f$ coefficient could be set at 20 for all numerical experiments. This simplification allows focusing on the architecture of the neural networks.
The \textit{PINN}$_i$s sharing the same batch index while handling different phases should use common collocation data (common batches). This introduces an additional stage of optimization. 
Further details on this can be found in the Supplementary Material (SM), section C. 
\subsubsection{Denoising Loss}
\begin{figure*}
    \centering
    \includegraphics[width=0.7\linewidth]{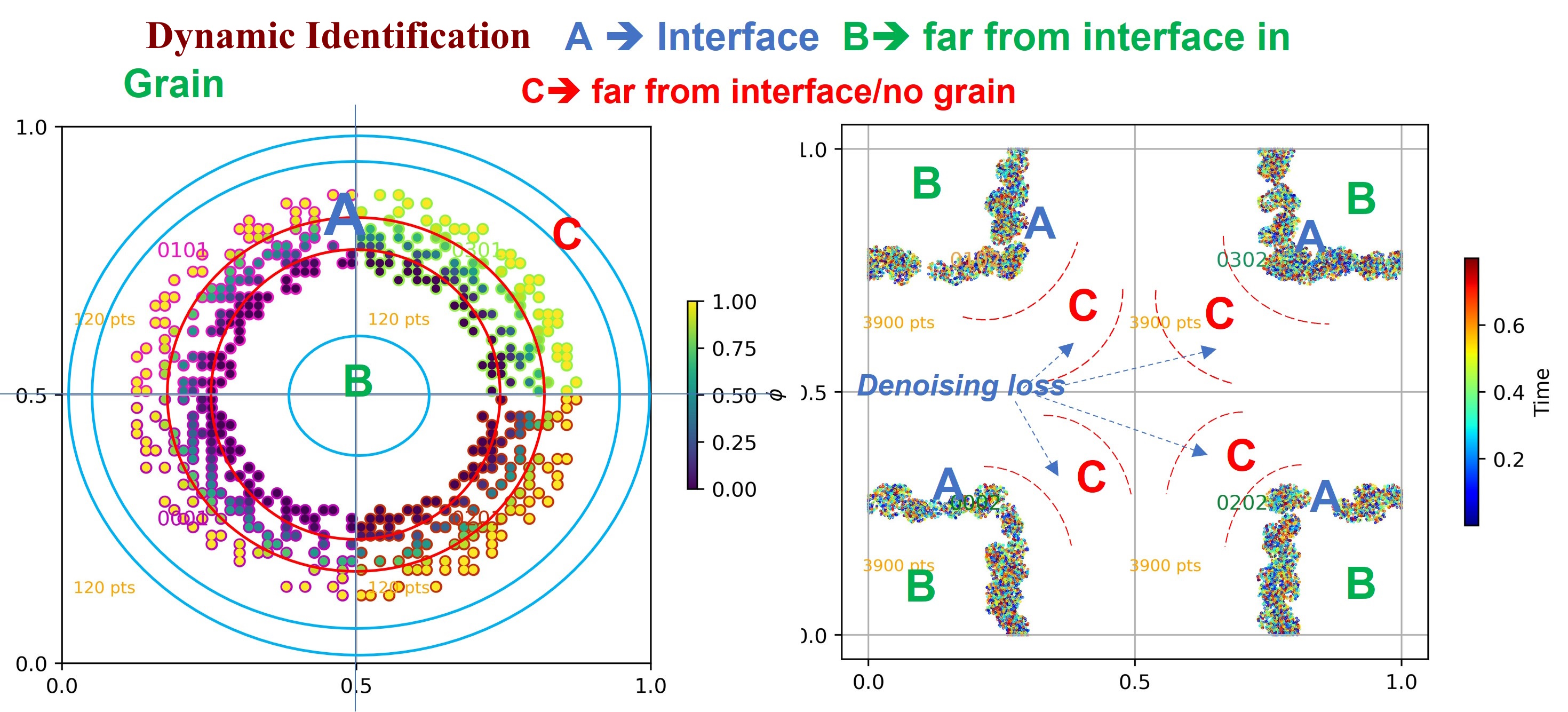}
    \caption{Illustration of the denoising loss which is computed in the identified areas B and C, far from the interface .}
    \label{fig:denoising_loss}
\end{figure*}

Focusing on the diffuse interfacial zones gives an efficient and precise approximation of the MPF solution, but still does not exclude random prediction away from the interfaces.
To address this issue, a denoising loss, functioning as a corrector, is introduced. A batch is typically made of three distinct regions: A, B, and C, where region A corresponds to the interface, region B is the grain-containing area, and region C encompasses areas far from the interface (no-grain), see Figure \ref{fig:denoising_loss}). By dynamically identifying areas of interest, as depicted in Figure \ref{fig:denoising_loss}), the additional loss term is applied within unlabeled regions that are free of collocation points, such that, each \textit{PINN}$_i$ operates to minimize the prediction of phase-field values in regions identified as 'no grain' while maximizing predictions in the grain-containing regions. Eq. \ref{eq:denois} describes this loss, where the Mean Squared Error (MSE) is calculated based on the prediction $\phi_{\text{pred}}$ and is conditioned on whether the region is identified as 'no grain' or 'grain,' denoted by $\text{flag}_{\text{no grain}}$ and $\text{flag}_{\text{grain}}$, respectively:
\begin{equation} \label{eq:denois}
\text{loss}_{\text{denoising}} = 
\begin{cases} 
~\text{MSE}~(\phi_{\text{pred}}, 0), & \text{where no grain } (flag_{~no ~grain} = 0) \\~ +~ 
\text{MSE}~(\phi_{\text{pred}}, 1), & \text{where grain } (flag_{ ~grain} = 1)
\end{cases}
\end{equation}
\subsubsection{Dynamic Resampling}
Resampling, applied after each Scipy optimization, serves as a dynamic process that benefits both converged and non-converged \textit{PINN}$_i$s. The collocation points undergo automatic resampling. This approach offers a dual purpose: converged \textit{PINN}$_i$s are subjected to testing with new but relatively similar populations. This testing mechanism safeguards against overfitting and ensures the robustness of the predictions as confirmed by previous PINN studies \cite{WU2023115671,raissiphysicsinformed2019,JAGTAP2020109136,li2022dynamic, daw2023mitigating}. Simultaneously, the non-converged \textit{PINN}$_i$s are exposed to new, unexplored samples to have a better chance to enrich their training and therefore a better convergence.

\subsection{Propagation of BC across Sub-domains}
As shown in Figure \ref{fig:flowchart}(b), the propagation of BC along the neighbors in our decomposed space is required to ensure continuity and therefore a reliable prediction of the MPF solution. To this end, a customized BC loss is introduced that concerns the spatial internal boundaries across subdomains/batches/\textit{PINN}$_i$s handling a given phase-field.
Here, each \textit{PINN}$_i$ optimizes its weights by identifying its neighbors while considering the boundary predictions of its neighbors that are used to minimize the additional BC loss. 
The BC loss function is set as the sum of differences between the prediction of the current \textit{PINN}$_i$ and those of the neighbors, depicted also in Figure \ref{fig:flowchart}b.
A detail that enhances the handling of the boundary condition is to consider the \textit{PINN}$_{00xx}$ (in batch 0 for each phase) as an absolute reference after addressing its PDE and IC losses. 
Subsequently, for a given \textit{PINN}$_i$, the west and inner neighbors are regarded as temporal references. 
This leads to a systematic propagation of the boundary conditions, ensuring that all \textit{PINN}$_i$ instances update their weights based on the \textit{PINN}$_{00xx}$.
These interactions create mutual dependencies among neighboring domains/batches/\textit{PINN}$_i$s, fostering spatial-temporal continuity during the training process.

\subsection{Handling Multi-Phases (Vertical Optimization)} \label{vertical_optim}
\begin{figure}
    \centering
    \includegraphics[width=0.75\linewidth]{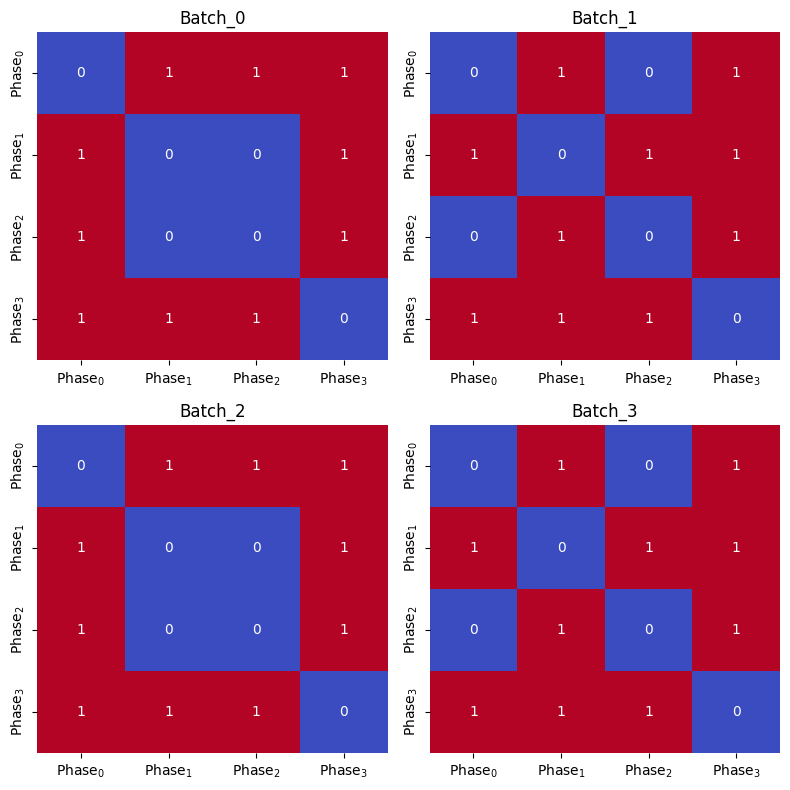}
    \caption{Illustration of the interaction between different phases in different batches for a 2D triple junction case: each phase, in each batch interacts with up to three phases.}
    \label{fig:interactios}
\end{figure}

The full discrete construction of the current framework in space, time and phase-fields allows the instant communication between \textit{PINN}$_{i}$s, through the \textit{MASTER PINN}, that is required to perform the multi-phase studies.
This is achieved in two steps: In the first step, one is entailed to identify the possible interaction(s) between any two phase-fields. This must be done for every \textit{PINN}$_i$. Obviously, with every interaction found, an associated \textit{PINN}$_i$ for that phase-field is identified as well.
A fundamental assumption here is that two given phase-fields interact when their diffuse interfaces intersect. This results in obtaining a batch-phase space. In 2D, a maximum of three phase-fields intersect, forming a triple junction. An example of this idea is illustrated in Figure \ref{fig:interactios} where we find that out of the 48 possible batch-phasе combinations, 8 exhibited no interactions.
This approach has a significant impact when handling a large number of phase-fields.
Hereafter, the term 'grain' is also used interchangeably with 'phase' once reflects the same entity.
\par The \textit{MASTER-WORKER PINN}$_{i}$ structure enables a computational- and memory-efficient calculation of the Laplacian term in Eq. (\ref{eq:phiPunktMP}). Each \textit{PINN}$_{i}$ responsible for a specific phasе-field communicatеs its computеd Laplacian tеrm instantly to thе nеighboring \textit{PINN}$_{i}$s working on the same batch but handling different phase-fields. 
This is essential for avoiding the computation of multiple Laplacian terms by each \textit{PINN}$_{i}$ and enhancing the computational efficiency. 
The numerical implementation of the MPF model (Eq. (\ref{eq:phiPunktMP})) is simplified into the parallel computation of the interaction terms for each phase within each batch. Subsequently, a robust communication protocol ensures synchronization of the computation for the correct equation terms, as detailed in Algorithm \ref{pde_algorithmic}. 
To ensure synchronization, a given \textit{PINN}$_i$ needs to wait to receive all the required information from the \textit{PINN}$_i$s in interactions. As a reminder (c.f. paragraph \ref{horizontal_optim}), an optimization detail involves the use of a shared collocation dataset for all \textit{PINN}$_i$s handling different phases within the same batch. For simplification, the collocation dataset corresponding to the first phase in the initialization (phase $\alpha$ in algorithms \ref{general_algorithmic} and \ref{pde_algorithmic}) is consistently selected as a reference.
\begin{algorithm}
    \caption{The vertical optimization along phases to optimize the computing of the PDE loss for a given PINN$_{i}$s (prediction a given phase ($\alpha$) in a given batch). Here "self" denotes the considered PINN$_{i}$ while $batch_{Xf}$ represents the associated dataset of collocation points. The dynamic correction of the phase summations is done within this block. }
     \label{pde_algorithmic}
     
    \DontPrintSemicolon

  \SetKwInput{Input}{Input1}
  \SetKwInput{InputTwo}{Input2}
    
    \Input{PDE equation $\leftarrow$ Eq. \ref{eq:phiPunktMP} }
    \InputTwo{Interaction infos (dictionary)}
     \If{  phase 1 }
        {
        \textbf{get} $\leftarrow$ self ${batch_{Xf}}$ \;
        }
    \Else 
        {  
        \textbf{identify} the other \textit{PINN}$_{i}$s in interaction with self.   \;   
        \textbf{get} $batch_{Xf}$ of the \textit{PINN}$_{i}$ handling the phase 1.\;
        
         }
    \textbf{compute} $\dot{\phi}_{\alpha}$ \; 
    \textbf{compute} self interaction term ($I_{\alpha}$)  $\leftarrow$ Eq. \ref{eq:Inter_term} \;
    \For{\textit{PINN}$_{i}$$_{ ~beta}$ in  \textit{PINN}$_{i}$s in interaction }
        { \tcp*[l]{$ \alpha \neq  \beta $} 
         \While{$I_{\beta}$ is not updated  }
                {
                \textbf{wait} \;
                }
        
        \For{\textit{PINN}$_{i}$$_{ ~beta}$ in \textit{PINN}$_{i}$s in interaction }
            {
            \tcp*[l]{$\gamma \neq \alpha, \beta $}
                \While{$I_{\gamma}$ is not updated  }
                {
                \textbf{wait} \;
                }
            }
        All interaction terms are computed \;
        compute the right side of Eq. \ref{eq:phiPunktMP}    \;
        \textbf{compute} the PDE  loss for the actual batch $\leftarrow$ Eq. \ref{eq:theta_PDE}  \;
        \textbf{correct} the phase predictions ; $\phi_{\alpha} \gets \frac{\phi_{\alpha}}{\sum_{i=1}^{N} \phi_i}$   \;
        \textbf{compute} the sum constraint loss for the entire grid at $t= t_{max} $ $\leftarrow$ Eq. \ref{eq:theta_sum_phi} \;
        \textbf{return} the total loss \; 
        }
\end{algorithm}

\subsection{The Wheel of Optimizers} \label{wheel_optimizers}
The framework effectively combines Adam's stochastic optimization with Scipy's L-BFGS, fostering a collaborative approach. Adam operates at each epoch, contributing its stochastic optimization capabilities, while Scipy, notably the L-BFGS optimizer, is periodically selected. During these selections, L-BFGS efficiently leverages information from the Hessian matrix to optimize weights. 
When a given \textit{PINN}$_i$ successfully converges, reaching a loss value below the predetermined threshold, it temporarily steps back, entering a wait state. Meanwhile, other \textit{PINN}$_i$s, selected from a shared 'basket of \textit{PINN}$_i$s,' proceed with their L-BFGS optimization. This metaphorical basket illustrates the practice of choosing specific PINNs for weight optimization and then reintegrating them, symbolizing a cyclic process; hence, the terminology 'wheel of optimizers'.  
The purpose of this cyclic process is to ensure that optimized learning results in synchronized motion between batches, leading to accurate predictions at boundaries, and effectively captures the dynamics across phases, including terms from PDE. 
\subsection{Transfer of Learning}
The highly symmetric and parallel construction of the current implementation enables the transfer of learning in multiple directions. Within the same simulation, we transfer the learning spatially through the pyramidal training (between the levels of pyramids) and across boundary conditions. The transfer also occurs temporally, from one spatial domain to its image in the next time step. These allow us to create checkpoints to continue/restart a simulation without losing the accumulated learning. 
A particular application of such transfer is to optimize the parallel performance of the worker \textit{PINN}$_i$s. 
When a worker \textit{PINN}$_i$ fails to converge after a certain number of Scipy optimizations, it is then warm-started by receiving the learning from the nearest neighbor already converged. This is to enhance its convergence rate and overall performance. 
Such a situation could arise indeed due to various factors, including insufficient data or training samples, numerical instabilities, singularities or discontinuities in the data and inadequate regularization.
Additionally, the framework allows real-time processing, facilitating easy debugging and dynamic adaptability. 

\subsection{Phase correction}
As a requirement to handle MPF problems, a dynamic correction mechanism for phase predictions is hereafter introduced. This algorithm conserves the summation of the phase fields of different phases within interfacial regions. PINNs often fail to predict correct solutions, necessitating correction algorithms to satisfy conservation laws \cite{HUANG2021103727}. Therefore, incorporating conservative correction algorithms and adequate optimization methods is essential for accurate predictions \cite{Zheng2022}.


\section{Methods } \label{method_sec}

\subsection{Multi-Phase-Field Method}
The free energy functional for a polycrystalline domain is written as 
\begin{equation}
    \label{fAllgmein}
    F = \int_\Omega f^{GB} + f^{Ext}
\end{equation}
with $f^{GB}$ is the grain boundary free energy density [$J.m^{-3}$] and the $f^{Ext}$ is to emphasize the presence of other energetic contributions (e.g. chemical and elastic energy densities) which are not specifically considered here. 
The grain boundary free energy density that is given as \cite{steinbachphasefield2009}
\begin{equation}
    \label{fGB}
    f^{GB} = \sum_{\alpha,\beta=1..N,\alpha > \beta} \frac{4\sigma_{\alpha\beta}}{\eta} \Bigm\{ -\frac{\eta^2}{\pi^2} \nabla\phi_\alpha \cdot \nabla\phi_\beta  + \phi_\alpha \phi_\beta \Bigm\} 
\end{equation}

Here $\phi_\alpha$ is the non-conservative phase-field variable corresponding to the phase (grain) $\alpha$ (idem for $\beta$), $\sigma_{\alpha\beta}$ [$J.m^{-2}$] is the grain boundary energy and $\eta$ is the width of the grain boundary interface [m]. 
The product $\nabla\phi_\alpha \cdot \nabla\phi_\beta $ accounts for the diffuse nature of the grain boundary and its sum asymptotically grows when going from a simple grain boundary to three and higher order junctions. 

The evolution of the different grains is determined by the following  system of equations:
\begin{equation}
\dot{\phi}_\alpha = - \sum_{\beta=1..N} \frac{\pi^2}{4\eta N} \mu_{\alpha \beta}
(\frac{\delta F}{\delta\phi_\alpha}-\frac{\delta F}{\delta\phi_\beta})
\end{equation}
with $\mu_{\alpha\beta}$ defined as pairwise grain boundary mobilities. 
Inserting the free energy functional $F$, one obtains 

\[
\dot{\phi}_\alpha = \sum_{\beta=1..N} \frac{\mu_{\alpha \beta}}
N \Bigm[\Bigm\{ \sigma_{\alpha\beta}(I_\alpha - I_\beta) + \sum_{\gamma=1..N, \gamma \ne \alpha, \gamma \ne \beta} (  \sigma_{\beta\gamma}-\sigma_{\alpha\gamma} ) I_{\gamma}\Bigm\}  + \frac{\pi^2}{4\eta} \Delta g_{\alpha\beta}\Bigm] 
\]

Here the additional term $\Delta g_{\alpha\beta}$ accounts for the non-interfacial contributions to the free energy functional, referred to as a driving force throughout this paper.
Further, we have
\begin{equation}
\label{eq:Inter_term}
I_{\alpha} = \nabla^2 \phi_\alpha + \frac{\pi^2}{\eta^2}\phi_\alpha
\end{equation}

The phase-field variables are assumed to respect the sum constraint:
\begin{equation} \label{eq:summConstraint}
\sum_{\alpha=1..N} \phi_\alpha = 1 
\end{equation}
which is included in the training as an additional but critical loss term.
Indeed, the phase prediction should be dynamically divided by the sum of the order parameters of all phases.

\par In our simulations, we have considered isotropic interfacial mobility ($\mu$) and energy ($\sigma$) also a constant driving force. 
This results to the simplified form of $\dot \phi$, presented in Eq. (\ref{eq:phiPunktMP}).
The latter consideration simplifies $\Delta g_{\alpha\beta} = \Delta g . \nabla \phi$ where $\Delta g$ is a constant value.
In case of a single grain boundary (two phase-fields system):
\begin{equation}
\dot{\phi} = \mu \left[ \sigma\left(\nabla^2 \phi + \frac{\pi^2}{2 \eta^2}(2\phi-1)\right) +  \frac{dh}{d\phi}  \Delta g \right]
\label{eq:dual_form}
\end{equation}
with
\begin{equation}
\frac{dh}{d\phi}=\frac{\pi \sqrt{\phi(1-\phi)}}{\eta}
\label{eq:double_obstacle}.
\end{equation}
and the kinetic coefficient of the interface $L$: $\mu=\frac{8\eta L}{\pi^{2}} $ 

For a triple junction, the Young's law becomes
\begin{align}
    \frac{\sin \theta_{12}}{\sigma_{12}} = \frac{\sin \theta_{13}}{\sigma_{13}} = \frac{\sin \theta_{23}}{\sigma_{23}},
\end{align}
%
\subsection{Dimensionless form of the PDEs}
The convergence and training speeds of the model are strongly linked to the chosen physical parameters. The use of dimensionless representations for the PDEs is recommended when dealing with slow training.
The non-dimensional time, spatial coordinate, temporal derivative, and dimensionless gradient as proposed as follows.
\begin{equation}
\tau^* = \frac{t}{T}; \quad x^* = \frac{x}{\eta}; \quad \dot{\phi}^* = \frac{\dot{\phi}}{T}; \quad \nabla^* = \frac{1}{\eta}\frac{\partial}{\partial x}\mathbf{i}
\end{equation}
where the time characteristic \textit{T} is identified:
\begin{equation}
T = \frac{\eta^2}{\mu \sigma}
\end{equation}
while $\eta$ represents the characteristic length.
Thus, the MPF model, as represented by Eq. (\ref{eq:phiPunktMP}), can be simplified for numerical purposes:
\begin{equation}
 \label{eq:phiPunktMP_dim}
\dot{\phi}_\alpha ^{*}= \sum_{\beta=1..N \neq \alpha} \frac{1}
N \Bigm[\Bigm\{ ({I_\alpha}^{*} - {I_\beta}^{*}) + \frac{\pi^2}{4} {\Delta g_{\alpha\beta}}^{*}\Bigm]
\end{equation}
with:
\begin{equation}
{I_{\alpha}}^{*} = {{\nabla}^{*}}^2 \phi_\alpha + \pi^2 \phi_\alpha
\label{eq:Inter_term_dim}
\end{equation}
Similarly, the dual-phase case (Eq. (\ref{eq:dual_form})) could be simplified to:
\begin{equation}
\dot{\phi}^* = \left[  \left( \nabla^{*2} \phi + \pi^2 (2\phi - 1) \right) \right] + \frac{dh}{d\phi}  { \Delta g}^{*}
\label{eq:dual_form_dim}
\end{equation}
\section{Results}

\subsection{Simulation Results}
Analogous to the MPF approach, the reliability and performance of a PINNs-MPF for studying the interface dynamics in polycrystalline materials need to be benchmarked capturing three basic aspects: 
First, the curvature-driven motion of an interface, second, the correct motion of the interface under any additional driving force and third, the force balance at the junctions of interfaces \cite{darvishikamachali2013grainPhD}.
The first two requirements in this set can be generally written as:
\begin{align}
    \label{Eq:Vel_Relation}
    v_n = \frac{\dot \phi}{\nabla \phi} = M \left( \sigma \kappa + \Delta g \right),
\end{align}
with $v_n$ the local interface velocity normal to its plane, $\kappa$ the local curvature of the interface, and $\Delta g$ a constant driving force, where with consider $\Delta G = \nabla \phi \Delta g$ in Eq. (\ref{eq:phiPunktMP}).
The third requirement concerning interface junctions necessitates the MPF to fulfill Young's law, which relates the angles between interfaces at the junction to the interfacial energy.
For a triple junction and isotropic interface energy assumed, the three interfaces at equilibrium meet at a 120$^\circ$ angle.
To test the PINNs-MPF framework, certain simulation scenarios related to the above requirements are studied as listed in Table \ref{tab:table1}, along with related physical parameters and applied hyper-parameters in the PINNs implementation in Table \ref{tab:table2}. 
We note that for the grain shrink scenarios, the number of IC points per batch $N_{\text{ini per batch}}$ serves as an input parameter, and Equation \ref{eq:per_int_points} is used for its generation (with $N_{\text{ini min}}$ and $N_{\text{ini max}}$ varied up to 90). However, for the triple junction scenario, this process is automated through Equation \ref{eq:per_int_points_chi}, where 
\( \chi \) is set to 40 in order to dynamically populate the batches during training, depending on the evolution of the solution.

\begin{table}
\centering
\caption{Numerical and physical parameters of the testing benchmarks. The corresponding units of the interface width, mobility, and interfacial energies are respectively m, $m^{4} J^{-1} s^{-1}$, and $J m^{-2}$.}
\label{tab:table1}
\begin{tabular}{|c|c|c|c|c|c|c|c|c|c|} 
\cline{2-10}
\multicolumn{1}{c|}{}                                                                                    & \multicolumn{5}{c|}{\begin{tabular}[c]{@{}c@{}}\textbf{Spatio-temporal}\\\textbf{parameters (dimensionless)}\end{tabular}} & \multicolumn{4}{c|}{\textbf{Physical parameters}}                                                                                                                                                 \\ 
\hline
\textbf{Benchmark}                                                                                       & \textbf{Nx} & \textbf{dx} & \textbf{Ny} & \textbf{dy} & \textbf{Nt}                                                        & \begin{tabular}[c]{@{}c@{}}\textbf{Interface}\\\textbf{width}\end{tabular} & \textbf{Mobility} & \begin{tabular}[c]{@{}c@{}}\textbf{Interfacial}\\\textbf{energy}\end{tabular} & \textbf{Phases}  \\ 
\hline
\begin{tabular}[c]{@{}c@{}}\textbf{Traveling}\\\textbf{wave}\\\textbf{interface}\end{tabular}            & 100         & 0.01        & 100         & 0.01        & 1000                                                               & 10dx                                                                       & 1e-4              & 1                                                                             & 1                \\ 
\hline
\begin{tabular}[c]{@{}c@{}}\textbf{Grain shrinkage}\\\textbf{under}\\\textbf{driving force}\end{tabular} & 64          & ..          & 64          & ..          & 100                                                                & 7dx                                                                        & 1e-4              & 1                                                                             & 1                \\ 
\hline
\begin{tabular}[c]{@{}c@{}}\textbf{Curvature-driven}\\\textbf{grain shrinkage}\end{tabular}              & 64          & ..          & 64          & ..          & 150                                                                & 4-9dx                                                                      & 1e-4              & 1                                                                             & 1                \\ 
\hline
\textbf{Triple juncion}                                                                                  & 64          & ..          & 64          & ..          & \begin{tabular}[c]{@{}c@{}}up to\\~200\end{tabular}                & 6dx                                                                        & 1e-4              & 1                                                                             & 4                \\
\hline
\end{tabular}
\end{table}

\begin{table*}[htbp]
\caption{Hyper-parameters of the implemented PINN for the different testing benchmarks. LR: Learning Rate, Per.: Periodicity/Epochs }
\label{tab:table2}
\centering
\begin{tabular}{c|cc|c|cccc|}
\cline{2-8}
                                                                                                             & \multicolumn{2}{c|}{\textbf{NN Architecture}}                                                            & \textbf{}                                                                            & \multicolumn{4}{c|}{\textbf{Optimizers}}                                                                                                                                                                                            \\ \hline
\multicolumn{1}{|c|}{\multirow{2}{*}{\textbf{Benchmark}}}                                                    & \multicolumn{1}{c|}{\multirow{2}{*}{\textbf{NNs}}}                     & \multirow{2}{*}{\textbf{HL/NN}} & \multirow{2}{*}{\textbf{\begin{tabular}[c]{@{}c@{}}Nodes\\ /\\  layer\end{tabular}}} & \multicolumn{2}{c|}{\textbf{Adam}}                                                                                              & \multicolumn{2}{c|}{\textbf{L-BFGS}}                                                              \\ \cline{5-8} 
\multicolumn{1}{|c|}{}                                                                                       & \multicolumn{1}{c|}{}                                                  &                                 &                                                                                      & \multicolumn{1}{c|}{\textbf{LR}} & \multicolumn{1}{c|}{\textbf{\begin{tabular}[c]{@{}c@{}}Activation\\  function\end{tabular}}} & \multicolumn{1}{c|}{\textbf{\begin{tabular}[c]{@{}c@{}}max\\  Iter\end{tabular}}} & \textbf{Per.} \\ \hline
\multicolumn{1}{|c|}{\textbf{\begin{tabular}[c]{@{}c@{}}Traveling \\ wave interface\end{tabular}}}           & \multicolumn{1}{c|}{1}                                                 & 6                               & 32                                                                                   & \multicolumn{1}{c|}{1e-3}        & \multicolumn{1}{c|}{\begin{tabular}[c]{@{}c@{}}tanh + sigmoid\\  (last layer)\end{tabular}}  & \multicolumn{1}{c|}{1000}                                                         & 100           \\ \hline
\multicolumn{1}{|c|}{\textbf{\begin{tabular}[c]{@{}c@{}}Grain shrink\\ under driving\\  force\end{tabular}}} & \multicolumn{1}{c|}{4}                                                 & 6                               & 32-128                                                                               & \multicolumn{1}{c|}{1e-4}        & \multicolumn{1}{c|}{idem}                                                                    & \multicolumn{1}{c|}{1000}                                                         & 50            \\ \hline
\multicolumn{1}{|c|}{\textbf{\begin{tabular}[c]{@{}c@{}}Grain shrink with \\ natura motion\end{tabular}}}    & \multicolumn{1}{c|}{4}                                                 & 6                               & 128                                                                                  & \multicolumn{1}{c|}{1e-4}        & \multicolumn{1}{c|}{idem}                                                                    & \multicolumn{1}{c|}{1000}                                                         & 50            \\ \hline
\multicolumn{1}{|c|}{\textbf{\begin{tabular}[c]{@{}c@{}}Triple juncion \\ (*)\end{tabular}}}                 & \multicolumn{1}{c|}{\begin{tabular}[c]{@{}c@{}}16\\ (**)\end{tabular}} & 6                               & 128                                                                                  & \multicolumn{1}{c|}{1e-5}        & \multicolumn{1}{c|}{idem}                                                                    & \multicolumn{1}{c|}{500}                                                         & 50            \\ \hline
\end{tabular}
\begin{tablenotes}
\item (*) For the triple junction, the pyramidal approach was tested for the early stages of the training.
\item (**) When using the pyramidal training, 64 NNs are inititalizated, while 16 NNs are selected for training using the the concept of the basket of PINNs.
\end{tablenotes}

\end{table*}

\par To further highlight the effectiveness of multi-networking in conjunction with a combination of optimizers for training PINNs, an additional numerical experiment is proposed as follows: reproducing the benchmark of grain shrinkage under constant driving forces with varying numbers of hidden layers and neurons per layer. Specifically, a fixed number of six hidden layers with the number of neurons ranging from 64 to 512 was applied. This was achieved by employing a cyclic optimization strategy alternating between the Adam optimizer and the L-BFGS optimizer (referred to as the wheel of optimizers). Associated findings and comments are provided in the discussion section. 
\newline
Additionally, we illustrate in Table \ref{tab:opt_tech_tab} the progressive increase of the optimization techniques to deal with the induced complexity by the tackled scenarios.

\begin{table}[htbp]
\caption{Enumeration of applied optimization techniques with increasing complexity of target problems, as well as the associated number of model trainable parameters.}
\label{tab:opt_tech_tab}

\centering
\begin{tabular}{|c|l|c|c|}
\hline
\textbf{Benchmark}                                                                & \multicolumn{1}{c|}{\textbf{Activated Techniques}}                                                                                                                                                                                & \textbf{\begin{tabular}[c]{@{}c@{}}Number of\\ optimization \\ Techniques\end{tabular}} & \textbf{\begin{tabular}[c]{@{}c@{}}Number of \\ trainable \\ parameters\end{tabular}} \\ \hline
\begin{tabular}[c]{@{}c@{}}Travelling wave\\  interface\end{tabular}              & \begin{tabular}[c]{@{}l@{}}Wheel of optimizers (Adam and LFBGS), \\ resampling, discrete or continuous resolution\\  (optional), transfer of learning\end{tabular}                                                                & 4                                                                                       & 3,297                                                                              \\ \hline
\begin{tabular}[c]{@{}c@{}}Grain shrinkage \\ under\\  driving force\end{tabular} & \begin{tabular}[c]{@{}l@{}}Wheel of optimizers, discrete resolution,\\  extended domain decomposition, \\ Adaptive Mesh-Free Optimization (AMFO),\\  transfer of learning\end{tabular}                                            & 5                                                                                       & 416,005                                                                            \\ \hline
\begin{tabular}[c]{@{}c@{}}Curvature-driven \\ grain shrinkage\end{tabular}       & \begin{tabular}[c]{@{}l@{}}Wheel of optimizers, discrete resolution,\\  extended domain decomposition, \\ AMFO, transfer of learning\end{tabular}                                                                                 & 5                                                                                       & 416,005                                                                            \\ \hline
Triple junction                                                                   & \begin{tabular}[c]{@{}l@{}}Wheel of optimizers, discrete resolution, \\ Extended domain decomposition, \\ transfer of learning, AMFO, Basket of PINNs, \\ Handling multi-phases, pyramidal training \\ correction of phases predictions\end{tabular} & 9                                                                                       & 1,414,417                                                                           \\ \hline
\end{tabular}
\end{table}

Furthermore, for the scenarios involving grain shrinking and triple-junctions, the MPF solutions from OpenPhase calculations are used \cite{darvishikamachali2013grainPhD}. The associated scheme is provided in the SM, section A. All numerical implementations were coded using TensorFlow and performed on standard workstations with an AMD Ryzen Threadripper PRO 5975WX 32-Cores CPU (64 threads). 


\subsection{Traveling Interface under Constant Driving Forces}
As a key starting point, the behavior of a planar interface traveling under a given driving force is explored. In the absence of the curvature, this can be simplified to a 1D problem in which the phase-field profile propagates according to \cite{steinbachphasefield2009,darvishikamachali2013grainPhD}: 
\begin{equation}
 \label{eq:traveling_wave}
\phi(x, t)=\left\{\begin{array}{lll}
1 & \text { for } & x<v_n t-\frac{\eta}{2} \\
\frac{1}{2}-\frac{1}{2} \sin \left(\frac{\pi}{\eta}\left(x-v_n t\right)\right) & \text { for } & v_n t-\frac{\eta}{2} \leq x<v_n t+\frac{\eta}{2} \\
0 & \text { for } & x \geq v_n t+\frac{\eta}{2}
\end{array}\right.
\end{equation}

\begin{figure}[!ht]
    \centering
    \begin{tabular}{cc}
        \begin{subfigure}{0.35\textwidth}
        \raisebox{0.17\height}{ 
            \includegraphics[width=\linewidth, height=4cm]{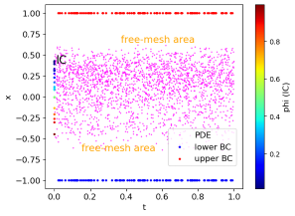}
            }
            \caption{Initial mesh}
        \end{subfigure} &
        \begin{subfigure}{0.45\textwidth}
            \includegraphics[width=\linewidth, height=5cm]{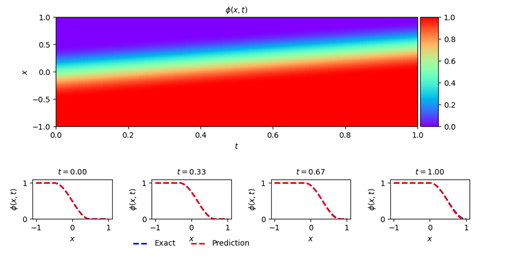}
            \caption{Travelling wave solution}
        \end{subfigure} \\
        
        \begin{subfigure}{0.35\textwidth}
            \includegraphics[width=\linewidth]{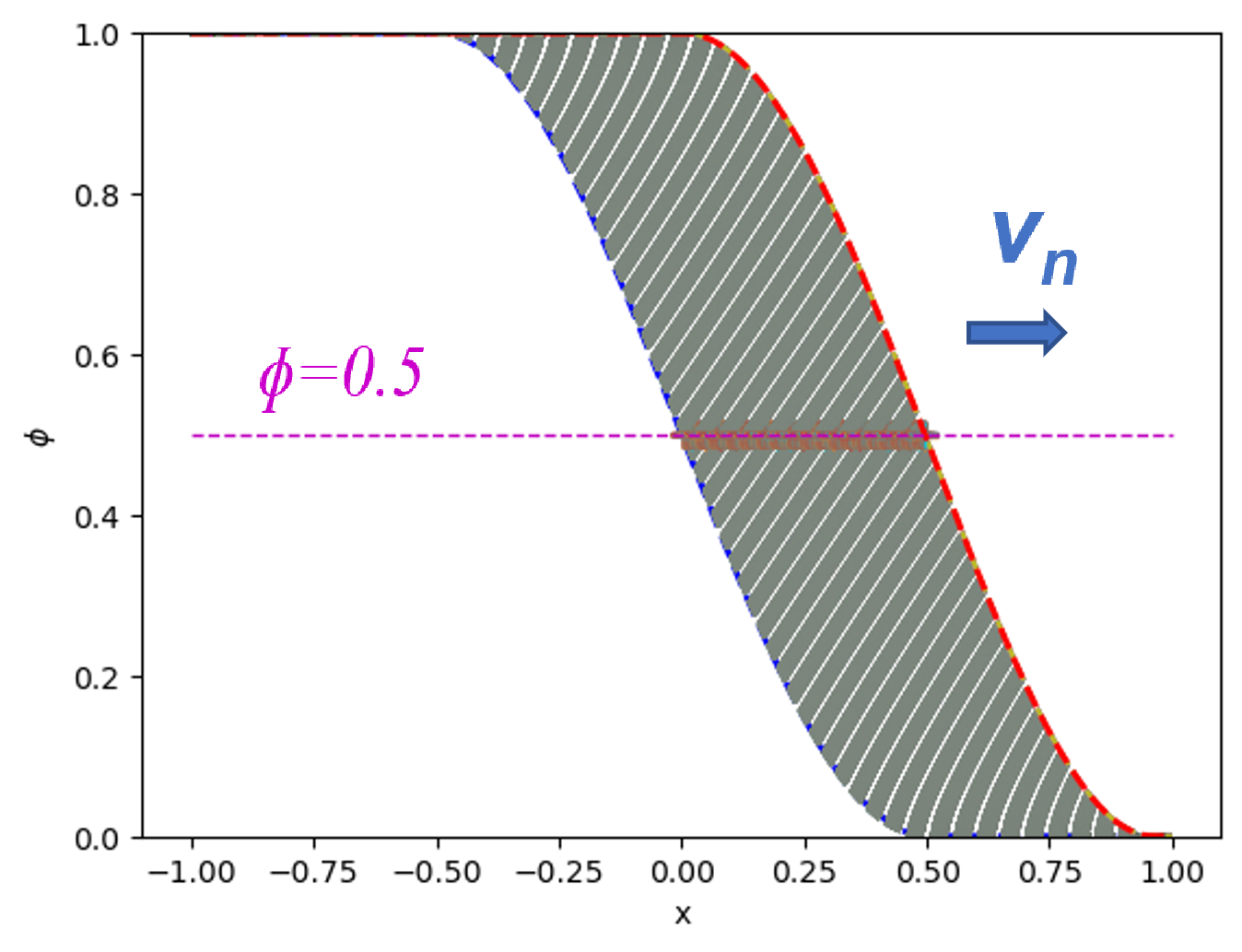}
            \caption{Spacial motion of the interface}
        \end{subfigure} &
        \begin{subfigure}{0.35\textwidth}
            \includegraphics[width=\linewidth]{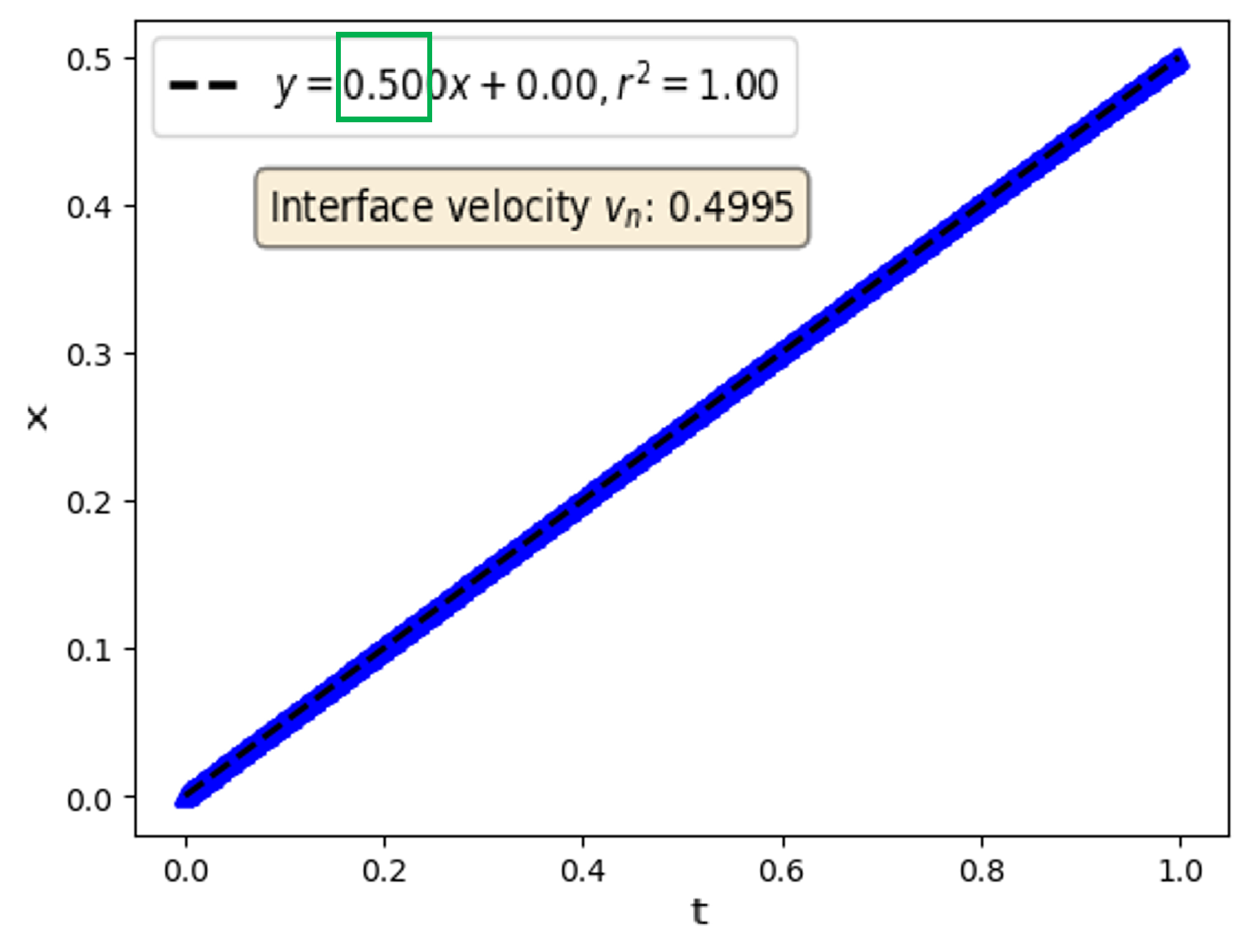}
            \caption{Identification of the interfacial velocity}
        \end{subfigure} \\
        
        \begin{subfigure}{0.45\textwidth}
            \includegraphics[width=\linewidth, height=5cm]{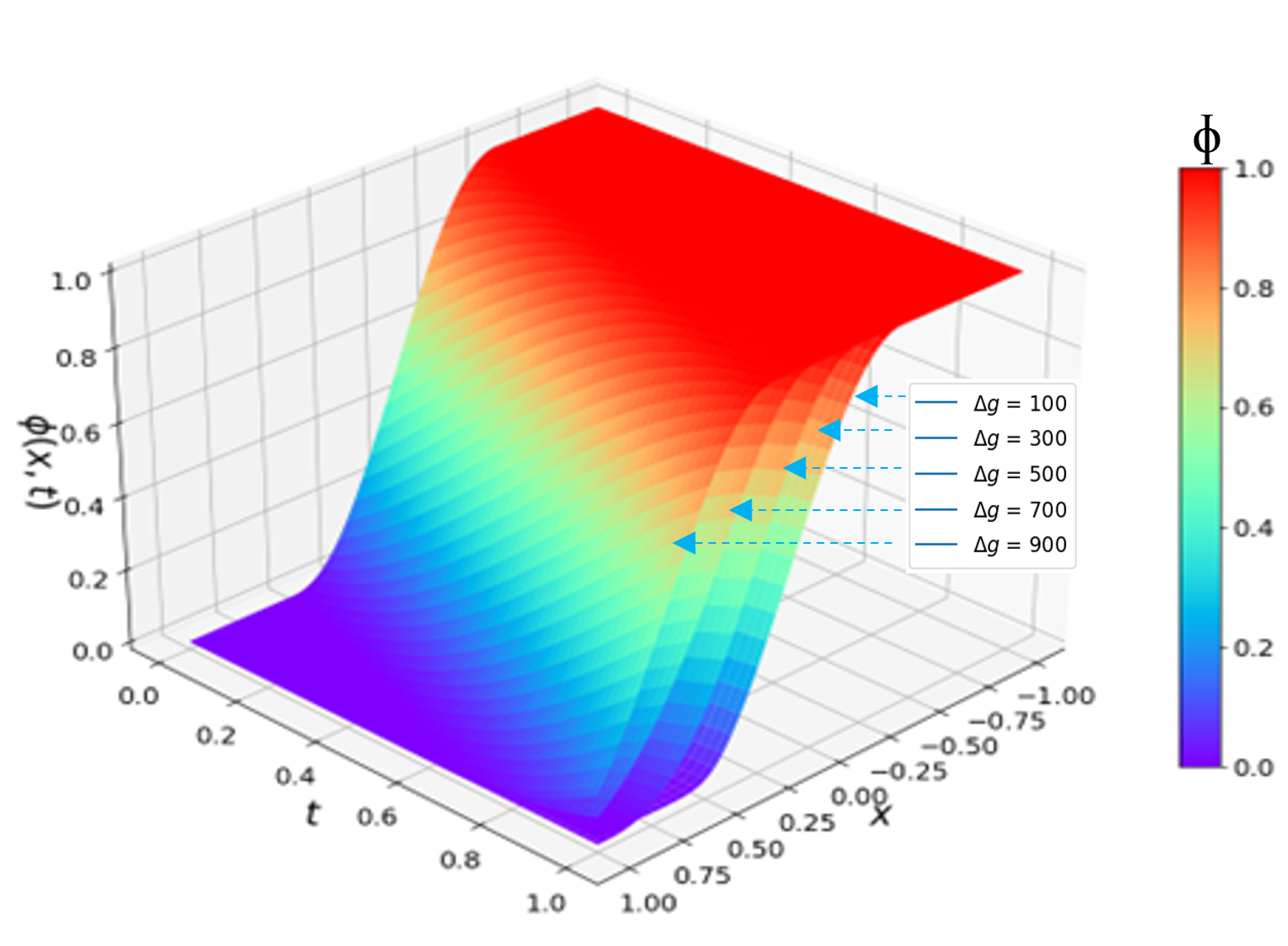}
            \caption{Impact of external driving force}
        \end{subfigure} &
        \begin{subfigure}{0.35\textwidth}
            \includegraphics[width=\linewidth]{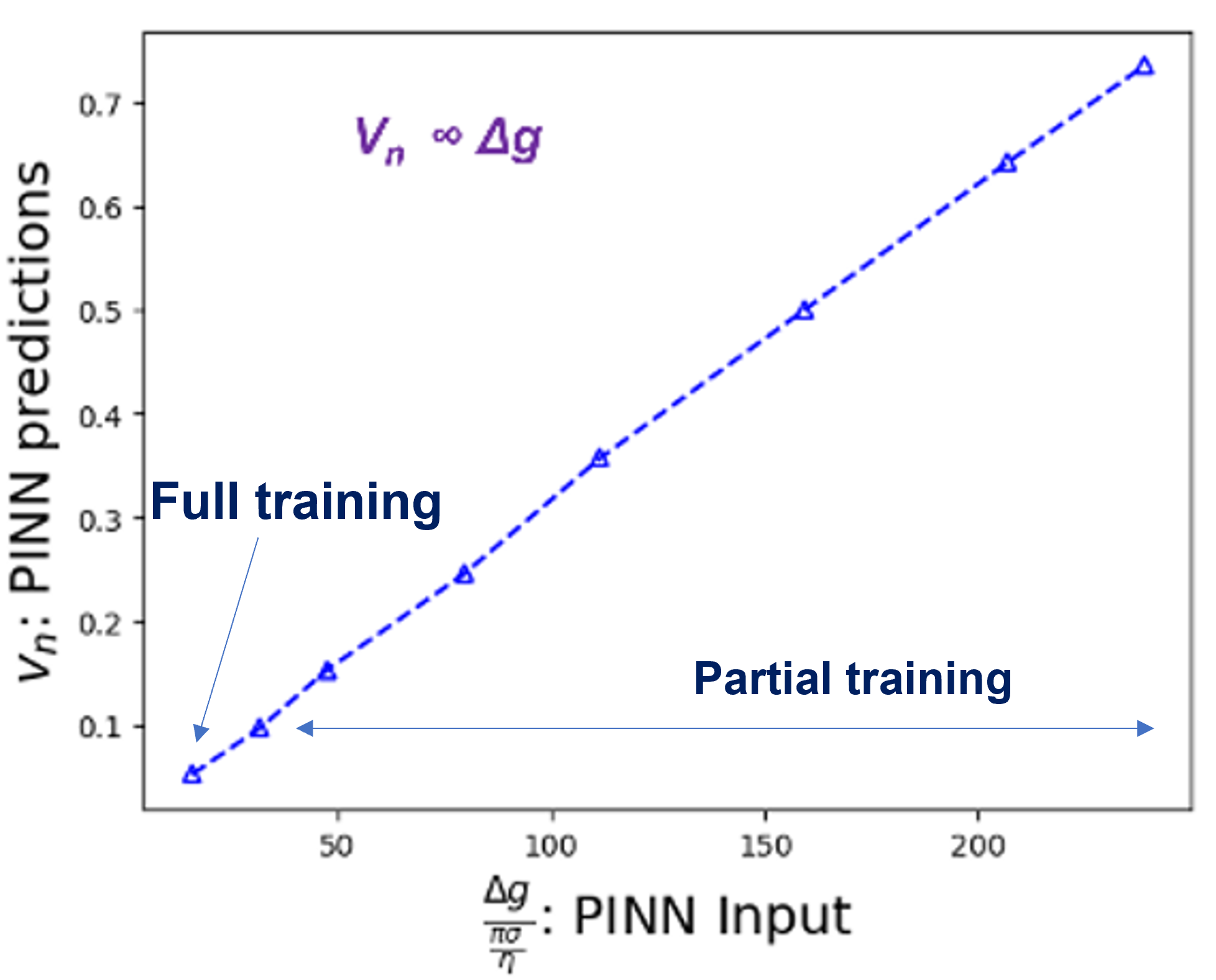}
            \caption{PINN extrapolations}
        \end{subfigure}
    \end{tabular}
    \caption{PINN solution for the traveling wave interface problem against the theoretical solution.}
    \label{fig:traveling_wave_int}
\end{figure}

The results of this campaign are gathered in Figure \ref{fig:traveling_wave_int}.
The utility of analyzing a traveling wave solution lies in its ability to accurately predict the shape and velocity of the phase-field profile.  
Figure \ref{fig:traveling_wave_int}(a) presents the setup of the BC and IC points, as well as the placement of PDE collocation points for this problem. 
As described in paragraph \ref{horizontal_optim}, the meshing is predominantly focused within the interface regions, where the PINN model is trained to predict the solution.
Excellent agreement is obtained between the PINNs-MPF and the theoretical solutions at various time instants, Figure \ref{fig:traveling_wave_int}(b).
Figure \ref{fig:traveling_wave_int}(c) shows the spatial evolution of the interface resolved by the PINNs-MPF. 
To trace the interface motion and its velocity, a representative point with a phase-field value $\phi=0.5$ at $x,t=0,0$ is selected. 
Figure \ref{fig:traveling_wave_int}(d) compares the theoretical and predicted velocities, revealing an accuracy approaching unity after 2500 epochs of training, with a total $\approx$10 minutes of computation time.

For a planar interface (no curvature), the interface velocity is expected to linearly scale with the input driving force (Eq. (\ref{Eq:Vel_Relation})).
Figure \ref{fig:traveling_wave_int}(e) shows the interface velocities obtained for various driving forces, perfectly matching the theoretical expectation. 
Here the PINNs-MPF were first trained for a single $\Delta g$ value (Figure \ref{fig:traveling_wave_int}(f)) and the learning is then transferred to predict the interface velocity for various $\Delta g$ values. 
Such transfer of learning is typically carried out within a significantly reduced number of (maximum 100) epochs of training, as the training is not always evitable \cite{weinan_deep_2018}. 
From a technical perspective, this analysis tests localized meshing, the transfer of learning and full-discrete resolution. 
It is worth noting that here a single neural network effectively handles the 1D scenario. 

\begin{figure}[!ht]
    \centering
    \begin{tabular}{cc}
        \begin{subfigure}{0.3\textwidth}
            \includegraphics[width=\linewidth]{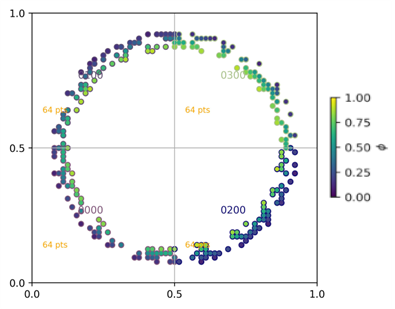}
            \caption{Initial Condition}
        \end{subfigure} &
        \begin{subfigure}{0.3\textwidth}
            \includegraphics[width=\linewidth]{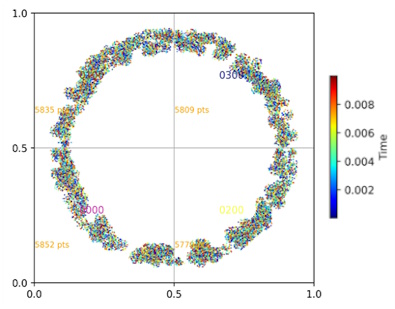}
            \caption{Localized mesh }
        \end{subfigure} \\
        
        \begin{subfigure}{0.3\textwidth}
            \includegraphics[width=\linewidth]{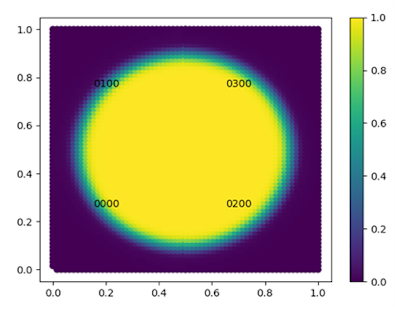}
            \caption{$\phi$ at Time (1 \%)}
        \end{subfigure} &
        \begin{subfigure}{0.3\textwidth}
            \includegraphics[width=\linewidth]{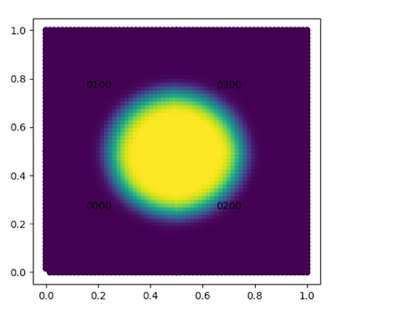}
            \caption{$\phi$ at Time (51 \%)}
        \end{subfigure} \\
        
        \begin{subfigure}{0.3\textwidth}
            \includegraphics[width=\linewidth]{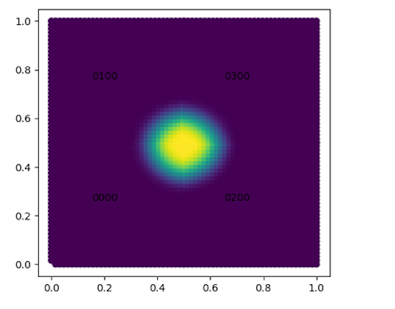}
            \caption{$\phi$ at Time (86 \%)}
        \end{subfigure} &
        \begin{subfigure}{0.35\textwidth}
            \includegraphics[width=\linewidth]{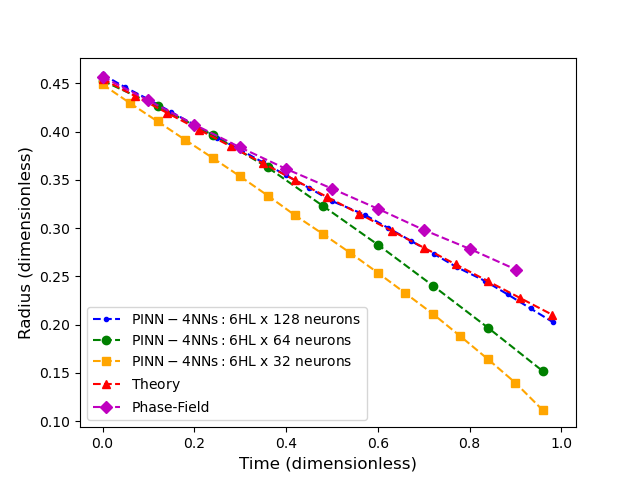}
            \caption{Radius vs. time for different architectures. }
        \end{subfigure}
    \end{tabular}
    \caption{Grain prediction of PINN against theoretical and PF solutions}
    \label{fig:motion_with_driving_forc}
\end{figure}

\begin{table}[bp]
\caption{Quantitative comparison between each of the tested architectures and the theoretical solution.}
\label{tab:table_quant}

\centering
\begin{tabular}{|c|c|c|c|c|c|}
\hline
\textbf{Configuration} & \multicolumn{1}{l|}{\textbf{Neural Networks}} & \textbf{HL} & \textbf{Neurons/HL} & \textbf{MAE} & \textbf{MSE} \\ \hline
1 & 4 & 6 & 32 & $4.72\times 10^{-2}$ & $3.06\times 10^{-3}$ \\ \hline
2 & 4 & 6 & 64 & $3.12\times 10^{-2}$ & $1.47\times 10^{-3}$ \\ \hline
3 & 4 & 6 & 128 & $2.82\times 10^{-3}$ & $1.14\times 10^{-5}$ \\ \hline
\end{tabular}

\end{table}
One can immediately expand the driving-force-driven interface kinetics to 2D where the planar interface is replaced by a circular interface.
In this scenario, the interface has a curvature, but its effect is negligible under the condition that $\Delta g \gg \sigma \kappa$ (Eq. (\ref{Eq:Vel_Relation}).
Although the linear scaling of the interface velocity holds, this scenario imposes a challenge for the PINNs-MPF going from a cartesian to a spherical coordinate.
PINNs-MPF predictions versus theoretical solution from OpenPhase are shown in Figure \ref{fig:motion_with_driving_forc}.
Here initial trials involving a single NN (used for the 1D case above) revealed limitations marked by training instabilities and noise (c.f. the SM for qualitative and quantitative comparison between PINN predictions and ground truth solution for a single NN).
However, stable training was obtained when applying four neural networks.
Indeed, subsequent refinements increased the prediction accuracy by progressively increasing the number of neurons starting from 32, resulting in an optimal configuration of 128 neurons and 6 hidden layers, as observed in Figure \ref{fig:motion_with_driving_forc}. To quantify the differences between each PINN prediction and the theoretical solution, it is proposed in Table \ref{tab:table_quant} measures of the MSE and Mean Absolute Error (MAE) for each prediction compared to the ground truth solution. 
The architecture with 128 neurons not only accurately matched the theoretical solution (with MAE and MSE values of $2.82\times 10^{-3}$ and $1.14\times 10^{-5}$ respectively) but also outperformed the phase-field solution obtained using an explicit scheme.
Thus, it is hereafter proposed to keep this architecture for subsequent benchmarks as it allowed to deal with the double-well potential term and the non-convex potential term  in Eq. \ref{eq:dual_form} on a one hand, and to fix the related set of parameters of this architecture, especially for the multi-phase field scenario. 

It is worth noting that changing from the 1D to 2D setup, the PINNs-MPF is well capable of capturing and maintaining the correct interface thickness throughout the simulation.

\subsection{Curvature-driven Interface Motion}
\label{Curv_driv_int_Motion}
\begin{figure}[!ht]
    \centering
    \begin{tabular}{ll}
        \begin{subfigure}{0.32\textwidth}
            \includegraphics[width=\linewidth, height=4.2cm]{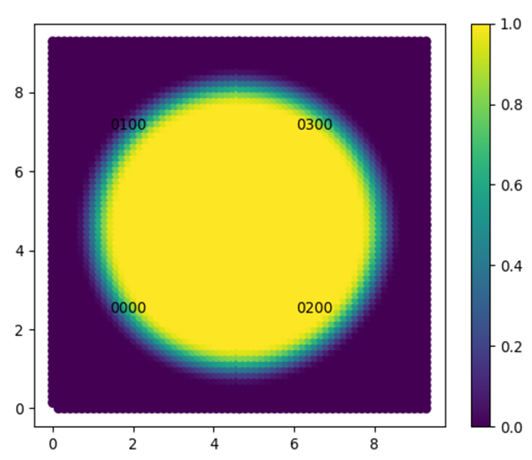}
            \caption{$\phi$ at Time (1 \%)}
        \end{subfigure}
        &
        \begin{subfigure}{0.4\textwidth}
            \includegraphics[width=\linewidth, height=4.2cm]{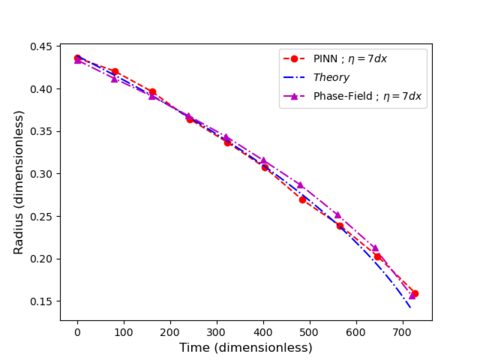}
            \caption{Radius vs. time for $\eta = 7dx$}
        \end{subfigure} \\
        
        \begin{subfigure}{0.32\textwidth}
            \includegraphics[width=\linewidth, height=4.2cm]{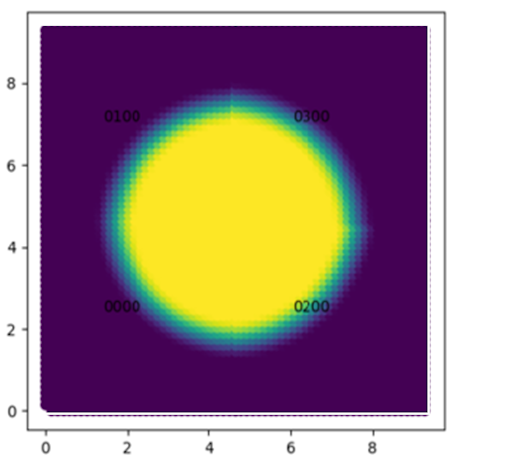}
            \caption{$\phi$ at Time (40 \%)}
        \end{subfigure}
        &
        \begin{subfigure}{0.4\textwidth}
            \includegraphics[width=\linewidth, height=4.2cm]{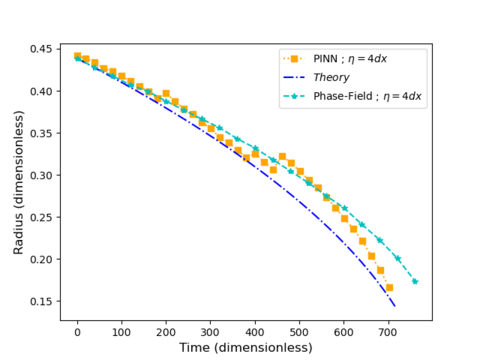}
            \caption{Radius vs. time for $\eta = 4dx$}
        \end{subfigure} \\
        
        \begin{subfigure}{0.32\textwidth}
            \includegraphics[width=\linewidth, height=4.2cm]{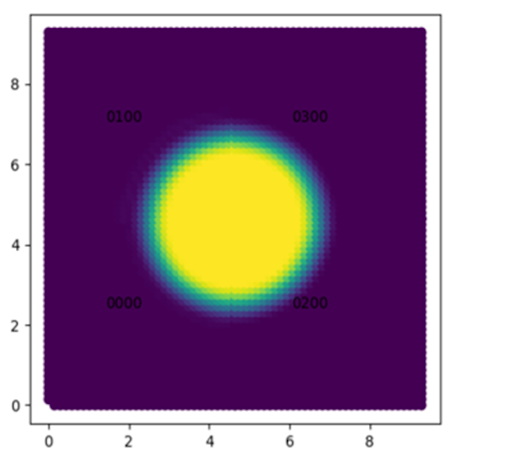}
            \caption{$\phi$ at Time (65 \%)}
        \end{subfigure}
        &
        \begin{subfigure}{0.4\textwidth}
            \includegraphics[width=\linewidth, height=4.2cm]{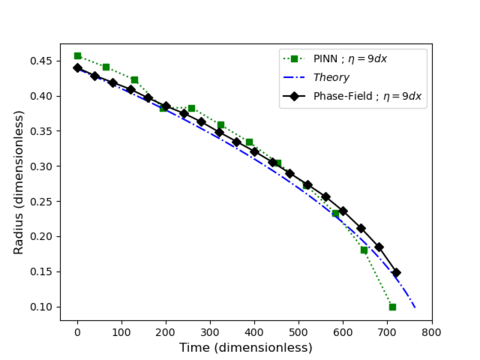}
            \caption{Radius vs. time for $\eta = 9dx$}
        \end{subfigure}
    \end{tabular}
    \caption{Grain radius prediction of PINN against theoretical and PF solutions (Open-Phase) for a natural motion. The plots correspond for PINN prediction for an interface width $\lambda = 7dx$. }
    \label{fig:motion_without_driving_forc}
\end{figure}

Capturing the natural curvature-driven motion of an interface is the most central capability of a phase-field model, encoded into the gradient energy term and related Laplacian in the governing free energy functional and equations of motion, respectively.
When the Laplacian terms take control over the interfacial motion, i.e., $\sigma \kappa \gg \Delta g$ in Eq. (\ref{Eq:Vel_Relation}), nonlinear kinetics arise: 
Here we study the shrinkage of a circular phase (grain) with $\kappa = 1/R$, giving $v_n = dR/dt = M\sigma/R$.
This gives another level of complexity to test the PINNs-MPF framework.
The number of time intervals required for the resolution of this case was 150, corresponding to ${\Delta} t^{*} =10^{-3}$, while ${\Delta} t^{*}=10^{-4}\textbf{}$ is required to resolve the same problem using the PF scheme (c.f. the SM, section A). 
The results are presented in Figure \ref{fig:motion_without_driving_forc}. Here, the PINN solution fits the theoretical one. However, the phase-field linearized scheme results in a premature shrinkage of the grain. The same result is obtained even when trying with smaller time steps. This case, where the motion is fully governed by the interface and no external driving force is applied, demonstrates that PINNs can efficiently handle non-linearities in the phase-field context.
Indeed, it is worth reminding that we use the tanh activation function in hidden layers to capture diverse non-linear patterns in physics-informed input data. For the output layer, the sigmoid activation ensures bounded predictions, making it suitable for interpreting outputs. The specific limits (the order parameter) remain between 0 and 1. These tailored non-linear activations allow PINN to effectively handle the grain shrinkage scenario where the motion is controlled by the interfacial energy and curvature. 

The pattern of the phase-field solution depends on the interface width $\lambda$; for a thin interface ($\eta = 4dx$), there is a slight deviation from the theoretical solution, while the results improve with increasing the width. This behavior is systematically investigated  \cite{darvishikamachali2013grainPhD}, showing that it arises from the numerical limitations of calculating the Laplacian and consequently the interface curvature.
This is a feature of the diffuse phase-field interfaces. 
The PINNs-MPF predictions were studied for three interface widths of $\eta =$ 4dx, 7dx and 9dx. 
In all configurations, a consistent alignment is evident between the PINNs-MPF and the OpenPhase solutions. This is best visible for the thinnest interface shown in Figure \ref{fig:motion_without_driving_forc}b. 
This set of benchmarks demonstrates that the PINNs-MPF is not only capable of handling the nonlinear interface kinetics but also has a good sense of the interface width and its impact on the solution. The latter is crucial for future developments of the model to deal with interfacial phenomena, such as interfacial elasticity and solute segregation \cite{wang2021incorporating,zhou2022revealing}.

\subsection{Establishing Equilibrium Triple Junctions}
\label{triple_junction_sec}
\begin{figure*}[htbp]
    \centering
    \begin{subfigure}{0.4\textwidth}
    \includegraphics[width=\linewidth]{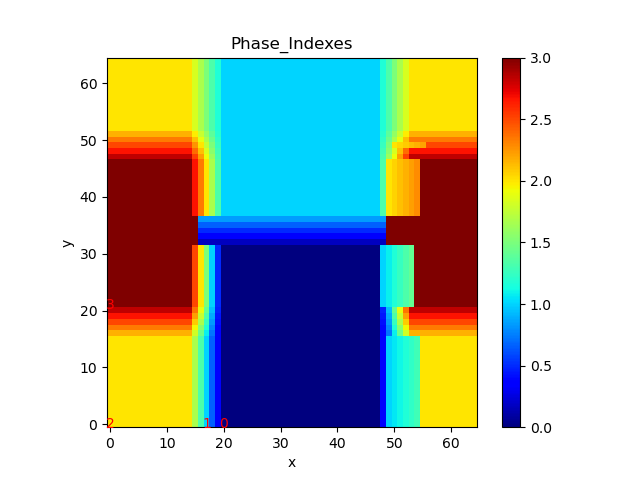}
        \caption{Phase Indexes}
    \end{subfigure}
    
    \begin{subfigure}{0.35\textwidth}
    \includegraphics[width=\linewidth]{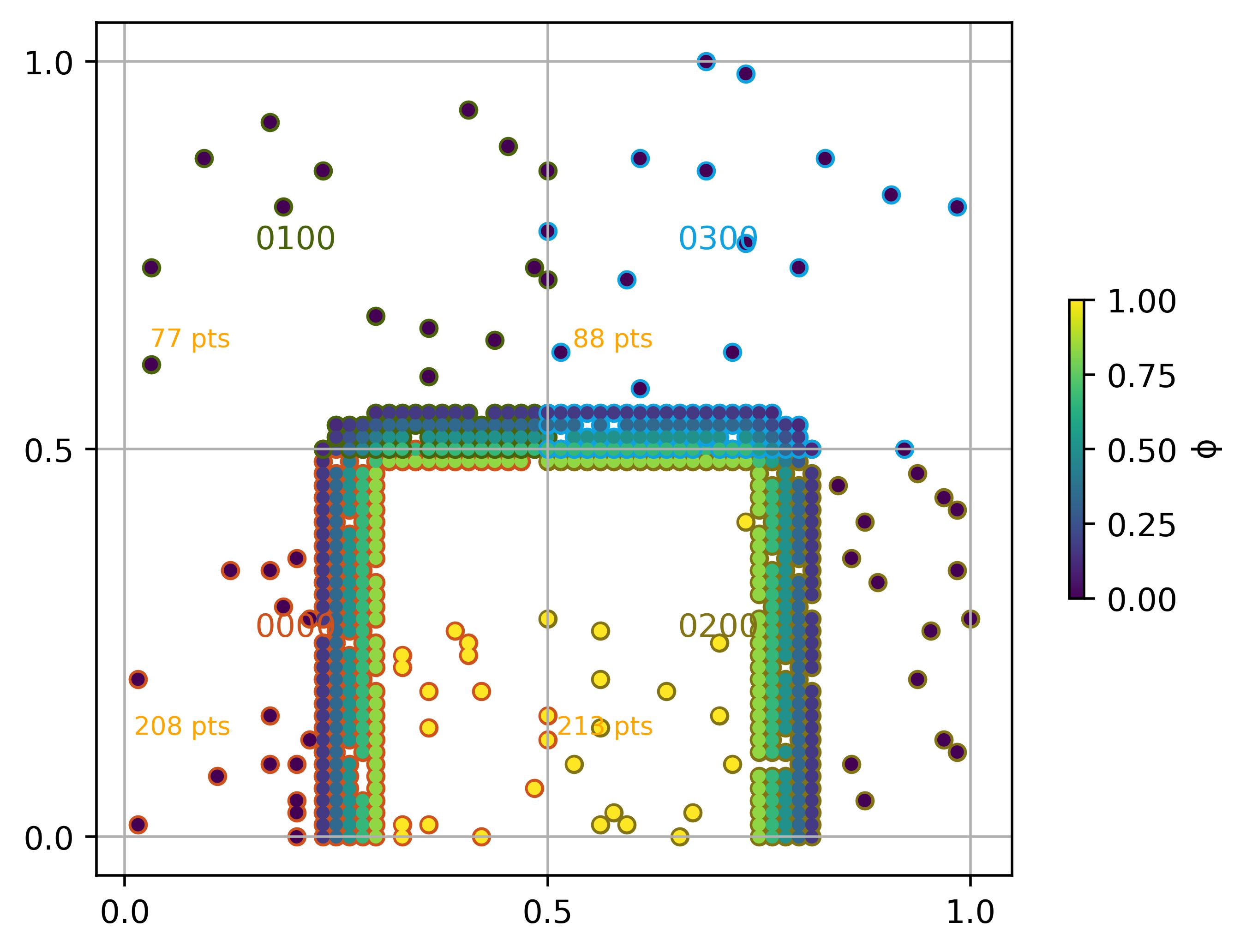}
        \caption{ Phase 0}
    \end{subfigure}
     \vspace{5pt} 
        \begin{subfigure}{0.35\textwidth}
    \includegraphics[width=\linewidth]{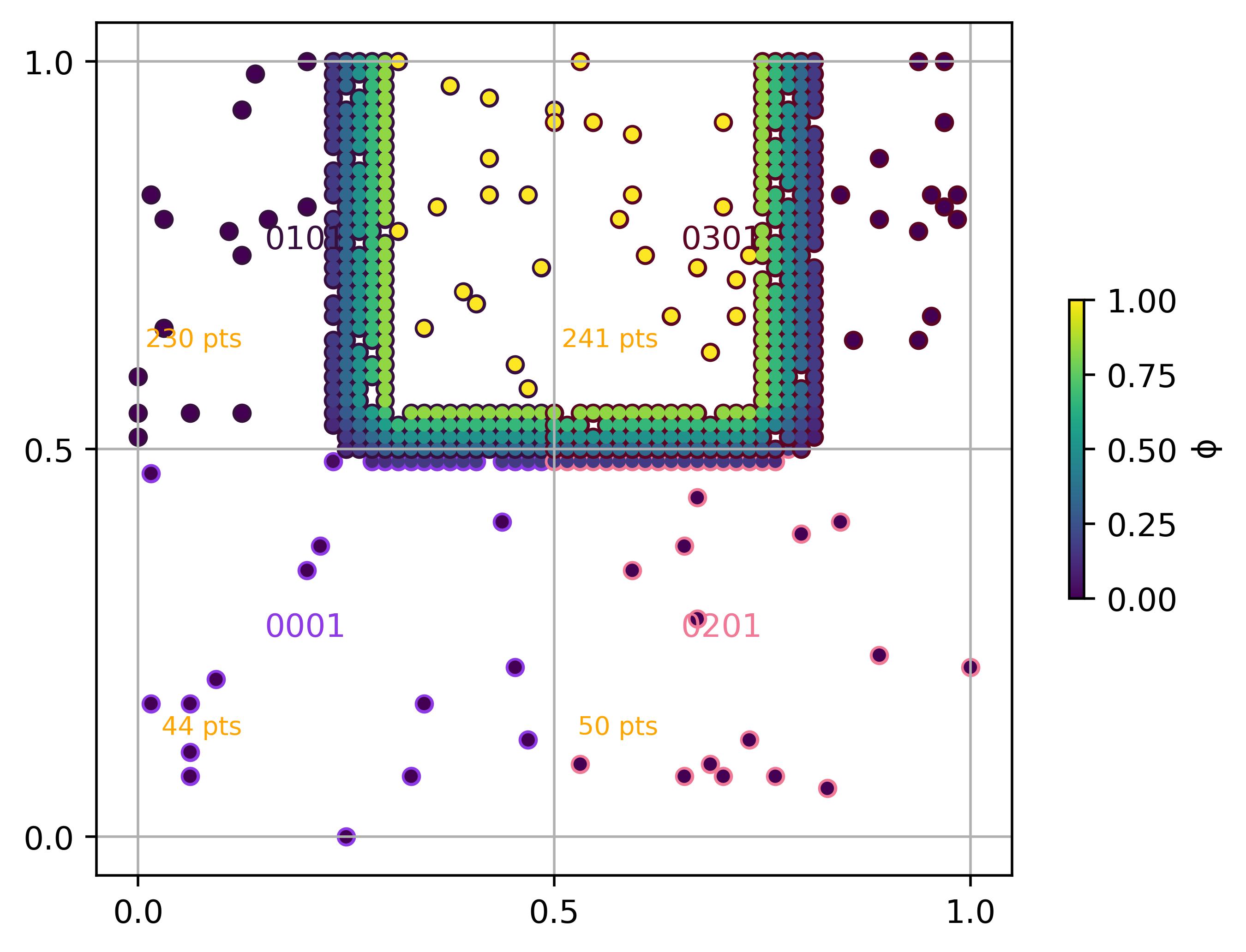}
        \caption{ Phase 1}
    \end{subfigure}
    
    \begin{subfigure}{0.35\textwidth}
    \includegraphics[width=\linewidth]{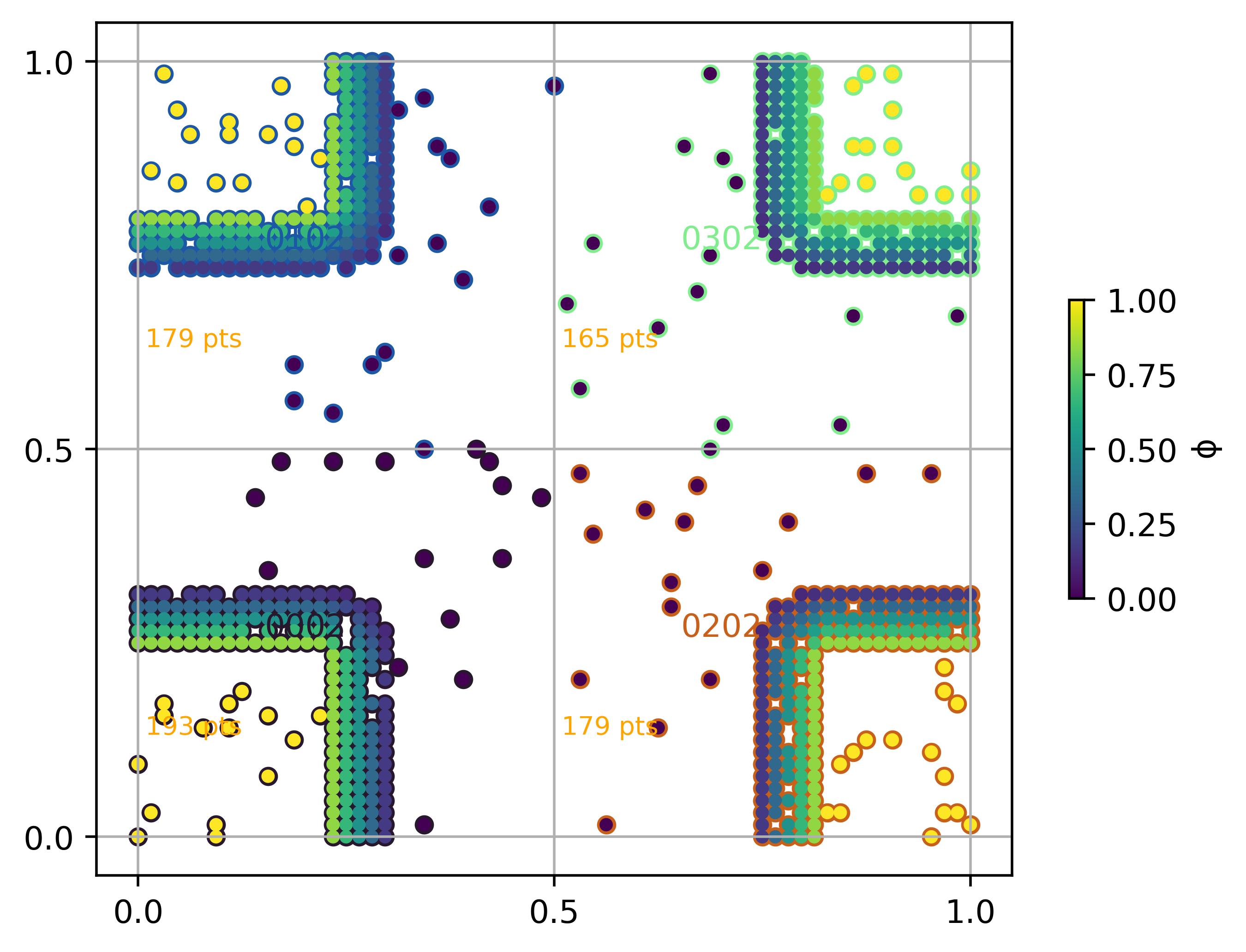}
        \caption{Phase 2 }
    \end{subfigure}
     \vspace{5pt} 
        \begin{subfigure}{0.35\textwidth}
    \includegraphics[width=\linewidth]{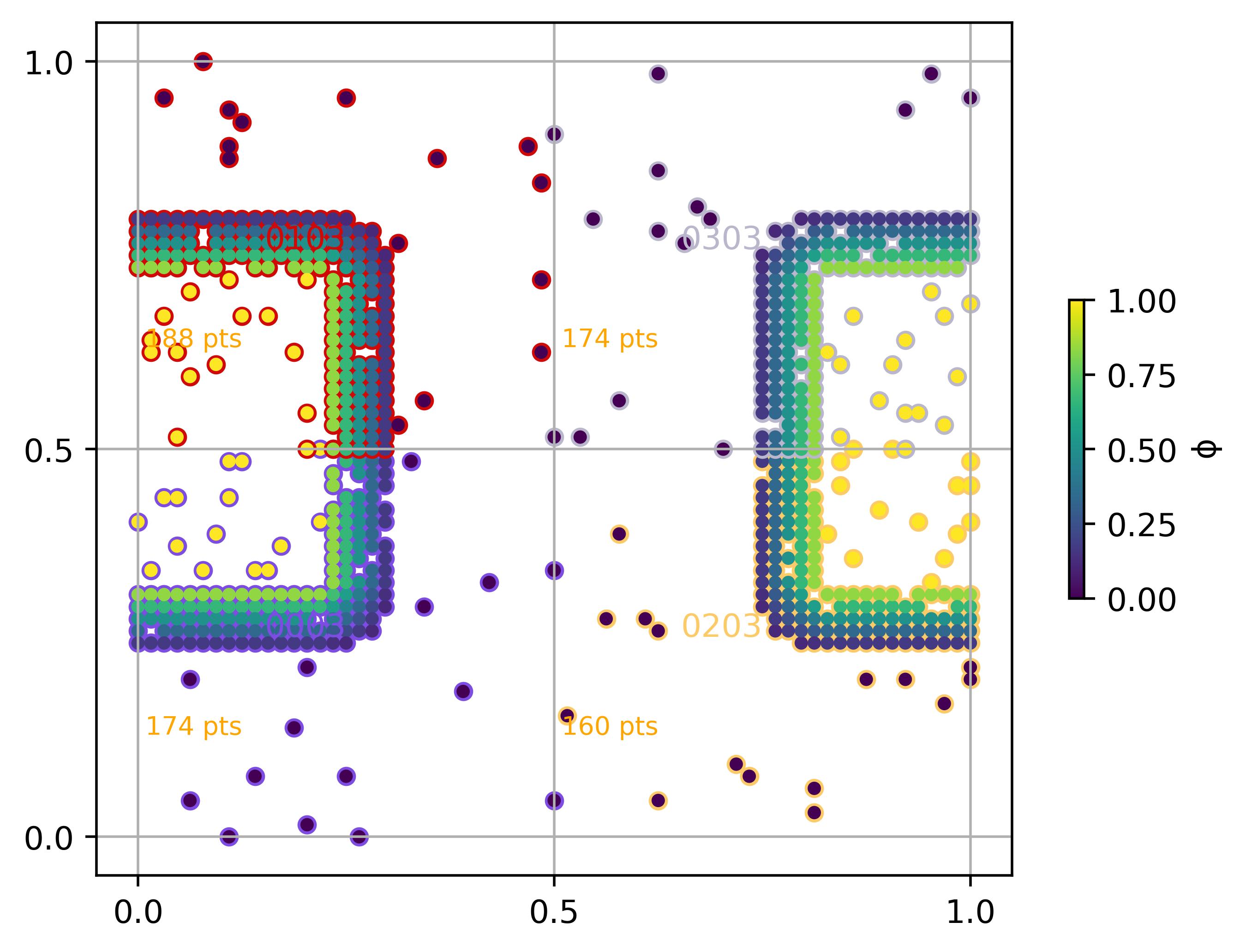}
        \caption{ Phase 3}
    \end{subfigure}
    
    \caption{Initial condition points for the different NNs handling the four phases in different batches. The corresponding mesh is given in the SM, section C. Reminder: the numbers on each batch correspond to the corresponding \textit{PINN}$_i$ index (cf. Figure \ref{fig:flowchart}). The IC points in each batch are surrounded by a circle with a common color for checking purposes.
    }
    \label{fig:triple_junction_IC}
\end{figure*}

A major capability of MPF is to handle interfacial junctions. 
This is not only to fulfill Young's law but also to ensure the evolutionary path leading to a configuration with minimum energy.
To test our PINNs-MPF framework, a system with four phase-fields (grains) and six triple junctions was considered. 
The initial mesh is visualized in Figure \ref{fig:triple_junction_IC}. 
Using our method (c.f. the SM, section C), the IC points are densified within the interfacial region, while some random points were also selected away from the interfaces to allow PINNs to learn better.  
To streamline computations, BC are selectively addressed along the interface lines between different batches.
It is worth highlighting that the initialization having sharp angles presents a challenge for PINNs to produce accurate solutions. 
This is because of the curvature-driven nature of the interface dynamics, thus, initiating away from smooth curves/shapes results in complex evolutionary scenarios.

\begin{figure*}[htbp]
    \centering

    \begin{minipage}{\textwidth}
    \centering
    \thead{Initialization \quad \quad \quad \quad MPF Solution \quad \quad \quad \quad PINN Predictions}
    \end{minipage}

     \vspace{0.5cm}
     
    \begin{subfigure}{0.2\textwidth}
        \includegraphics[width=1.16\linewidth, height=\textheight, keepaspectratio]{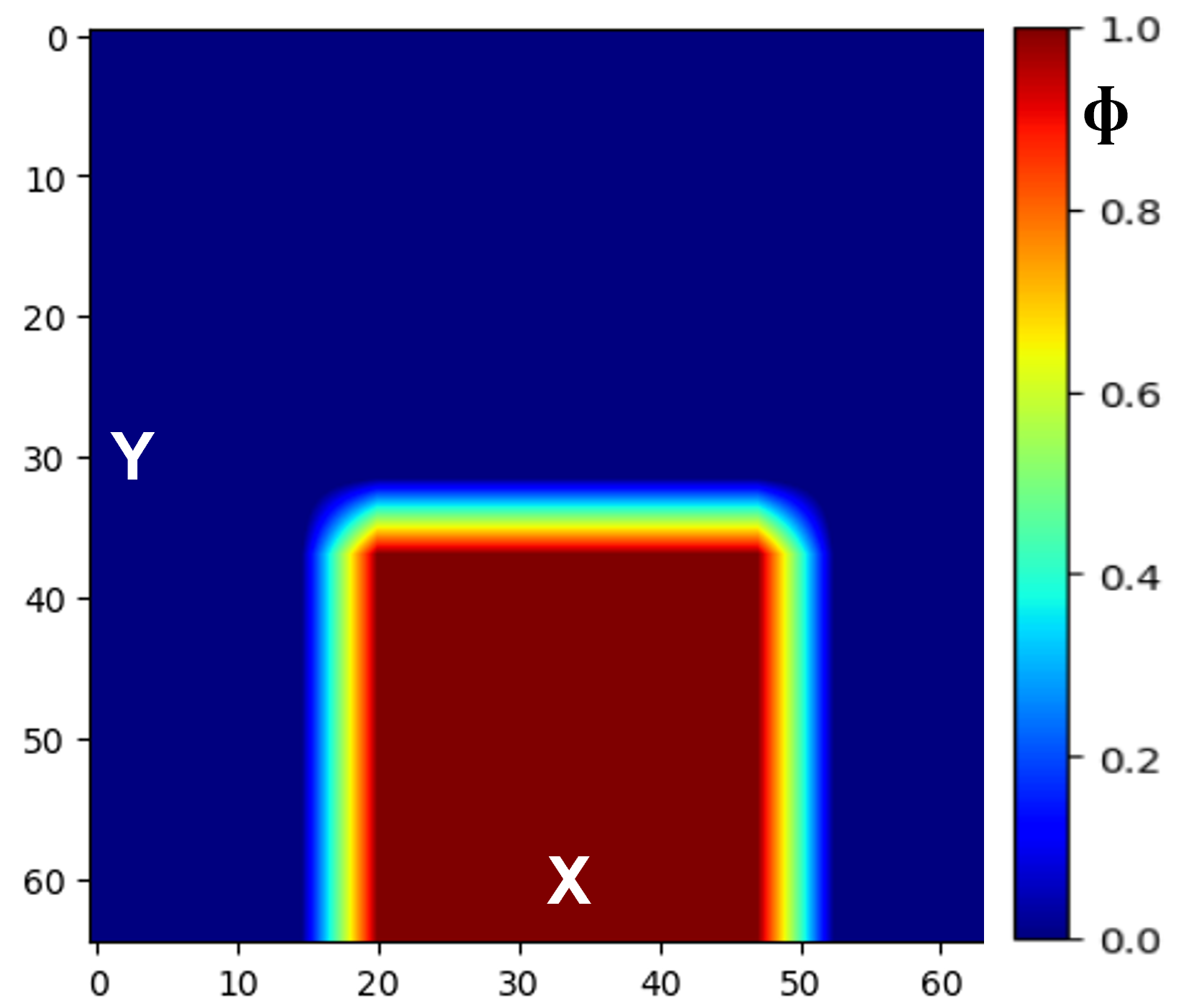}
        \caption{ grain 0}
    \end{subfigure}
    \hspace{0.5cm}
    \begin{subfigure}{0.2\textwidth}
        \includegraphics[width=\linewidth, height=\textheight, keepaspectratio]{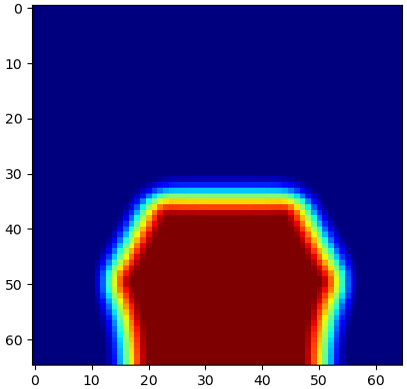}
        \caption{ }
    \end{subfigure}
    \hspace{0.5cm}
    \begin{subfigure}{0.2\textwidth}
        \includegraphics[width=\linewidth, height=\textheight, keepaspectratio]{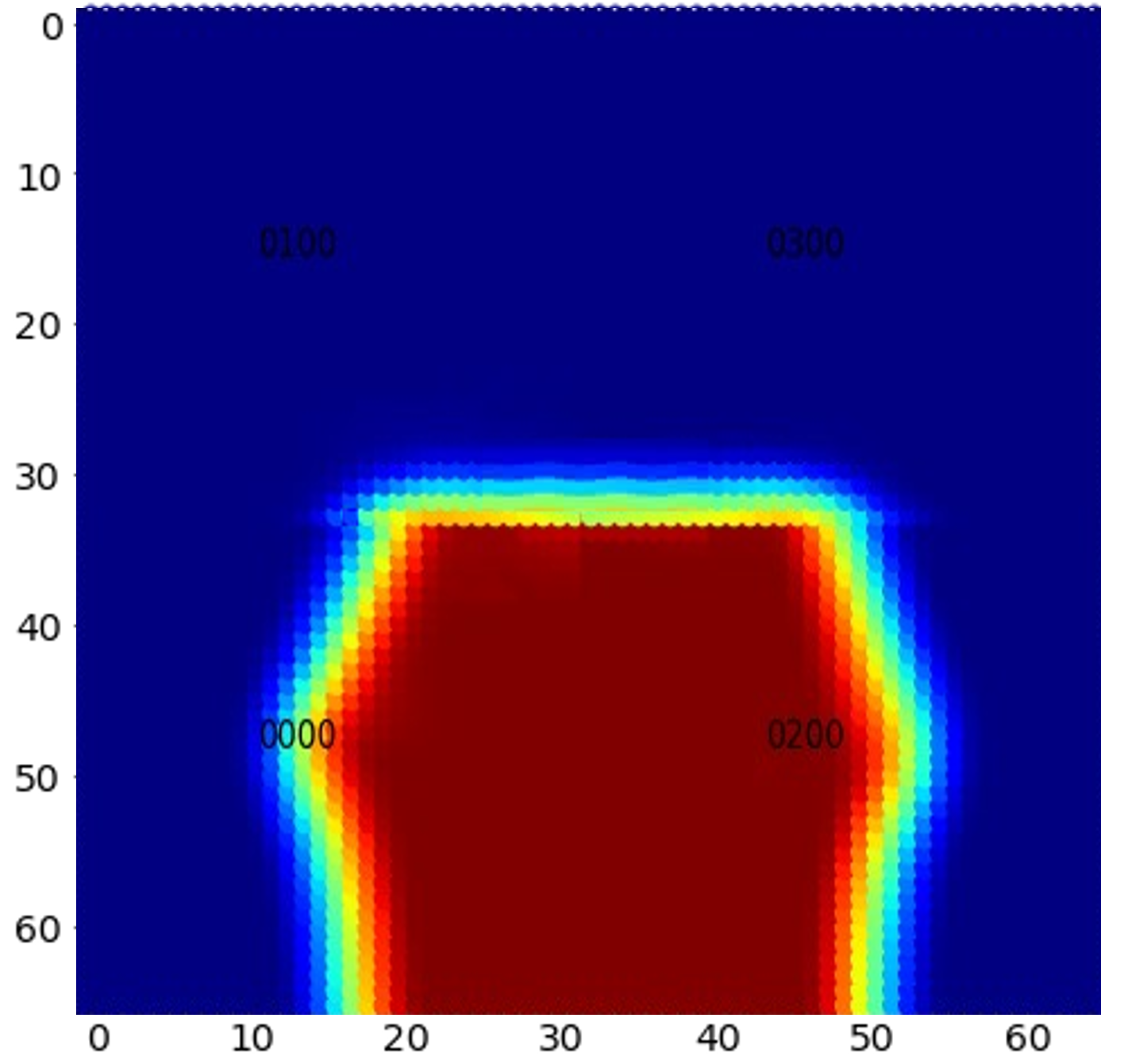}
        \caption{ }
    \end{subfigure}
     \vspace{0.5cm}
 
    \begin{subfigure}{0.2\textwidth}
    \includegraphics[width=0.75\textwidth,keepaspectratio]{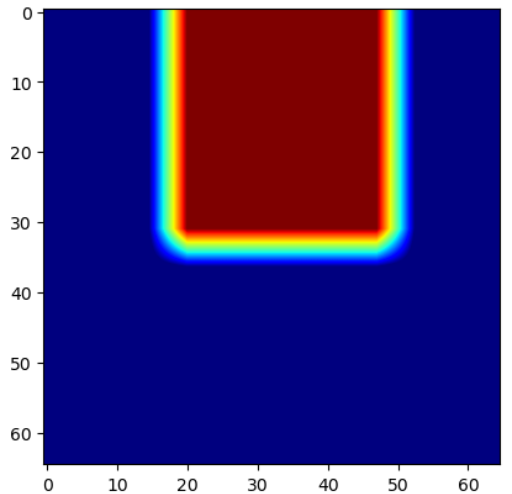}

        \caption{ grain 1}
    \end{subfigure}
    \hspace{0.5cm}
    \begin{subfigure}{0.2\textwidth}
        \includegraphics[width=\linewidth, height=\textheight, keepaspectratio]{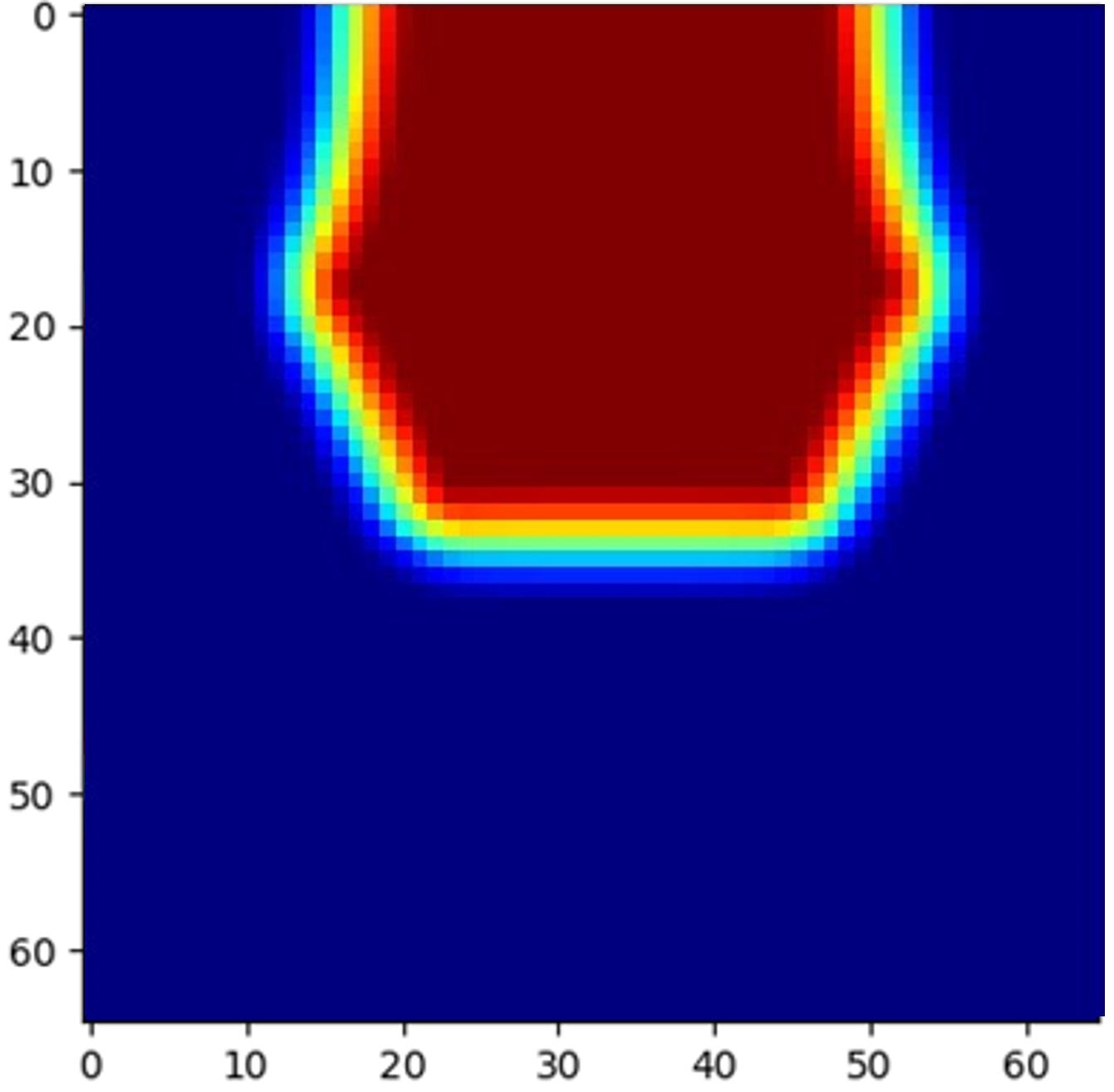}
        \caption{ }
    \end{subfigure}
    \hspace{0.5cm}
    \begin{subfigure}{0.2\textwidth}
        \includegraphics[width=\linewidth, height=\textheight, keepaspectratio]{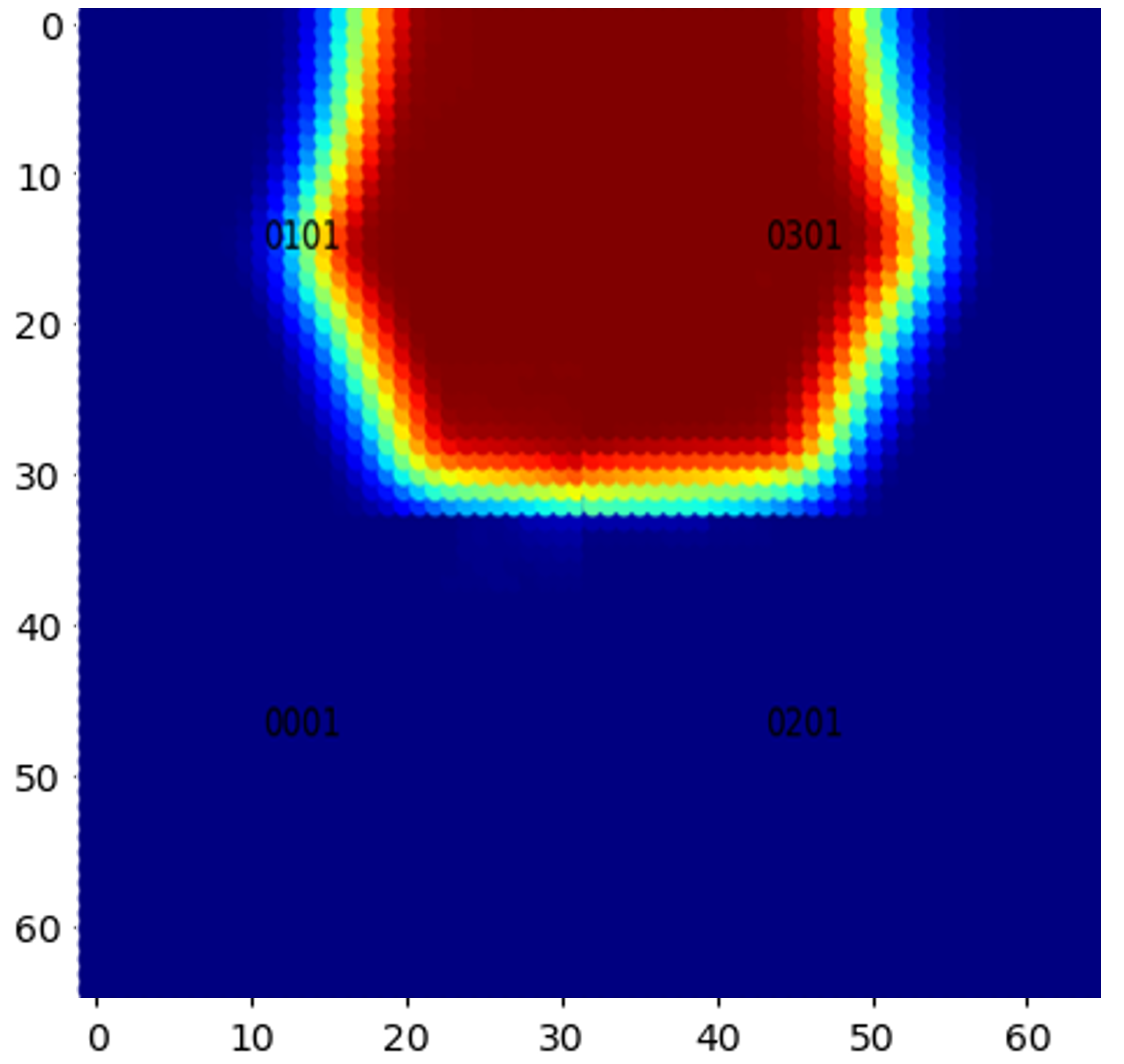}
        \caption{ }
    \end{subfigure}
     \vspace{0.5cm}
 
    \begin{subfigure}{0.2\textwidth}
        \includegraphics[width=\linewidth, height=\textheight, keepaspectratio]{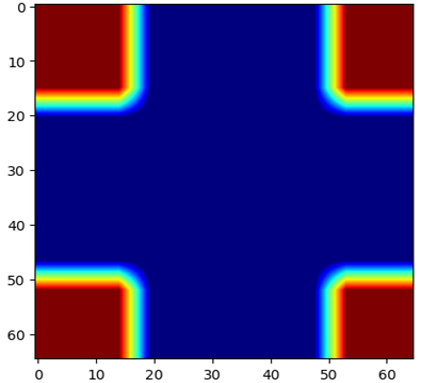}
        \caption{ grain 2}
    \end{subfigure}
    \hspace{0.5cm}
    \begin{subfigure}{0.2\textwidth}
        \includegraphics[width=\linewidth, height=\textheight, keepaspectratio]{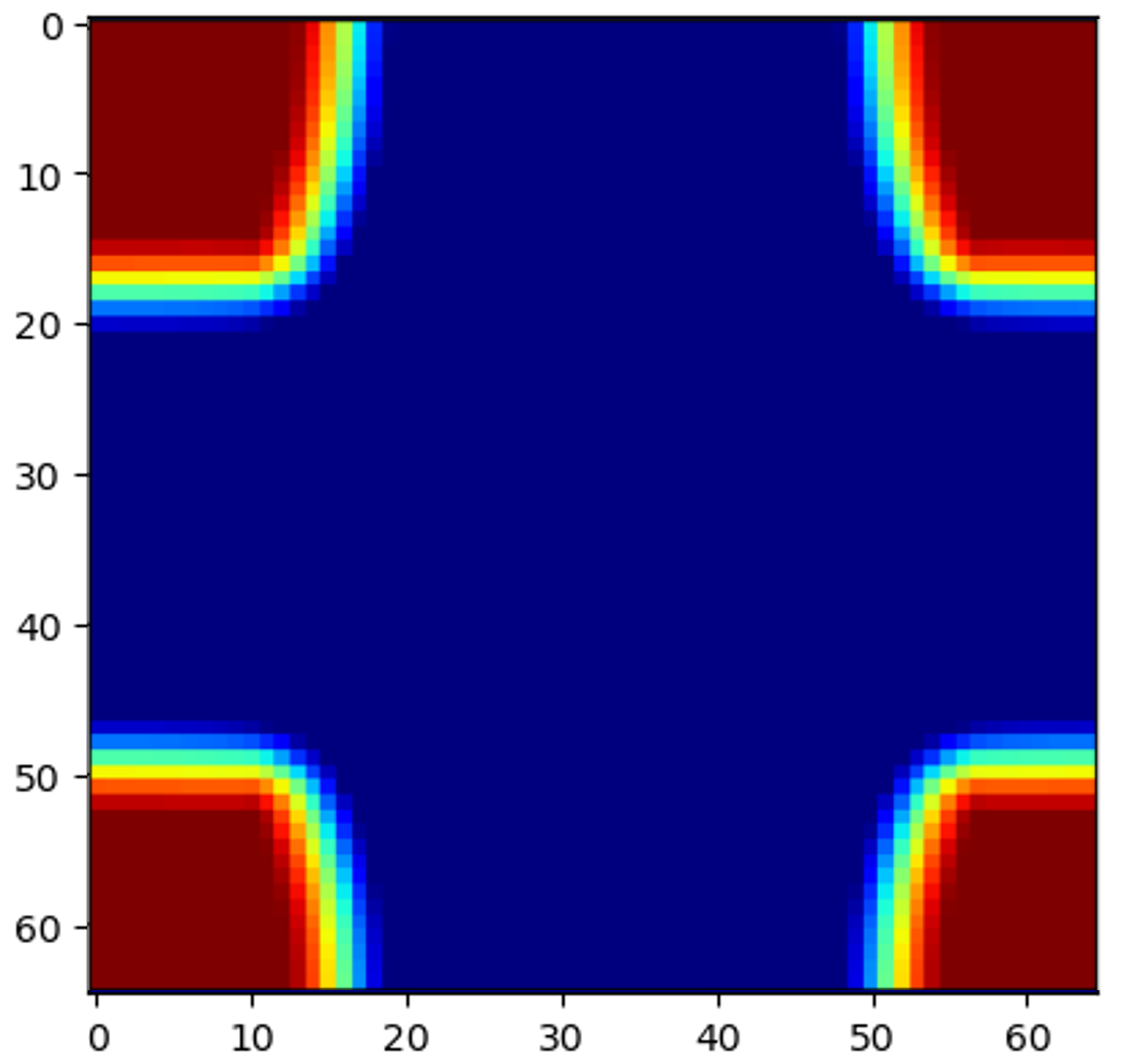}
        \caption{ }
    \end{subfigure}
    \hspace{0.5cm}
    \begin{subfigure}{0.2\textwidth}
        \includegraphics[width=\linewidth, height=\textheight, keepaspectratio]{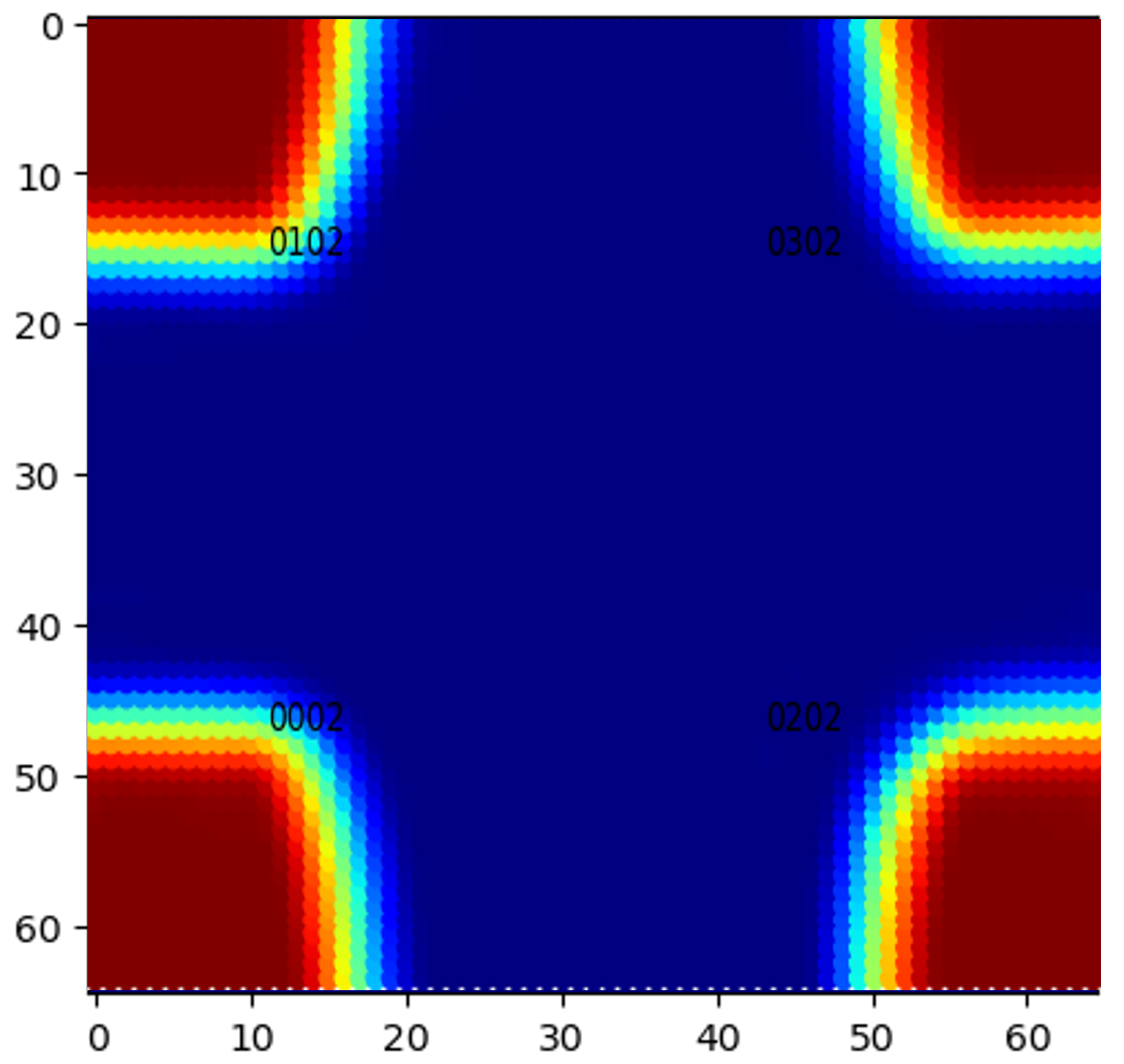}
        \caption{ }
    \end{subfigure}
     \vspace{0.5cm}
     
    \begin{subfigure}{0.2\textwidth}
        \includegraphics[width=\linewidth, height=\textheight, keepaspectratio]{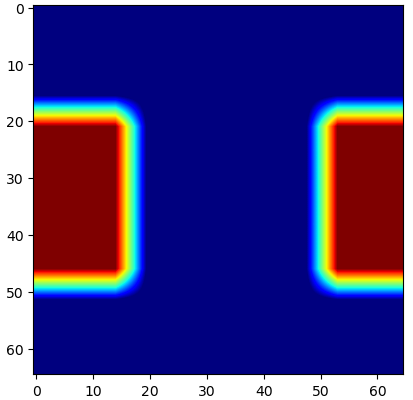}
        \caption{ grain 3}
    \end{subfigure}
    \hspace{0.5cm}
    \begin{subfigure}{0.2\textwidth}
        \includegraphics[width=\linewidth, height=\textheight, keepaspectratio]{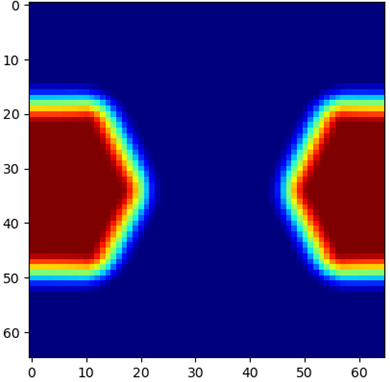}
        \caption{ }
    \end{subfigure}
    \hspace{0.5cm}
    \begin{subfigure}{0.2\textwidth}
        \includegraphics[width=\linewidth, height=\textheight, keepaspectratio]{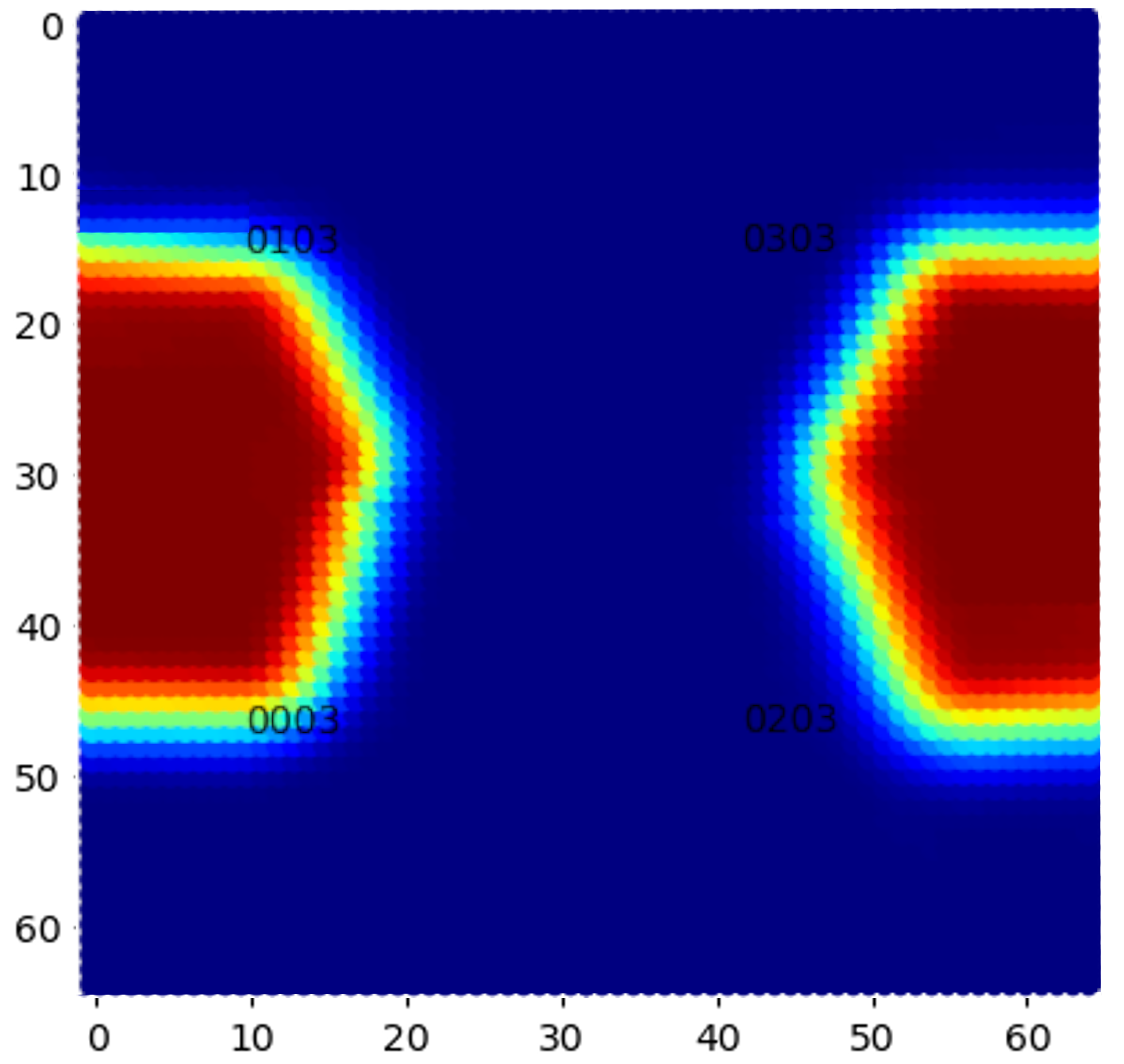}
        \caption{ }
    \end{subfigure}
     \vspace{0.5cm}
     
    \begin{subfigure}{0.2\textwidth}
    
    \includegraphics[width=\linewidth]{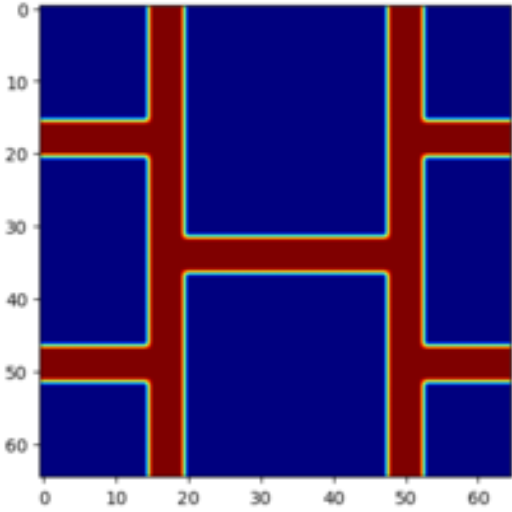}
        \caption{Sum of Interfaces}
    \end{subfigure}
    \hspace{0.5cm}
    \begin{subfigure}{0.2\textwidth}
    \includegraphics[width=\linewidth]{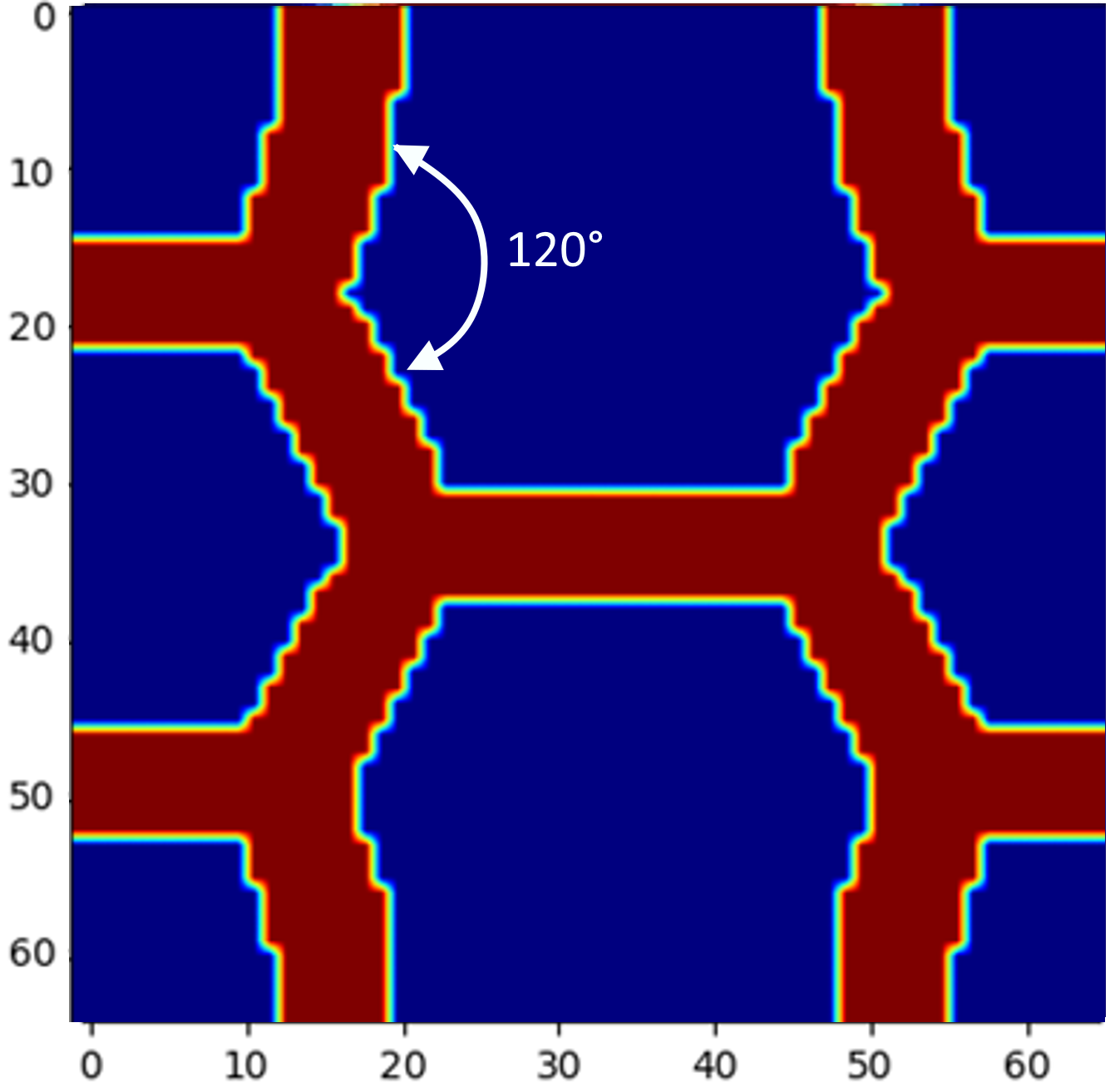}
        \caption{ }
    \end{subfigure}
     \hspace{0.5cm} 
    \begin{subfigure}{0.2\textwidth}
    \includegraphics[width=\linewidth]{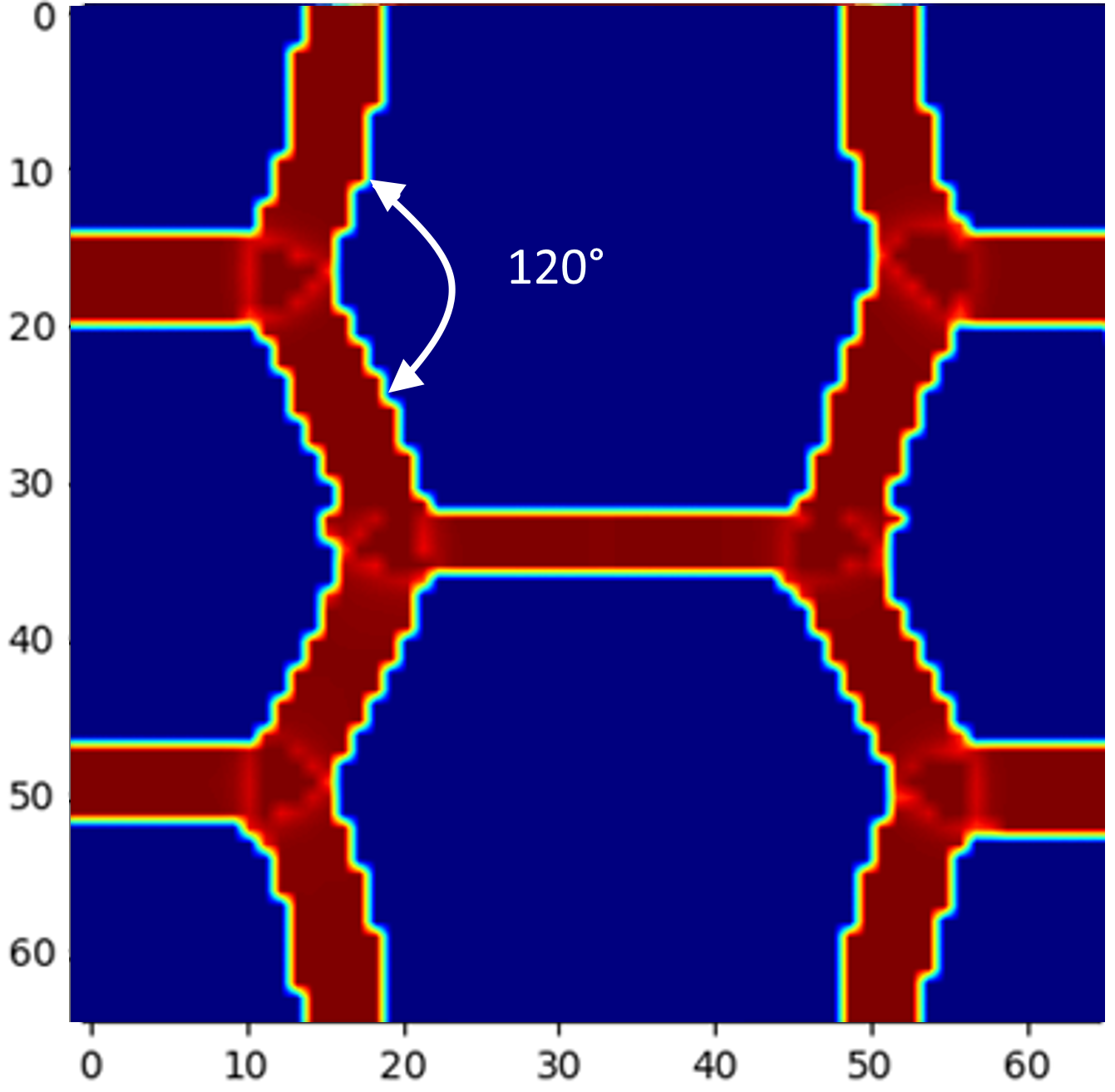}
        \caption{ }
    \end{subfigure}
    
    \caption{Pure Phase-field solution (phase-field $\phi$) against PINN predictions for a triple-junction problem involving four grains/phases in the system.}
    \label{fig:triple_junction}
\end{figure*}

The simulation results are visually depicted in Figure \ref{fig:triple_junction}.
Here, each grain is shown in its initial and final state of the simulation, compared to the simulation results from OpenPhase.
Figure \ref{fig:triple_junction}(m)-(o) show the total interfacial region in the initial and final state.
The comparison between PINNs-MPF and OpenPhase reveals an excellent agreement.
The microstructure evolves into an equilibrium state where the angles at the triple junctions approach 120 $^\circ$.
One can see that the motion of the phases is well synchronized, allowing the global solution to converge to the equilibrium state. 
Note that the application of boundary conditions on the outer boundary of the simulation box requires the interfaces on both ends to adjust in an energy-minimizing manner. 

The difference between the PINNs-MPF solution and OpenPhase lies in the interface thickness. This difference may be justified by the gradient topology of each methodology: 
It is indeed noted that the gradient calculation employed in the implementation, specifically using the "gradient.tape()" module in TensorFlow, differs from the Laplacian used in the regular grid computation for MPF calculations. 
Additional details about the difference in the computation of the Laplacian terms are given in the SM, section A. 
This difference explains the minor discrepancies related to the interface width. 
However, it can be asserted that PINNs predictions exhibit greater fidelity compared to OpenPhase when preserving similar interface widths to the initial state. 
We also note that, due to the curvature-driven nature of grain growth problems, handling an initialization with sharp angles (90$^\circ$) is challenging. 
This demonstration of the correct evolution of such a critical initialization consolidates the use of such a framework in increasingly difficult contexts. 
From a technical standpoint, it's worth noting that this results in relatively slow training in the beginning until addressing the sharp angles, after which the training accelerates.

\section{Discussion} \label{discussion_part}
\paragraph{Analyzing the training dynamics through the loss functions }
\mbox{}
\par
\begin{figure}[htbp]
    \centering
    \begin{subfigure}{0.75\textwidth}
     \includegraphics[width=0.75\linewidth]{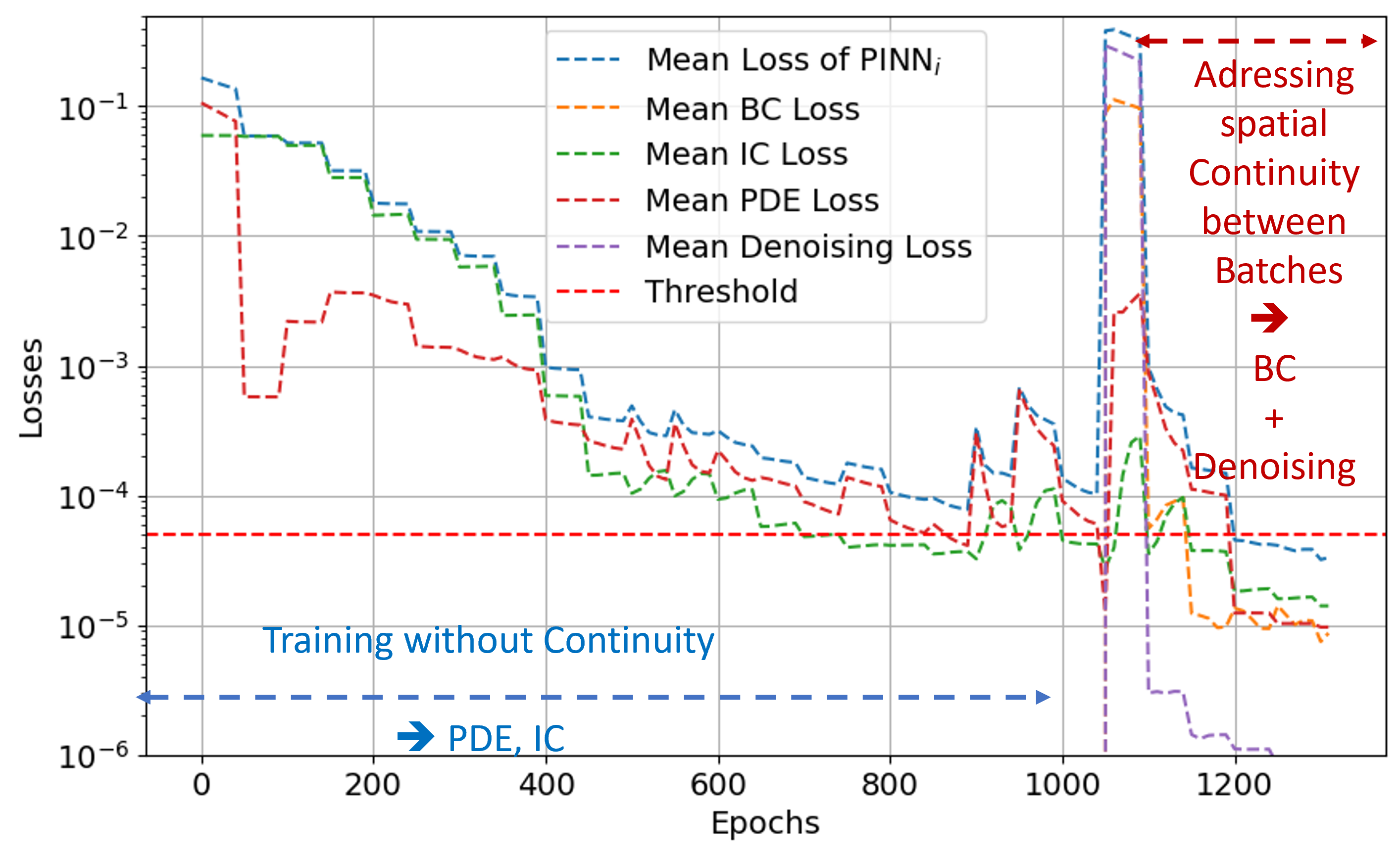}
        \caption{Training on the first time interval $t_0$-$t_1$ }
    \end{subfigure}
    \quad
    \begin{subfigure}{0.8\textwidth}
     \includegraphics[width=0.75\linewidth]{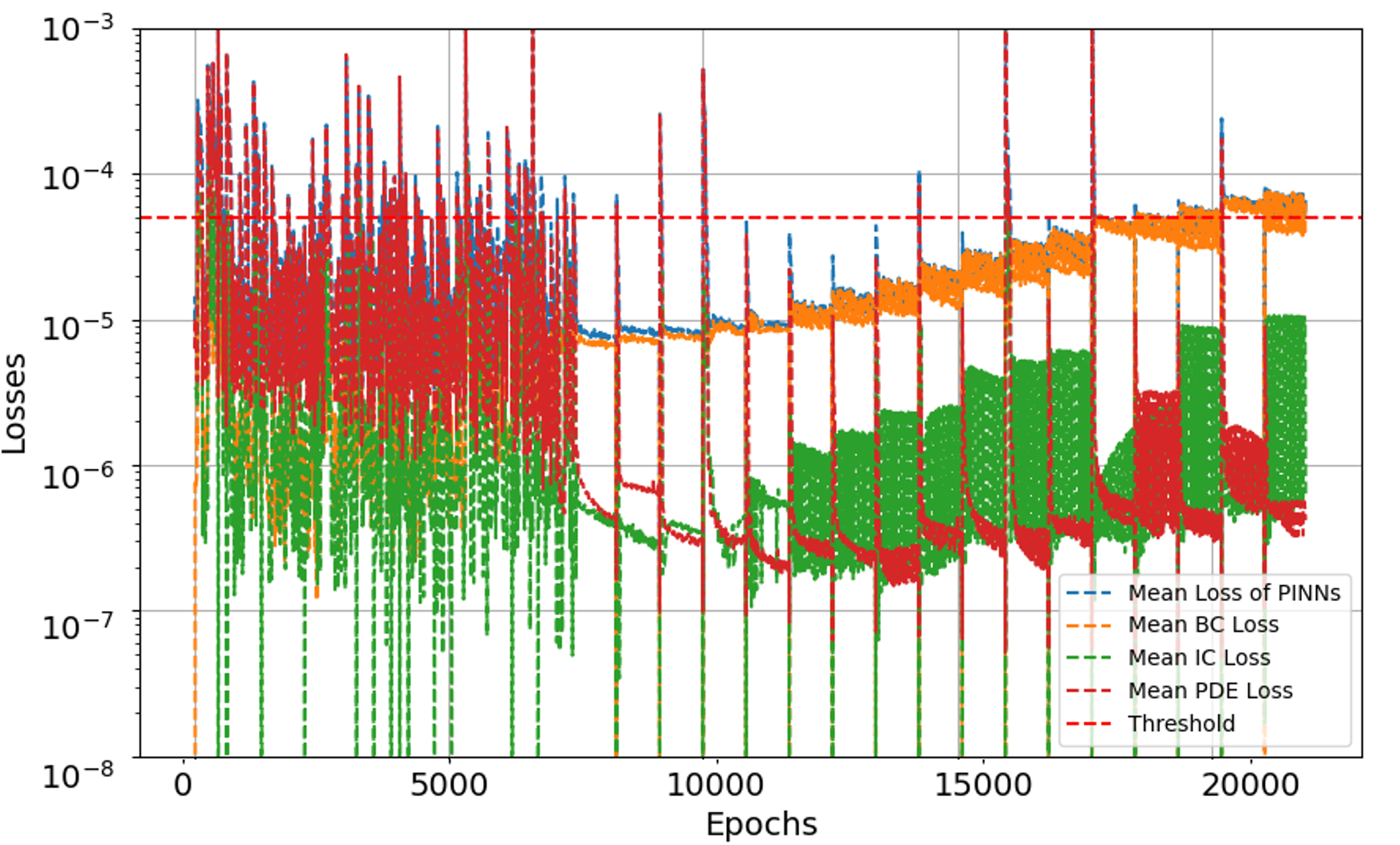}
     \caption{Training on the whole time domain) }
    \end{subfigure}
    
    \caption{Evolution of the different losses (PDE, IC, BC, denoising and mean loss) for the four \textit{PINN}$_i$s over the epochs for a grain shrink without driving force (for an interface width $\lambda= 7dx$). The plots are displayed in a logarithmic scale. In (b), values below $10^{-8}$ and above $10^{-3}$ are excluded from the plot for visualization purposes. Additionally, the denoising loss is omitted from the plot due to its low magnitude and fluctuating nature, which may affect the visualization of other losses.}
    \label{fig:losses_grain_schrink}
\end{figure}

 The training of PINNs-MPF involves various loss terms, described through Eqs. \ref{eq:theta_PDE} to \ref{eq:theta_sum_phi}.
To understand and optimize the training process necessitates a comprehensive analysis of each loss in terms of their individual behavior and collective synchronization. 
\newline For the grain shrinking scenario (paragraph \ref{Curv_driv_int_Motion}), the evolution of various mean losses for the four \textit{PINN}$_i$s across epochs is shown in Figure \ref{fig:losses_grain_schrink}.
Here, (i) each loss curve corresponds to the average of the four corresponding losses from the four neural networks (NNs) and (ii) the entire training process is technically conducted in a single execution. However, for the sake of clarity in illustrating the evolution of losses, we choose to present the simulation in two stages, such that, after convergence in the initial time interval, the acquired learning is then applied to restart the simulation. 

Figure \ref{fig:losses_grain_schrink}a shows the evolution of the different losses for the first time interval ($t_0$-$t_1$). 
The threshold is set at $5 \times 10^{-5}$, marking the transition point to the next time interval.
The training initially focuses on the interfacial regions, activating only the PDE and IC losses. Once the solution is computed in these regions, BC and denoising losses are subsequently activated. This aims to establish spatial continuity between batches and eliminate noise within both grain and non-grain areas, respectively. The discrete activation of BC and denoising loss has proven to be more efficient in enhancing training compared to continuous activation. It is proposed to categorize the losses into intrinsic losses, including partial PDE and IC losses, representing core elements for the NN. On the other hand, extrinsic losses, here BC and denoising, play a role in fine-tuning the solution and establishing continuity with neighboring NNs. These extrinsic losses could be subjected to optimized integration.

Figure \ref{fig:losses_grain_schrink}(b) shows the local minimums corresponding to this transition, where all maximum losses of \textit{PINN}$_i$s (as defined in Eq. (\ref{theta_theori})) fall below this threshold, indicating the onset of training for the next time intervals. The gaps between peaks correspond to the length of each time interval. 
The simulation could be then divided into two distinct phases. Initially, when the grain radius is substantial, the training progresses relatively fast. 
Subsequently, in the second phase with a smaller grain size, the BC loss exhibits an inverse relationship with the grain radius, becoming the most influential. The seamless transfer of learning between time intervals becomes evident as the PDE loss initially tends to decrease and then stabilizes in the later stages of training. Notably, the magnitudes of losses appear reasonable, with the PDE loss being approximately ten times greater than that of IC. If the BC loss remained activated throughout the entire simulation, in the subsequent benchmark (triple junction), it was managed through discrete activation, mirroring the approach adopted during the initial time interval.

\begin{figure}[!ht] 
    \centering
    \begin{subfigure}{0.75\textwidth}
     \includegraphics[width=0.9\linewidth]{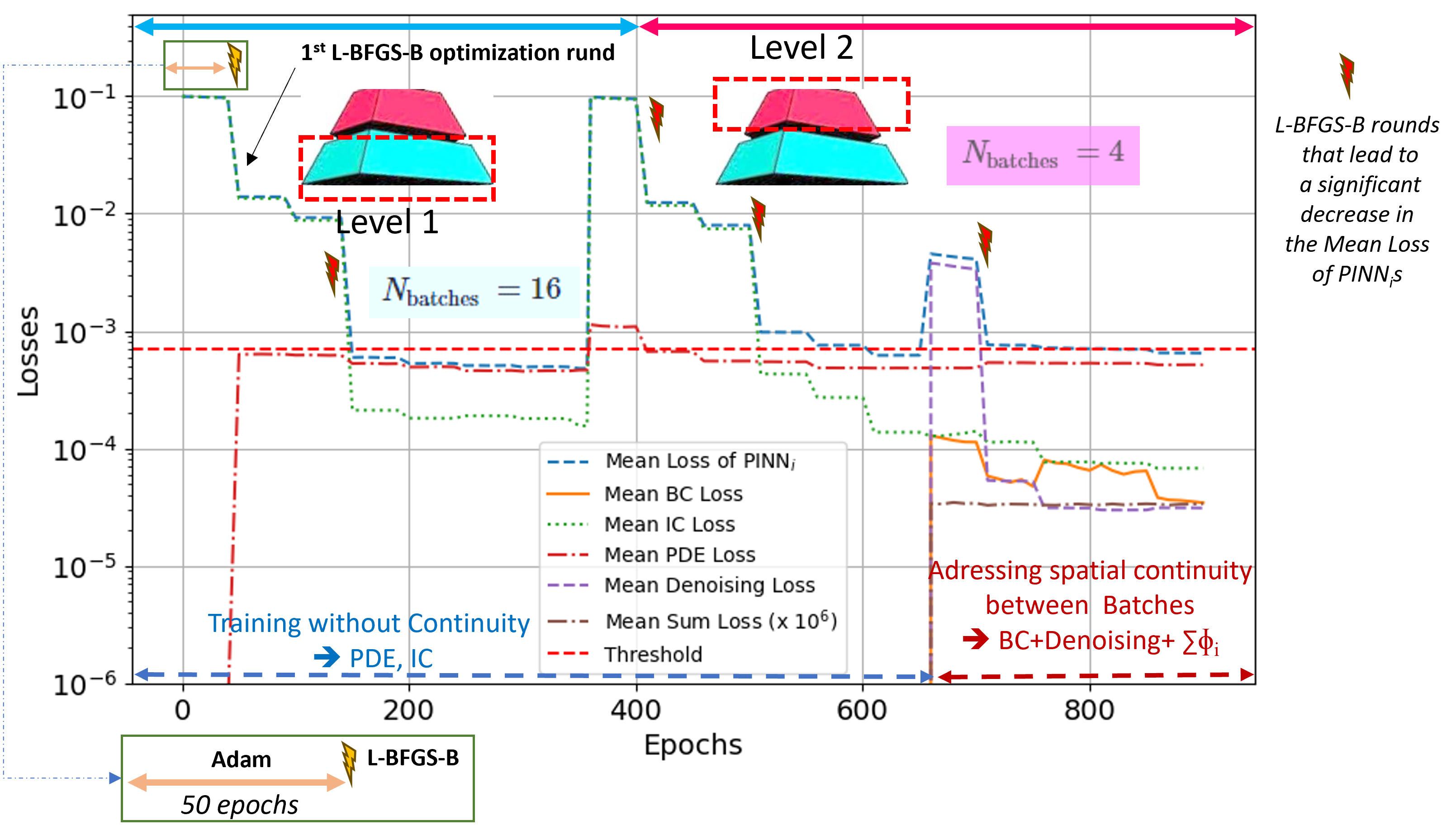}
\caption{\colorbox{white}{\textcolor{black}{Training on the first time interval $t_0$-$t_1$ using the pyramidal approach.}}}
    \end{subfigure}
    \quad
        \centering
    \begin{subfigure}{0.75\textwidth}
     \includegraphics[width=0.75\linewidth]{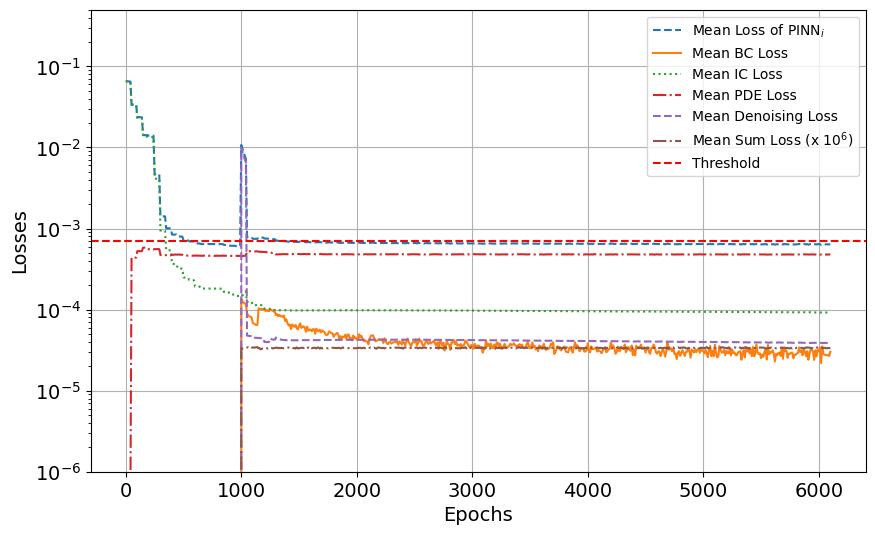}
      \caption{Training on the first time interval $t_0$-$t_1$ without the pyramidal approach. }

    \end{subfigure}
    \quad
        \centering
    \begin{subfigure}{0.75\textwidth}
     \includegraphics[width=1\linewidth]{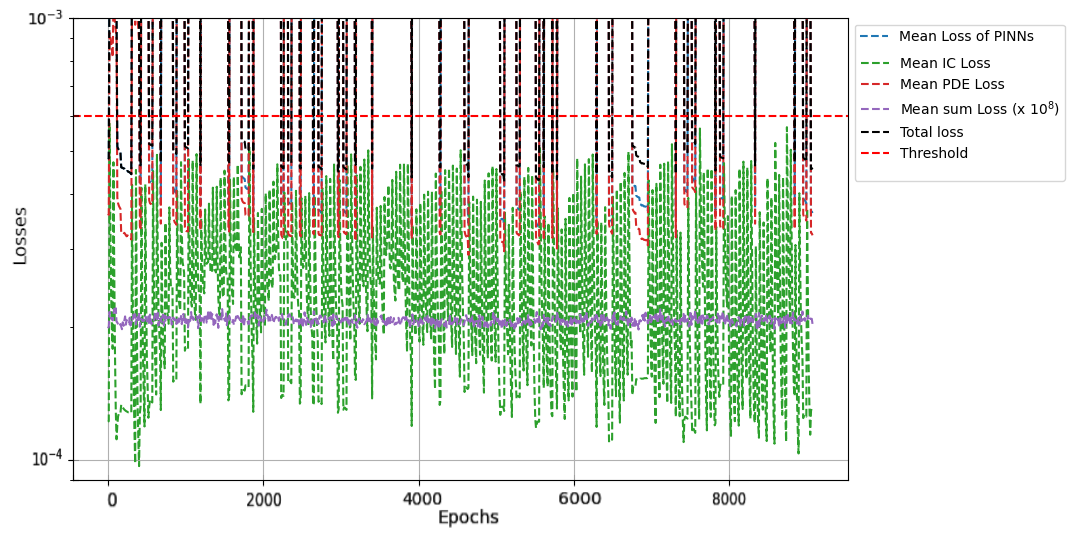}
        \caption{Training on the whole time domain. }
    \end{subfigure}
     \caption{Evolution of the different losses for the 16 \textit{PINN}$_i$s over the epochs for the triple junction scenario. The plot is displayed in a logarithmic scale. 
 }
    \label{fig:triple_junction_loss}
\end{figure}

For the triple junction scenario (paragraph \ref{triple_junction_sec}), the evolution of various losses (PDE, IC, BC, and mean loss) for the 16 NNs across epochs is illustrated in Figure \ref{fig:triple_junction_loss}. \\ 
In this context, the threshold is now set at $6 \times 10^{-4}$.

Similar to the grain shrinking scenario, the first time interval is addressed, and then the cumulative learning is used to restart the simulation. For this purpose, two approaches are compared, with and without a pyramidal approach (respectively in Figures \ref{fig:triple_junction_loss} a and b). It is noted that the only difference between the two simulations is the number of batches for the fine subdivision $N_{\text{batches min}}$ (c.f. Algorithm \ref{general_algorithmic}). Specifically, $N_{\text{batches min}}$=16 and $N_{\text{batches}}$=4 for the pyramidal approach, while $N_{\text{batches min}}$=$N_{\text{batches }}$=4 for the second configuration.
In Figure \ref{fig:triple_junction_loss}(a), the impact of the optimization techniques is subsequently showcased: wheel of optimizers and the pyramidal training. The wheel of optimizers is activated with a period of 50 epochs; L-BFGS-B initially decreases the global loss from a magnitude of $10^{-1}$ to $\times 10^{-2}$ MSE, then a progressive decrease is achieved with subsequent rounds to below the threshold of $5 \times 10^{-4}$ MSE. Then, Adam allows the transfer of learning from level 1 to level 2 of the pyramid, where the whole spatial domain is handled. Even if the global loss increases initially, it quickly falls below the target threshold (the 16 NNs converge together within 850 epochs), demonstrating the model's capacity to generalize the learnable solution to the whole domain. However, the non-adoption of such a strategy makes it hard to enforce the convergence of all the NNs, as seen in Figure \ref{fig:triple_junction_loss}(b), and the global loss stagnates above the threshold even after 6000 epochs of training, using the same set of parameters as in Figure \ref{fig:triple_junction_loss}(a) except for the non-activation of the pyramidal training. It is worth noting that such a comparison was subjected to excessive re-execution for repeatability with the same outputs; the 16 NNs failed to converge simultaneously without the activation of pyramidal training during the first time interval.
It is also suggested that this finding is also attributed to two main impacts of pyramidal training: the progressive data augmentation basis of this concept, and the facilitation of addressing boundary conditions, as accurate boundary predictions are a direct consequence of good convergence within the domain.
Additional details about the domain decomposition related to the pyramidal approach are provided in the SM, section D. 

The black dashed line in Figure  \ref{fig:triple_junction_loss}c corresponds to the total mean loss, exhibiting oscillations that dip below the threshold (red line) for the enhancement of the next training interval. Here, the boundary condition loss is not visualized due to its discrete activation; however, its magnitude is noted to be in the range of  $10^{-5}$.
It is noteworthy that all losses for all NNs should fall below this limit (not just the mean values). 
Previous observations regarding the behavior of the IC and PDE losses remain relatively valid in terms of the stabilization of losses across epochs. 
However, the declining aspect observed in the previous scenario is not evident here. 
This discrepancy could be attributed to the complexity of the triple junction evolution and dynamic interactions between different phases when compared to the single-grain shrinkage scenario. 
The sum loss (Eq. (\ref{eq:summConstraint})) has the lowest magnitude ($10^{-12}$), and its constant values reflect the algorithm's capacity to dynamically correct predictions, ensuring adherence to the summation constraint.
\paragraph{Addressing Nonlinearities in Microstructure Prediction } 
\mbox{}
 \par The current implementation showcases the effectiveness of this method in addressing non-linearities such as in the MPF model, offering good predictions of microstructure evolution in different scenarios above.
The ability to capture both curvature-driven and driving-force-driven interface dynamics is well demonstrated.
Typical MPF simulations employ a time-stepping approach with a magnitude of up to 10$^{-4}$ and a minimum of 1000 steps across the given scenarios that is to ensure computational stability. 
In contrast, applying PINNs is shown here to allow for the selection of much larger time-stepping up to 200 time steps, except for the traveling interface simulation where computational speed was a focal consideration. 
This difference stems from the fact that while the conventional MPF is limited due to the spatial resolution, the PINNs-MPF focuses on the overall microstructural patterns, substantiating its ability to capture underlying physics while maintaining stable computational schemes.
\paragraph{Tuning the model hyper-parameters} 
\mbox{} \\
One of the crucial points when dealing with PINNs is the tuning of the hyper-parameters. Here, any simplification made for hyper-parameters directly impacted convergence and can reduce the efforts for trial and error. In terms of the training dataset, this is automatically set, simplifying user inputs to the initialization, physical, and geometrical parameters.
On the other hand, we have adopted major concepts from conventional MPF frameworks, namely the OpenPhase \cite{darvishikamachali2013grainPhD}. This includes considering our simulation domain heterogeneously, with a focus on the phase-field interfacial region, while parallelizing the handle of individual phase-field variables. This demonstrates that the PINNs-MPF can successfully inherit optimization techniques already applied in classical computing approaches. Especially, the PINN resolution can be then assimilated as a multi-variable sequential learning, showcasing its adaptability and integration capabilities.
The application of multi-networking and training by blocks demonstrates the feasibility of addressing diffuse interface problems with simplified hyper-parameters. 

\begin{figure}[!ht] 
    \centering
    \includegraphics[width=0.75\linewidth]{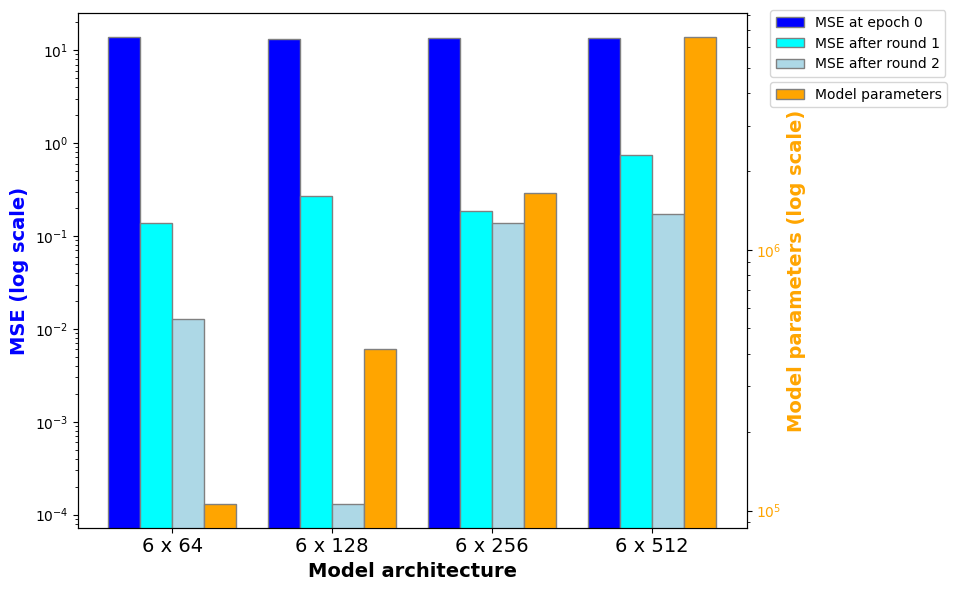}
    \caption{Evolution of MSEs and model parameters count for different architectures as variants of the benchmark 2.}
    \label{fig:virt_exp_fig}    
\end{figure}

For specific challenges handling moving boundaries, the implementation of adaptive weights (gradient scaling algorithm) was unavoidable \cite{Haghighat2022}.
It is therefore worth noting that the loss terms in the implemented PINNs-MPF are by default left unweighted, i.e. $\lambda_{\mathrm{pde}}=\lambda_{\mathrm{bc}} = \lambda_{\mathrm{ic}} =\lambda_{\sum \phi}  = 1$ in Eq. \ref{theta_theori}. 
This reduces the complexity of the current framework and the need for meticulous tuning, especially when considering the diverse nature of loss terms in Eq. \ref{theta_theori}.
Moreover, this kind of algorithm could be useful to integrate in more complex scenarios. 
We note that our numerical experiments have indicated that the extrinsic losses can be weighted to optimize performance; for instance, reducing the weights for the BC loss to expedite training, and conversely, increasing the weighting of the phase summation loss to allocate more attention to it. However, the intrinsic losses (IC and PDE) cannot be weighted.
\paragraph{Interfacial Attention for a Latent Microstructure Representation}
\mbox{} \\
The results of the conducted benchmark studies indicate that the global PINN solution in simulations involves a precise computation of solutions along interface regions throughout the entire domain, complemented by an extrapolation method to approximate solutions at the interface region. This hybrid approach proves beneficial for diffuse-interface problems using ML. 
In the triple junction scenario, the distributed '0 and 1' points, as depicted in Figure \ref{fig:triple_junction_IC}, were employed to enhance the precision of the solution, particularly for the application of phase summation (Eq. \ref{eq:MPF_sum}) and accurate addressing of boundary conditions between neighboring NNs. Similarly to the mono-phase benchmarks, it is possible to fully focus on interfacial regions within multi-phase simulations. First demonstrations of this are provided within Section E of the SM. 
This opens the way to resolve MPF simulations using NNs independently of the grid size, thus allowing for tackling more complex scenarios.
\paragraph{Dynamic correction mechanism for phase summation constraint} 
\mbox{} \\
Within the literature, a major limitation against the application of ML to resolve multi-phase problems is the phase summation criterion: 
Indeed, in previous works, a requirement for a bounding algorithm was obligatory as a correction step to the PINN prediction \cite{HUANG2021103727,Zheng2022}. 
Meanwhile, in the current work, the applied method introduces a dynamic correction mechanism for phase predictions. This involves dividing the prediction of each neural network (NN) responsible for a specific phase by the sum of predictions from other NNs handling other phases within the same batch. This dynamic correction not only ensures continuous training but also suppresses the need for an external correction algorithm. Instead, the correction of phase summation is seamlessly integrated as an additional constraint loss term, enhancing the efficiency of the training process.
\paragraph{Exploring model complexity and impact of the optimization strategy} \label{num_exp}
\mbox{} \\
The results about the additional numeric experiment are gathered in Table \ref{tab:virtual_exp}. Associated graphical illustration is provided in Figure \ref{fig:virt_exp_fig}.
The experiments revealed that while all architectures started with high and similar MSE values, the MSE decreased significantly after the first round of Adam optimization. The four models reach similar loss magnitude after the first round. Subsequent L-BFGS optimization further reduced the MSE for the $6 \times 64$ and $6 \times 128$ architectures, with the $6 \times 128$ model attaining the best overall MSE of $1.29 \times 10^{-4}$. Larger architectures like $6 \times 256$ and $6 \times 512$ failed to converge within the specified rounds, potentially due to overfitting or optimization challenges arising from their high parameter counts. The $6 \times 128$ architecture struck an optimal balance between model complexity and generalization ability, benefiting from the combined Adam and L-BFGS optimization strategy. While larger models had more parameters, they did not necessarily perform better and could be computationally expensive. 
This numerical experiment reveals a solid correlation between the model's parameter count, representing its complexity, and the impact of multi-networking. For instance, the $6 \times 256$ configuration, yielding 1,651,205 trainable parameters, challenges training when employing four NNs. Conversely, a reduced parameter count of 1,414,417 enables training 16 NNs (16 x (6 HL × 128 neurons)) simultaneously to effectively handle four phases (c.f. the triple junction benchmark). Moreover, even higher parallelism is feasible through pyramidal training, where initially 64 NNs are initialized, but only 16 are selected for further training using the basket of PINNs concept. That implies that using moderately-sized networks through the multi-networking context (the extended subdomain decomposition technique) allows to efficiently  alternate optimizers (the wheel of optimizers), while the transfer of learning facilitates the transition between time intervals. This underscores the rationale behind initially selecting a subset of PINN techniques, while the second subset (AMFO, phase correction, and vertical optimization) has proven crucial in prior MPF method studies and the current work. Pyramidal training and the comprehensive set of PINNs emerge as essential for progressively transferring learning to a unified domain when dealing with massive domain decomposition, thereby making the framework ready for up scaling, as hereafter detailed.
\begin{table}[htbp]
\caption{Evolution of MSE and model trainable parameters count for different architectures: optimization rounds with Adam-L-BFGS-Adam cycles. Comparisons are done at epoch 0, after first and second rounds.}
\label{tab:virtual_exp}
\centering
\begin{tabular}{|c|c|c|c|c|c|}
\hline
\textbf{\begin{tabular}[c]{@{}c@{}}HL\\  $\times$ \\ neurons/HL\end{tabular}} & \textbf{\begin{tabular}[c]{@{}c@{}}Trainable \\ parameters\end{tabular}} & \textbf{\begin{tabular}[c]{@{}c@{}}MSE\\  at epoch 0\end{tabular}} & \textbf{\begin{tabular}[c]{@{}c@{}}MSE \\ after round 1\end{tabular}} & \textbf{\begin{tabular}[c]{@{}c@{}}MSE\\  after round 2\end{tabular}} & \textbf{\begin{tabular}[c]{@{}c@{}}Rounds\\  required to\\converge $^(*)$\end{tabular}} \\ \hline
6 x 64                          & 105,605                                                              & $1.396 \times 10^{1}$                                              & $1.387\times 10^{-1}$                                              & $1.270^{-2}$                                                          & 5                                                                                 \\ \hline
6 x 128                         & 416,005                                                                & $1.33 \times 10^{1}$                                               & $2.712 \times 10^{-1}$                                                & $1.29 \times 10^{-4}$                                                 & 3                                                                                 \\ \hline
6 x 256                         & 1,651,205                                                              & $1.36 \times 10^{1}$                                               & $1.87 \times 10^{-1}$                                                 & $1.387 \times 10^{-1}$                                                & $\succ$ 10                                                           \\ \hline
6 x 512                         & 6,579,205                                                              & $1.351 \times 10^{1}$                                              & $7.517 \times 10^{-1}$                                                & $1.723 \times 10^{-1}$                                                & $\succ$ 10                                                           \\ \hline
\end{tabular}
\footnotesize{$^(*)$ It is worth reminding that convergence signifies that the global loss of the model has fallen below the threshold and the current time interval has been managed.}\\

\end{table}

\paragraph{Transfer of learning} 
\mbox{}
\par In the context of learning transfer, our framework demonstrates the feasibility of hierarchically transferring knowledge within subdomain decomposition. Through multi-networking and full parallel training, it employs a pyramidal training approach, categorized as 'Multiple to Multiple' transfer. This allows the transfer of learning, for example (but not limited to), from 4 NNs to 4 others, 16 to 16, 16 to 4, etc. From a technical standpoint, achieving 'Multiple to Single' transfer of learning is possible, where knowledge is transferred from multiple NNs to a single NN and vice versa. However, our first trials showcased that training a single NN  revealed challenges, leading to optimization failures. The synergy between ML and optimization algorithms is crucial for effectively exploiting patterns, but this task becomes challenging with large NNs. Multi-networking and pyramidal training until reaching a reasonable number of NNs proves efficient in managing such complexities. Nevertheless, we provide within the actual code repository a demonstration of this 'Multiple to Single' transfer, denoted as '${\textit{PINN}}_{i}s$ to \textit{Master}' transfer, that could be of interest for other physical applications. Additional details are provided in the SM, section F.
\paragraph{Exploring more ways for improvement of the framework} 
\mbox{}
\par In the current development phase, one aspect to consider is selecting the appropriate threshold to transition between time intervals in the full-discrete resolution. This threshold was found to be strictly related to each simulation and subjected to trial and error. For the magnitude of 10$^{-5}$ in single-phase benchmarks, it was slightly raised when handling the multiphase field (MPF) scenario to ensure a compromise between the speed of the training and the convergence of the solution. One possibility to deal with this limitation is to integrate an additional loss term related to the variation in the energy of the system. If the energy residual (and its derivative) reaches a steady state, it is then a physical indication that the system reaches its equilibrium within this time interval, and then the transition could be ensured \cite{GUILLENGONZALEZ2014821,WANG2023216}. 
Theoretical formulations and graphical illustrations related to the utility of integrating energy loss to enhance training are provided in the SM, section B. 

A potential point of improvement for the current PINNs-MPF framework is the integration of sequential training. Indeed, optimizing the evolution of losses related to the triple junction motion could be achieved if the model captures the temporal behavior of the solution. 
The flexibility of discrete resolution under the current implementation eases the integration of sequential learning through Recurrent Neural Networks. 
Optimizing runtimes against theoretical solutions or existing literature is beyond the scope of this study, as PINNs are inherently more time-consuming than classical approaches. Enhancing PINN solvers is a work in progress \cite{cuomoscientific2022,chuang2022experience} and studies on the use of PINNs for solving multi-phase problems remain uncommon \cite{Haghighat2022} . 
Therefore, the first goal of this study is to prioritize solution fidelity, training continuity, and efficient resource distribution. This approach is similar to recent studies focusing on multiphase phenomena, particularly in fluid dynamics. For instance, Haghighat et al. \cite{Haghighat2022}  have concentrated on the methodology and challenges associated with training PINNs for coupled flow and deformation in porous media. Zheng et al. \cite{Zheng2022} have aimed to predict the sequential motions of flow simulations in a discontinuous manner, alternating between PINN training and MCBOM corrections. Amini et al. \cite{AMINI2023112323} have focused on validating the inverse modeling approach targeting nonisothermal multiphase poromechanics.
To conduct running time studies on these multi-phase field frameworks, inspiration could be taken from the study of Shukla et al. \cite{SHUKLA2021110683}. Among other proposals, leveraging multiple GPUs and nodes and adopting a hybrid programming model could be beneficial.
\section{Conclusion}
An interconnected PINNs-MPF framework has been presented to resolve MPF simulation of interface dynamics. 
Our approach has been to address and benchmark most central requirements for a reliable reproduction of MPF simulations, namely, curvature- and driving-force-driven interface motions and the evolution of a triple junction.
In particular, the nonlinear evolution of curved interfaces and the effect of interface width on the interface kinetic were captured, demonstrating the capability of this framework in dealing with diffuse interfaces. 
The simulation of the triple junction established that the application of the sum constraint as a loss term and in combination with a parallelization scheme works precisely and efficiently. 
The rationale behind the current development has been to demonstrate the feasibility of PINNs in capturing key features of microstructure evolution, before increasing complexities. 
This ensures a well-prepared transition to more advanced dimensions while maintaining the robustness and reliability of the methodology.
All benchmark simulations are conducted in a single execution without the need for post-correction algorithms, such as intercalated phase correction algorithms.
Various advanced techniques were encompassed in the PINNs-MPF framework, including training optimization, extended domain decomposition, and boundary condition propagation.
These techniques have yielded accurate predictions and efficient training, establishing a robust foundation for future advancements in this research area. Potential avenues for further exploration include developing more unified solutions through sequential learning, conducting comparative studies with other PINN approaches in the literature, and integrating energy-based optimizations, which is a key component of the MPF Method.
As this progress step, the vision is to extend the capability of the developed framework to address more intricate scenarios involving a greater number of phases, larger grids, and three-dimensional dimensions. 
As the difficulty of the scenarios gradually increases, it is crucial to carefully address related issues when using machine learning to predict solutions, such as managing singularities, mitigating inaccurate predictions, and optimizing computation times.
\section*{Acknowledgments}
RDK acknowledges financial support from the German Research Foundation (DFG) within projects \emph{DA 1655/2-1} (Heisenberg program) and \emph{DA 1655/3-1}.

\section*{Supplementary material}
Additional results and discussions as well as a video presentation of the framework are given in the Supplementary Material attached.

\section*{Data availability}
The PINNs-MPF repository is available on GitHub at the following link: \\
\url{https://github.com/SFETNI/PINNs_MPF}

\section*{Competing financial interests}
The authors declare no competing financial interests.

\printbibliography


\section*{Supplementary material}

# Supplementary Material for
# PINNs-MPF: A Physics-Informed Neural Network Framework for Multi-Phase-Field Simulation of Interface Dynamics

Seifallah Elfetni and Reza Darvishi Kamachali

Federal Institute for Materials Research and Testing (BAM), Unter den Eichen 87, 12205 Berlin, Germany

In this document, we provide additional materials related to the current work. The section A focuses on showcasing the numerical scheme for pure phase-field resolution, presenting the sequence of governing equations, and highlighting the contrast in Laplacian computation between OpenPhase and the implemented PINN. Following that, section B introduces energy loss with theoretical demonstrations. Then, section C outlines the meshing methodology and associated optimizations for computational efficiency. Section D showcases additional details to better highlight the mechanism of the Pyramidal training. As a direct exploitation of this, the transfer of learning from multiple networks to a single NN and vice versa is described in section F. A demonstration for the possibility of handling multi-phase-field simulations while focusing only on interfacial regions is introduced in section E.

# Contents

# Specific terminologies of PINNs-MPF

As specific terminologies are introduced in this work, it is important to remind the definition of each one as follows.

- Extended domain decomposition: compared to the domain decomposition proposed in the literature (e.g., XPINN, PPINN), the current framework introduces an extension to handle the phase domain for Multi-Phase Field (MPF) simulations, as well as a specific synchronisation method to handle the phases domain. c.f. the related paper for more details.

- Discrete resolution: we propose to keep this terminology to be aligned with the MPF Method and classical approaches. Regarding the literature of PINNs, this approach makes our framework classed among the time-marching PINNs/temporal PINNs (causality literature).

- Wheel of optimizers: this reflects the training optimization approach by alternating two optimizers, Adam and L-BFGS, in a cyclic way to leverage their respective strengths for improved convergence.

- Round of optimization: a L-BFGS optimization.

- Basket of PINNs: training multiple neural networks simultaneously can be computationally expensive. This concept is technically implemented through the dynamic selection of networks that require further optimization at each L-BFGS round. In the case of pyramidal training, a neural network is selected from each block independently of the level of subdivision, leading to the simultaneous training of up to 16 neural networks for the triple junction benchmark (instead of 64 NNs).

- Pyramidal training: involves gradually subdividing the spatial domain, for each phase/grain, into finer blocks and training selected NNs at the finest level first. The training then transitions to coarser levels, with each selected PINN receiving a warm start from the previously trained PINN at the finer level.

- Adaptive Mesh-Free Optimization (AMFO): is an extreme mesh refinement process adapted from the Adaptive Mesh Refinement (AMR) technique in the phase-field and MPF literature, incorporating dynamic resampling and denoising processes..

- Horizontal optimization: refers to the AMFO.

- Vertical optimization: refer to the optimization of interactions between phases.

## A  Pure Multi-Phase-Field resolution

The Gibbs free energy of a multi-phase-field (MPF) system could be seen (from a first approach) as a summation of an interfacial free energy density $[J.m^{-3}]$ (also called Grain-Boundary (GB)) $f^{GB}$ and an external one $f^{Ext}$ as follows.

$$F = \int_\Omega f^{GB} + f^{Ext}$$

The other contributions could be added added to the MPF free energy formulation, such as the mechanical, chemical and magnetic ones.

$$f^{GB} = \sum_{\alpha,\beta=1..N, \alpha>\beta} \frac{4\sigma_{\alpha\beta}}{\eta} \left\{ -\frac{\eta^2}{\pi^2} \nabla\phi_\alpha \cdot \nabla\phi_\beta + \phi_\alpha \phi_\beta \right\}$$

$\phi_\alpha$ is is the non-conservative phase-field variable corresponding to the phase $\alpha$ (idem for $\beta$), N is the number of components/crystals present in the system, $\sigma_{\alpha\beta}$ is the interfacial energy while $\eta$ is the interface width.

For a system with two-phases (binary), the previous expression becomes:

$$f^{GB} = \frac{4\sigma}{\eta} \left\{ -\frac{\eta^2}{\pi^2} \nabla\phi . \nabla(1-\phi) + \phi(1-\phi) \right\}$$

or simply:

$$f^{GB} = \frac{4\sigma}{\eta} \left\{ \phi(1-\phi) + \frac{\eta^2}{\pi^2} (\nabla\phi)^2 \right\}$$

Th temporal evolution of the non-conserved order parameter $\phi$ follows the Allen-Cahn formulation :

$$\frac{\partial \phi}{\partial t} = -L . \frac{\delta F}{\delta \phi}$$

in 1D, for a dual interface, when considering only the interfacial energy $F = \int_\Omega f^{GB} dx$, the AC formulation gives:

$$\frac{\partial \phi}{\partial t} = -L \left\{ \frac{\partial f^{GB}}{\partial \phi} - \nabla \cdot \frac{\partial f^{GB}}{\partial (\nabla \phi)} \right\}$$

This expression could be also expressed as :

$$\dot\phi = \mu\sigma \left( \nabla^2 \phi + \frac{\pi^2}{2\eta^2} (2\phi - 1) \right)$$

where $\mu = \frac{8\eta L}{\pi^2}$ (for more details about the identification of $\mu$.)

We should note that in the previous equation, only the interface contribution was considered. However, in order to guarantee a moving interface, the non-interfacial contributions should be taken into account:

$$f^{dual} = f^{GB} + h(\phi) \Delta g$$

$\Rightarrow$ The evolution of the order parameter with respect to time is described by this equation:

$$\dot\phi = \mu \left[ \sigma \left( \nabla^2 \phi + \frac{\pi^2}{2\eta^2} (2\phi - 1) \right) + \frac{\pi\sqrt{\phi(1-\phi)}}{\eta} \Delta g \right]$$

where the left side describes the interfacial changes while the right one controls the moving of the interface. To solve this equation, we can use the following time-stepping scheme:

$$\frac{\phi_k^{n+1} - \phi_k^n}{\Delta t} = \mu \left[ \sigma \left( \nabla^2 \phi_k^n + \frac{\pi^2}{2\eta^2} (2\phi_k^n - 1) \right) + \frac{\sqrt{\phi_k(1-\phi_k)}}{\eta} \Delta g \right]$$

then

$$\phi_k^{n+1} = \frac{\phi_k^n + \Delta t \mu \left[ \sigma \left( \nabla^2 \phi_k^n + \frac{\pi^2}{2\eta^2} (2\phi_k^n - 1) \right) + \frac{\pi\sqrt{\phi_k(1-\phi_k)}}{\eta} \Delta g \right]}{1 + \Delta t \mu \sigma \left[ k^2 + \frac{\pi^2}{2\eta^2} \right]}$$

where $\phi_k^n$ and $\phi_k^{n+1}$ are the Fourier coefficients of $\phi$ at time steps $n$ and $n+1$ respectively, $\Delta t$ is the time step size, $k$ is the wave number, and $\eta$ is a constant parameter. The resolution could be performed using or not FFT. Hereafter using en Euler scheme:

$$\phi_i^{n+1} = \phi_i^n + \Delta t \mu \left[ \sigma \left( \nabla^2 \phi_i^n + \frac{\pi^2}{2\eta^2} \left( \phi_i^n - \frac{1}{2} \right) \right) + \frac{\pi\sqrt{\phi_i^n(1-\phi_i^n)}}{\eta} \Delta g \right]$$

Then

$$\phi^{n+1} i,j = \phi^n i,j + \Delta t \mu \left[ \sigma \left( \frac{\phi_{i+1,j}^n + \phi_{i-1,j}^n + \phi_{i,j+1}^n + \phi_{i,j-1}^n - 4\phi_{i,j}^n}{(\Delta x)^2} + \frac{\pi^2}{2\eta^2} \left( \phi_{i,j}^n - \frac{1}{2} \right) \right) + \frac{\pi\sqrt{\phi_i^n(1-\phi_i^n)}}{\eta} \Delta g \right]$$

Here , we note that the Laplacian is applied on a regular periodic grid :

$$\nabla^2 \phi_{i,j} = \frac{\phi_{i+1,j} + \phi_{i-1,j} + \phi_{i,j+1} + \phi_{i,j-1} - 4\phi_{i,j}}{(\Delta x)^2}$$

However, for the resolution of the same problem using PINN, another methodology is applied as follows, once the computing is applied on batches and not on a periodic grid.

**Algorithm A.1:** Compute Laplacian using PINN

**Data:** Input parameters: $x$, $y$, $t$
**Result:** Compute Laplacian using PINN

1 Initialize persistent gradient tape;
  /* Watch variables for differentiation                                        */
  /* Concatenate input variables                                                */
2 $g \leftarrow \text{concatenate}(x, y, t)$;
  /* Evaluate phase field at current location                                   */
3 $\phi_\alpha \leftarrow \text{evaluate}(g)$;
4 $\phi_\alpha \leftarrow \text{clip}(\phi_\alpha, 0.0, 1.0)$;
  /* Watch the phase field for differentiation                                  */
  // Compute gradients
5 $\phi_{t_\alpha} \leftarrow \text{tape.gradient}(\phi_\alpha, t)$;
6 $\phi_{x_\alpha} \leftarrow \text{tape.gradient}(\phi_\alpha, x)$;
  // Watch the phase field for differentiation
7 $\phi_{y_\alpha} \leftarrow \text{tape.gradient}(\phi_\alpha, y)$;
  // Watch the phase field for differentiation
8 $\phi_{xx_\alpha} \leftarrow \text{tape.gradient}(\phi_{x_\alpha}, x)$;
9 $\phi_{yy_\alpha} \leftarrow \text{tape.gradient}(\phi_{y_\alpha}, y)$;
  /* Compute Laplacian                                                          */
10 $\text{lap}(\phi_\alpha) \leftarrow \phi_{xx_\alpha} + \phi_{yy_\alpha}$;

# B Energy controlled problem

In this section, we showcase that the energy residual could be also integrated as an additional loss function and be an alternative to the threshold to move from one time interval to another. If we consider that $\Delta g$ is constant and h depends on $\phi$, then the expression for the energy evolution of the system at each time is given by:

$$\frac{dF}{dt} = \int_\Omega \left\{ \frac{4\sigma}{\eta} \left[ \frac{d\phi}{dt}(1 - 2\phi) + \frac{\eta^2}{\pi^2} 2(\nabla\phi) \cdot (\nabla\frac{d\phi}{dt}) \right] + \frac{dh}{d\phi}\frac{d\phi}{dt}\Delta g \right\}$$

Through this equation, one can differentiate two different energy residuals that should compensate each other during the motion: a term related to the energy derivative while the second one is related to the order parameters. We note that the negative sign in the energy values is only numerical and reflects the shrinking nature of the grain (in the grain shrink scenario).

# C Mesh methodology

In this section, we illustrate the collocation points generated to simulate the motion of the triple junction. As shown in Figure C.1, the meshing approach prioritizes interfacial points, neglecting areas outside the grains. The machine learning solution is computed within interfacial regions, while approximations are made for regions outside the grains.

**Common batches** In the current implementation, 16 neural networks ($PINN_i$) are trained, implying 16 sets of collocation points. To optimize training and reduce computational costs, these 16 sets are reduced to 4 as follows: The $PINN_i$ instances handling the same batch are assigned a common batch. This common batch is constructed based on the initial condition (IC) set of each $PINN_i$, as illustrated in Figure C.2. This approach operates under the assumption that if two phases/grains are in interaction, this occurs within interfacial regions. Consequently, these two phases are assigned a common batch, grouping their interfacial points. This process is then generalized to all interacting phases/grains within the same batch.

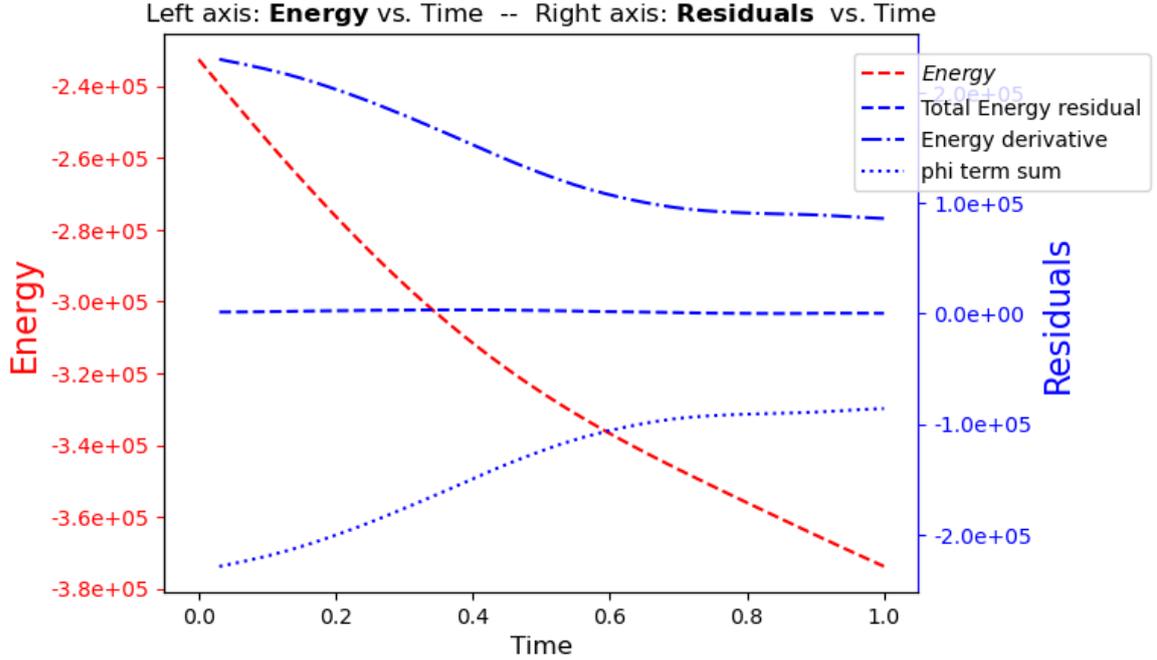

Figure B.1: Evolution of the energy loss terms.

## D  Pyramidal training

The pyramidal training is illustrated in Figures D.1 and D.2 respectively. The pyramidal approach functions as a preliminary training step, characterized by a meticulous subdivision into 16 batches within each phase domain.

**Block training**  Starting with Figure D.1, the bottom level marks the initiation of the pyramidal approach. To optimize training efficiency and avoid deploying 64 NNs/$PINN_i$s for the entire set of batches, here a progressive training is then adopted. Specifically, only highlighted batches from each Quarter/Block are trained in the initial round. This reduces the number of NNs to train to 16, offering the model an opportunity to focus on a smaller spatial area. This enhances the model's ability to capture features and dependencies between outputs and inputs. In this methodology, Block 0 serves as a reference, where the batch with the highest number of interfacial points is chosen for the pre-training process. For subsequent blocks, the selection criterion involves choosing each $PINN_i$ with the same batch number. This ensures consistency in the application of the concept of common batches. We note that the pyramidal training was conducted by meshing only the interfacial zones. Our experiments have indeed shown that incorporating additional points outside these areas proved to be beneficial in constraining the summation of the phases at advanced stages of the training.
Advancing to the upper level of training (Figure D.2), each pre-trained $PINNi$ from the lower level (fine subdivision) transfers its learning to the corresponding $PINNi$ within the same Block in the coarser subdivision. This progressive data augmentation has proven to be efficient in enhancing the training process.

## E  Toward enhanced Multi-Phase Field simulations: full focus on interfacial regions for scalable grid handling

First results for the triple junction without full attention to interfacial regions are available in Figure E.1. In the main benchmark of the associated manuscript, grain/no-grain regions were assigned a minimum number of IC points. This helps the model make accurate phase summation corrections on one hand, and efficiently handle the boundary interactions of neighboring networks on the other. Applying the denoising loss, it is possible to handle the simulation with full focus on interfacial regions. However, our initial trials have shown that the grain/no-grain regions require implicit con-

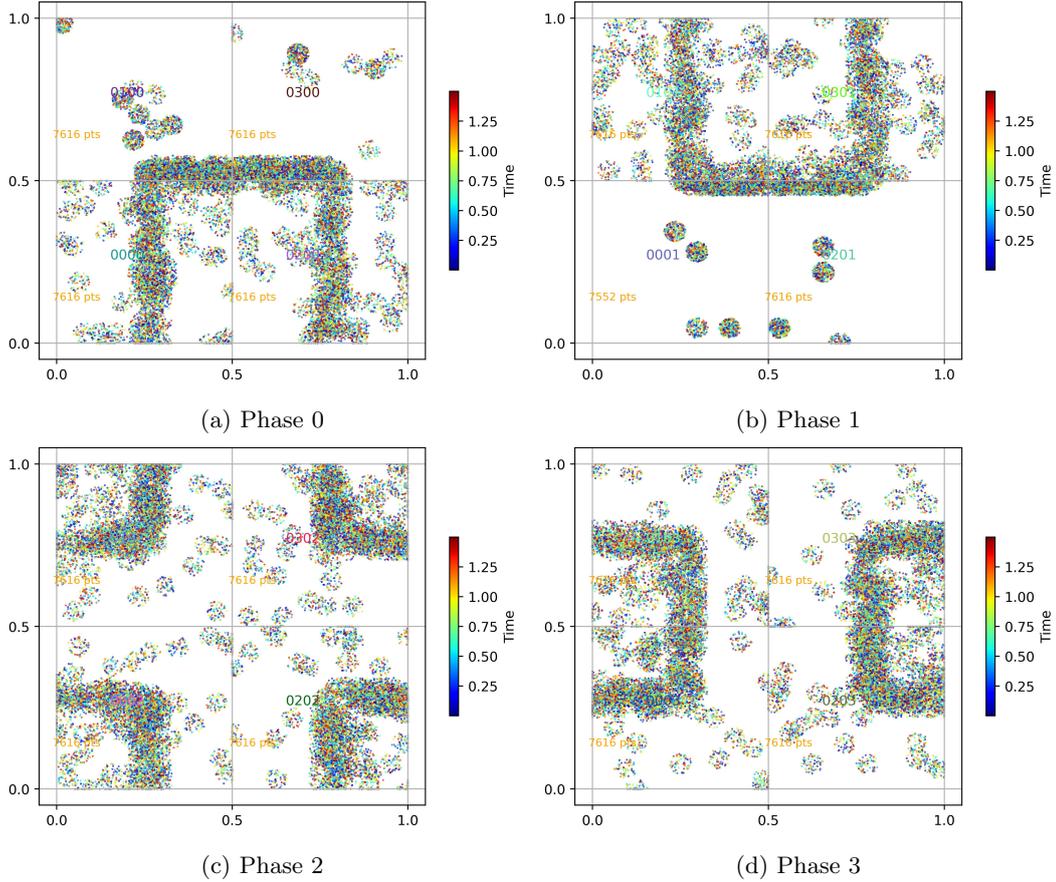

Figure C.1: Sets of the collocation points for the different NNs handling the four phases in different batches.

sideration and specific modifications within the code. These modifications concern functions handling the boundary interactions between NNs to incorporate consideration for the boundary prediction of the grain/no-grain lines, albeit implicitly (without training data). Once achieved, the grid size can be significantly increased, allowing for the exploration of more complex microstructures and the addition of a third dimension, thereby enhancing the dimensionality of the problem.

## F   $PINN_i$ to *Master* learning transfer

The $PINN_i$s to *Master* learning transfer is illustrated in Figure F.1 and can also be referred to as *Master* to $PINN_i$ since this transfer is circular. The knowledge accumulated over the entire domain can be redistributed to all NNs, enabling the exploration of similar physical problems. The learning is transferred between levels using a pyramidal approach. The base of the pyramid is set to 64, representing 64 subdivisions. The subdivision is then progressively increased until reaching 4 NNs. At this level, once the 4NNs are trained, one selected NN with the minimum loss is chosen as a candidate to handle the entire domain and becomes a Master *PINN*. This approach allows for continuous coverage of the entire domain. However, as previously reported in the Discussion section of the related manuscript, our experiments highlighted difficulties in optimizing weights, particularly using the LBFGS optimizer, due to the complexity of the problem for a single NN. Handling the upper level using 4 NNs, interacting at the boundaries, was found to be very efficient. In the technical demonstration found in the GitHub repository (link in the main paper), a demonstration with up to 16 batches/phase (64 initial $PINN_i$s) is successfully benchmarked for the simulation of the triple junction.

# G First findings about training using a single 1NN and the requirement for networking

Figure F.2 a PINN solution without multi-networking. This solution, configured with 1 NN, 6 Hidden Layers (HL), and 128 neurons per HL, without domain decomposition (employing one Network), illustrates a significant discrepancy between the prediction and the theoretical model. This discrepancy persists despite the simplified conditions compared to the benchmark 2 of grain shrinking (dimensionless grain radius of 0.3 instead of 0.4 and a driving force of -100 instead of -250). This highlights the necessity for domain decomposition to achieve accurate solutions for this physical problem.

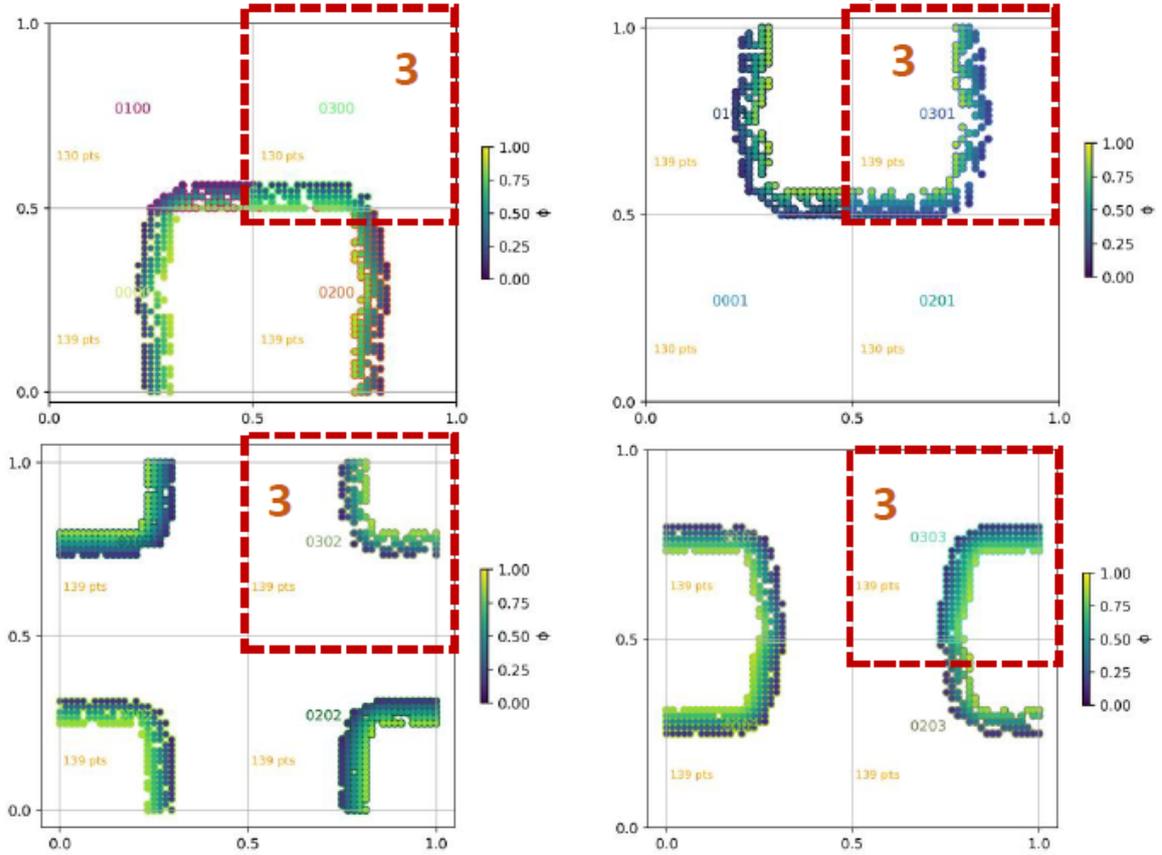

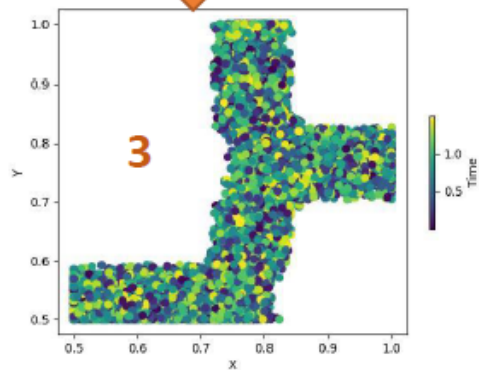

Figure C.2: Illustration of the common batch 3: In this example, the $PINN_i$ instances 0300, 0301, 0302, and 0303, each handling phases 0, 1, 2, and 3, respectively, are assigned a shared set of collocation points. This operation is concurrently performed for other batches. We highlight that out-interface points are not considered during the pyramidal training.

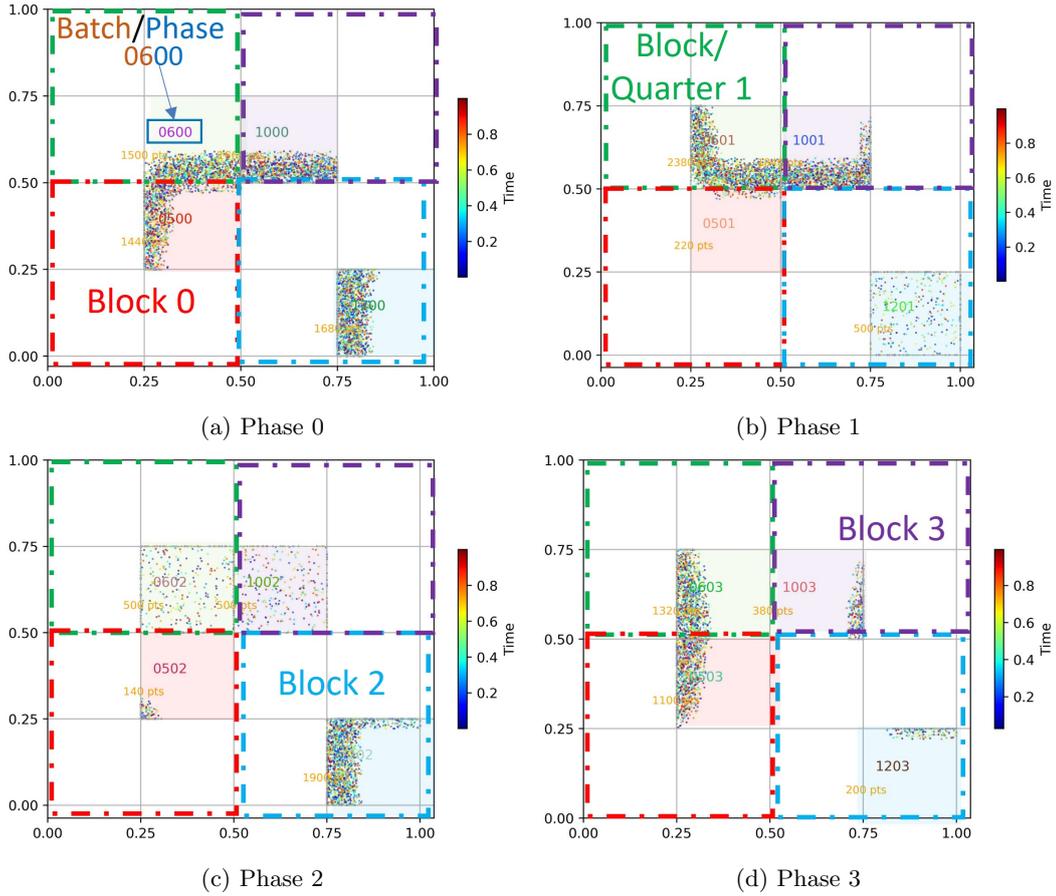

Figure D.1: Bottom level of the pyramidal approach and selected batches for the "pre-training". In Block 0, the $PINN_{0500}$ (Batch 05 - Phase 0) is chosen. In subsequent phases within Block 0, $PINN_i$s 0501, 0502, and 0503, are selected, ensuring they share the same batch number as $PINN_{0500}$, in order to systematically construct the common batches as shown in Figure C.2. This selection process is consistently applied to other Blocks as well. Consequently, a set of 16 $PINN_i$s is selected and placed in the Basket.

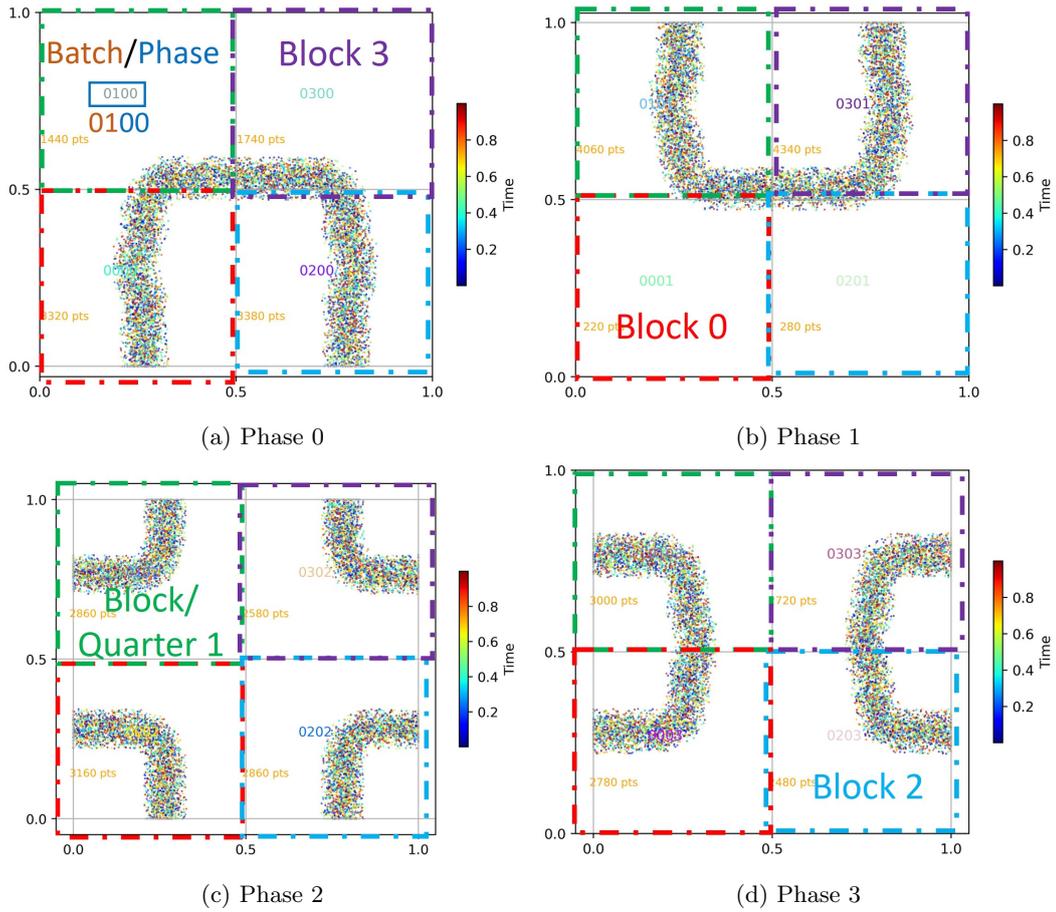

Figure D.2: Upper level of the pyramidal approach: the training is conducted within the full spatial domain for each Block. Subsequently, continuity between boundaries is addressed. The training is then warm-started, wherein each $PINN_i$ receives the weights (and thus accumulates learning) from the selected $PINN_i$ during the pre-training phase. For example, in Block 0, $PINN_{0001}$ (responsible for phase 1 in batch 0) receives the weights from $PINN_{0501}$.

10

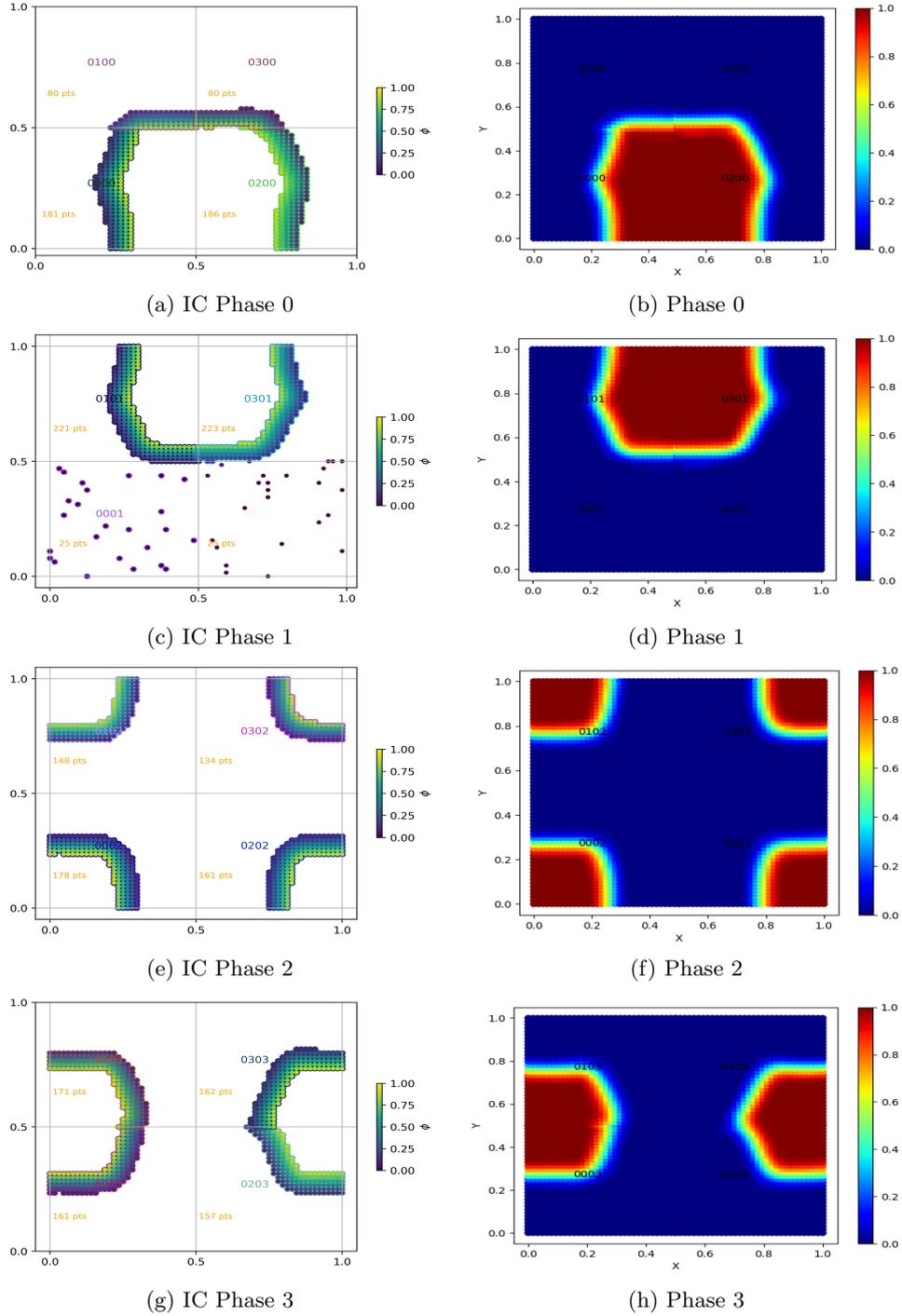

Figure E.1: State of the triple junction at approximately the middle of the simulation ($t_{min} = 39$), with full attention to interfacial areas (out-interface regions are excluded). Here, we note that the code, in its current state, allocates a minimum number of points for a batch where there are no interfacial points (as in batches 0 and 2 of phase 1). Additionally, the denoising loss plays a crucial role in addressing corrections for the extrapolations of the solution to the grain/no-grain regions. This benchmark is available within the GitHub repository (link in the main paper).

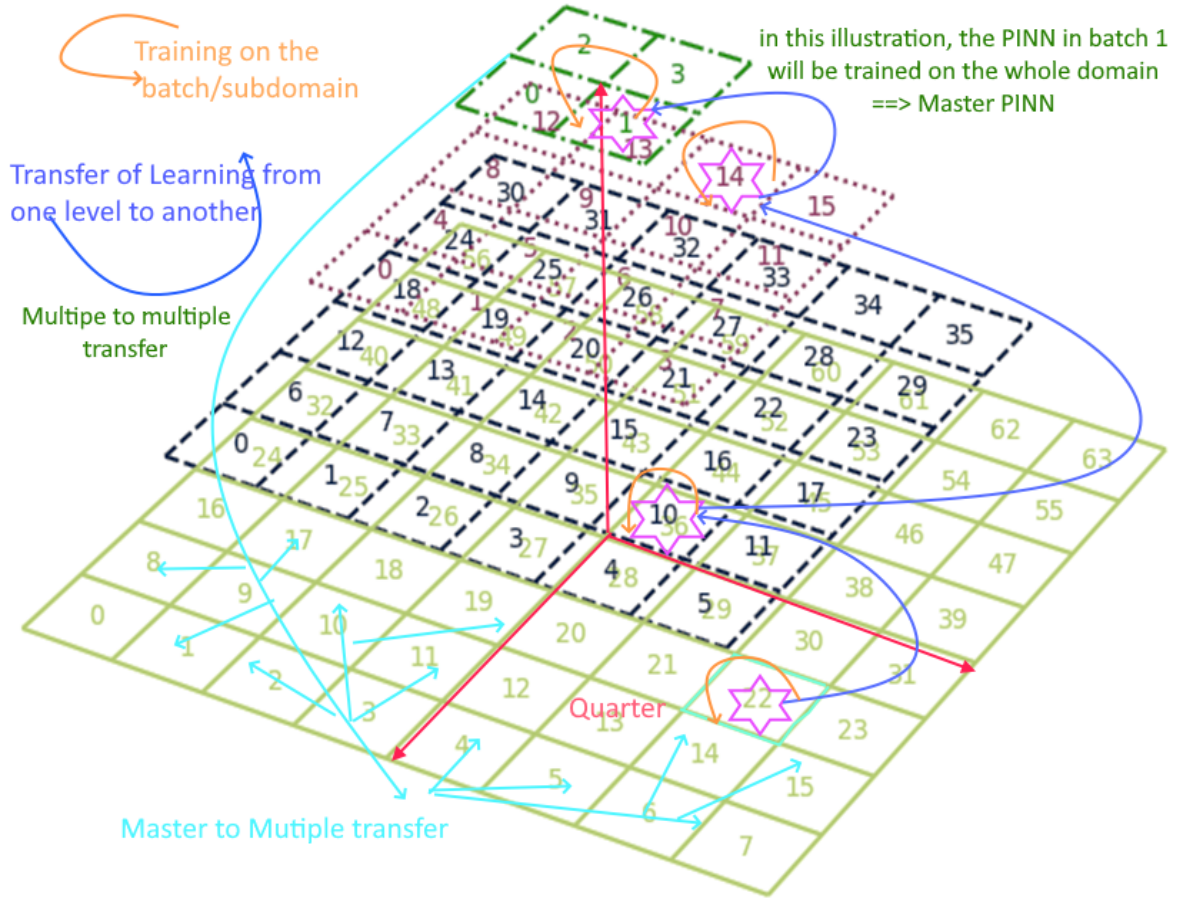

Figure F.1: Illustration of the 'Multiple to Master' and 'Master to Multiple' Transfer of the learning. In this illustrative example (could be also called "Master to $PINN_i$s - $PINN_i$s to Master" pyramid), the base consists of 64 batches, with each batch/ containing one $PINN_i$ for training. Each level is divided into four quarters or blocks. For each quarter batch, only the $PINN_i$s containing interfacial points are pre-selected. Then, a given $PINN_i$ (for example, in batch 22) is selected to transfer its learning to the subsequent level within the same quarter; this selection could be based on the percentage of interfacial points, or the $PINN_i$ with the minimum loss after training the whole set of $PINNi$. For the next level (36 batches), all $PINN_i$s within each quarter receive learning from the selected $PINN_i$ from the previous level within the same quarter. Then, the same principle of preselection occurs; for example, the $PINN_i$ with batch 10 is selected. This mechanism is repeated until reaching the upper level of the pyramid. Here, the selected $PINN_i$ will be trained not only in its batch but within the global domain, serving as the *Master PINN*. In the technical demonstration found in the GitHub repository (link in the main paper), a demonstration with up to 16 batches ($PINN_i$s) is successfully benchmarked for the simulation of the triple junction.

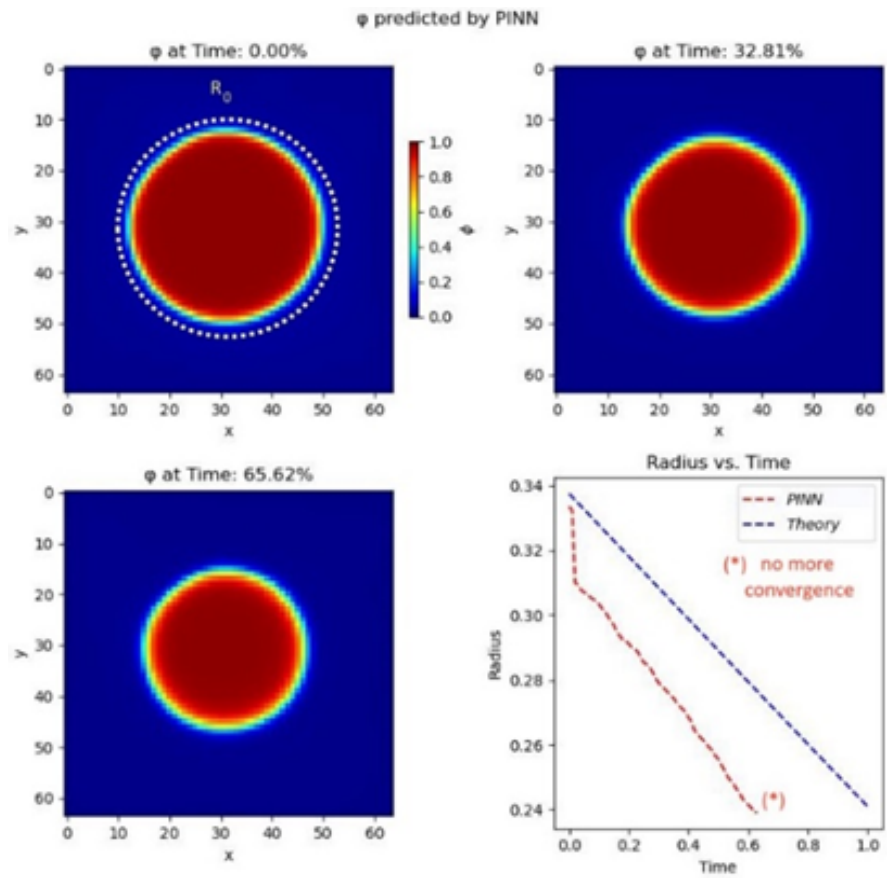

Figure F.2: Presenting to the reader a PINN solution without multi-networking in the supplementary material with reference to this in the article.



\end{document}